\documentclass[11pt]{article}
\usepackage[utf8]{inputenc}
\usepackage{mathtools}
\usepackage{amsmath}
\usepackage{amsthm}
\usepackage{amssymb}
\usepackage[margin=1.25in]{geometry}
\usepackage{setspace}
\usepackage{esint}
\usepackage[authoryear]{natbib}
\setcitestyle{round}

\usepackage[normalem]{ulem}
\usepackage[dvipsnames]{xcolor}

\usepackage[
 bookmarks=false,
 breaklinks=false,
 pdfborder={0 0 0},
 colorlinks=true,
 linkcolor=Maroon,
 citecolor=NavyBlue,
 pdfauthor={Miguel A.\ Delgado and Julius Vainora},
 pdftitle={Conditional Distribution Specification Testing Based on Data-Dependent Partitions}]{hyperref}

\makeatletter
\theoremstyle{plain}
\newtheorem{thm}{\protect\theoremname}
\newtheorem{lemma}{Lemma}

\theoremstyle{remark}
\newtheorem{rem}{\protect\remarkname}
\theoremstyle{definition}

\usepackage{setspace}
\usepackage{makecell}
\usepackage{calc}
\usepackage{rotating}
\usepackage{changepage}
\usepackage{multirow}
\usepackage{booktabs}
\usepackage{threeparttable}
\usepackage{array}
\newcolumntype{H}{@{}>{\setbox0=\hbox\bgroup}c<{\egroup}@{}}
\newcolumntype{R}{@{}>{\setbox0=\hbox\bgroup}c<{\egroup}}
\DeclareMathOperator{\Var}{\operatorname{Var}}
\DeclareMathOperator{\E}{\mathbb{E}}
\DeclareMathOperator{\rank}{rank}
\DeclareMathAlphabet{\mcc}{OMS}{cmsy}{m}{n}
\DeclarePairedDelimiterX{\UB}[1]{[}{]}{#1}
\newcommand*{\unit}{1}
\newcommand{\bs}{\boldsymbol}
\newcommand*{\bt}{\bs{\theta}}
\newcommand*{\mbb}{\mathbb}
\usepackage[normalem]{ulem}

\author{%
\begin{tabular}{cc}
Miguel A.\ Delgado\thanks{Department of Economics, Universidad Carlos III de Madrid. E-mail address:\nobreakspace\mbox{\href{mailto:delgado@est-econ.uc3m.es}{delgado@est-econ.uc3m.es}}. Research funding from Ministerio de Ciencia, Innovaci\'on y Universidades (MICIU), Agencia Estatal de Investigaci\'on (AEI, 10.13039/501100011033), grant numbers CEX2021-001181-M and PID2021-125178NB-I00, and Comunidad de Madrid, grant EPUC3M11 (V PRICIT), is gratefully acknowledged.}
&
Julius Vainora\thanks{Faculty of Economics, University of Cambridge. E-mail address: \href{mailto:jv429@cam.ac.uk}{jv429@cam.ac.uk}}
\tabularnewline
\tabularnewline
\end{tabular}}

\title{%
 Conditional Distribution Specification Testing \\ Based on Data-Dependent
 Partitions}

\providecommand{\theoremname}{Theorem}
\providecommand{\remarkname}{Remark}

\makeatother

\begin{document}
\maketitle 
\begin{abstract}
This article introduces a Pearson-type goodness-of-fit test for the
parametric specification of conditional distribution models with continuous
responses. Under correct specification, the Rosenblatt transform is
uniformly distributed on $[0,1]$ conditionally on the explanatory
variables. The test exploits this characterization by cross-classifying
the transformed observations and the explanatory variables according
to partitions of $[0,1]$ and their support, respectively. The resulting
Pearson statistic has a chi-squared limiting distribution with known
degrees of freedom and detects local alternatives converging to the null
at the $n^{-1/2}$ rate. These results remain valid for the class of
data-dependent partitions considered. Monte Carlo simulations
indicate accurate size control and favorable power relative to existing
bootstrap-based tests, particularly in higher-dimensional settings.
\end{abstract}



\section{Introduction}

We propose a Pearson-type goodness-of-fit test for assessing the parametric
specification of conditional cumulative distribution function (CDF)
models for continuous responses. The test exploits the fact that,
for any random vector $(Y,\boldsymbol{X}^{\text{\textsc{t}}})^{\text{\textsc{t}}}$
with continuous conditional CDF $F$, the Rosenblatt transform $U_{0}=F(Y\mid\boldsymbol{X})$
is uniformly distributed on $[0,1]$ conditionally on $\boldsymbol{X}$.
The proposed procedure examines whether the transform computed under
the specified model satisfies this characterization.

The data are grouped into an $L\times J$ contingency table by cross-classifying
the Rosenblatt-transformed observations into the $L$ subintervals
of a partition of $[0,1]$ and the observations of $\boldsymbol{X}$
into the $J$ subsets of a partition of its support. Under the null
hypothesis, the cell probabilities equal the products of the corresponding
marginal probabilities, while the marginal probabilities associated
with the Rosenblatt transform equal the lengths of the $[0,1]$ subintervals.
Consequently, the cell probabilities satisfy restrictions implied
by the conditional CDF model, and the resulting contingency-table
counts follow a multinomial distribution with restricted parameters.
The corresponding Lagrange multiplier (LM) statistic, which is equivalent to the Pearson statistic for these multinomial restrictions, is analogous to the classical Pearson goodness-of-fit test for marginal distributions.

The unrestricted multinomial model has $LJ-1$ free parameters, whereas
the restricted model has $J-1+p$: the $J-1$ marginal probabilities
that $\boldsymbol{X}$ falls into $J-1$ of the $J$ subsets, together
with the $p$ unknown parameters of the model. The LM/Pearson statistic
is therefore asymptotically distributed as $\chi_{J(L-1)-p}^{2}$
under correct specification.

We develop an LM/Pearson-type specification test for conditional CDF
models and establish its validity for the class of data-dependent
partitions considered here. Our analysis adapts the treatment of
data-dependent cells in \citet{pollard1979general} within the framework
of \citet{birch1964new}. General chi-squared specification tests with
covariates and data-dependent cells were developed by
\citet{andrews1988chi,andrews1988theory}. Our procedure exploits the
conditional uniformity of the Rosenblatt transform to express
conditional-CDF specification as multinomial restrictions on an
$L\times J$ contingency table. This structure leads to a feasible
Pearson statistic with a standard $\chi_{J(L-1)-p}^{2}$ null distribution,
with parameter estimation reflected in the limiting degrees
of freedom.

There is an extensive literature on specification testing of conditional distributions. Omnibus empirical-process tests include \citet{andrews1997conditional}, based on a conditional Kolmogorov statistic with bootstrap critical values, and the asymptotically distribution-free procedures of \citet{bai2003testing} for dynamic location-scale models and \citet{delgado2008distribution} for general conditional distributions. \citet{bai2003testing} uses the martingale transformation for univariate empirical processes proposed by \citet{khmaladze1982martingale}, whereas \citet{delgado2008distribution} exploit the \citet{Rosenblatt1952} transform and the martingale transformation for multivariate empirical processes of \citet{khmaladze1993}. Martingale transforms are difficult to implement, particularly in the multivariate case. An alternative smoothing approach was proposed by \citet{zheng2000consistent}, who used kernel estimation and a Kullback–Leibler criterion to compare parametric and nonparametric conditional distributions. Such smoothing-based tests typically detect local alternatives at rates slower than $n^{-1/2}$ and require a bandwidth choice. Their finite-sample size properties can also be sensitive to the degree of smoothing, and bootstrap methods are frequently used for implementation. They may, however, have advantages against increasingly localized or high-frequency departures; see \citet{Rosenblatt1975}.

More directly related to our grouped-data approach, \citet{mcfadden1974} used chi-squared procedures to assess multinomial choice models under product-multinomial sampling with fixed regressors. When parameters are instead estimated from ungrouped microdata, the conventional Pearson statistic does not generally retain its chi-squared null distribution. \citet{heckman1984chi} addressed this problem by constructing a Wald statistic based on the covariance-adjusted difference between observed and predicted frequencies. Related procedures were developed for discrete-choice models by \citet{horowitz1985}. In the general Wald
formulation of \citet{andrews1988chi,andrews1988theory}, the limiting degrees of freedom are given by the rank of the asymptotic covariance matrix of the frequency residuals, which is typically singular. Heckman and Andrews also discuss important cases in which this rank, and hence the degrees of freedom, can be characterized a priori. For the conditional-CDF problem considered here, we establish that when the parameters are estimated by conditional maximum likelihood from the ungrouped observations, the rank is $J(L-1)$: parameter estimation introduces no additional degeneracy beyond the $J$ adding-up restrictions. Consequently, under the conditions of Section~\ref{sec:wald-tests}, the Wald statistic has the standard $\chi_{J(L-1)}^{2}$ null distribution.

We also characterize the power of the feasible Pearson test under fixed and contiguous
alternatives. With fixed $L$ and $J$, the test is not omnibus: it is consistent against
fixed alternatives that change the selected cell probabilities, but alternatives that leave
those probabilities unchanged are not detected by that partition. This directional feature
is the cost of retaining a simple fixed-dimensional chi-squared test that does not
require resampling. The Monte Carlo results indicate that this trade-off can be favorable,
particularly when the dimension of the conditioning vector is moderately large. For
$n^{-1/2}$ local departures, the statistic has a noncentral chi-squared limit whose
noncentrality parameter is determined by the component of the induced cell-probability
drift that is not absorbed by parameter estimation. This representation also shows
how the choice of partition affects local power. The asymptotic theory treats the dimension
$k$ of the conditioning vector as fixed. Allowing $k$ to diverge with $n$ would require a
different asymptotic framework and is beyond the scope of the paper.

The remainder of the paper is organized as follows. Section~\ref{sec:testing-procedure}
develops the partition-based Pearson test and studies an infeasible
grouped-data statistic constructed from the Rosenblatt transform evaluated
at the unknown true parameter. Although infeasible in practice, this
procedure characterizes the first-order effect of parameter estimation
as an orthogonal projection and provides the theoretical benchmark
for the feasible implementation. Section~\ref{sec:data-dependent-grouping}
extends the analysis to data-dependent partitions, showing that the
additional randomness has no first-order effect on the asymptotic
distribution. Section~\ref{sec:feasible-test} develops a one-step
scoring procedure based on a preliminary $\sqrt{n}$-consistent estimator
and shows that the resulting feasible Pearson statistic is asymptotically
equivalent to its infeasible counterpart. Section~\ref{sec:wald-tests}
studies the Wald-type statistics. Section~\ref{sec:power} studies consistency under
fixed alternatives and derives the local asymptotic power of the feasible Pearson test.
Section~\ref{sec:monte-carlo} reports Monte Carlo evidence,
Section~\ref{sec:final-remarks} concludes, and the Appendix contains the proofs.

\section{The Partition-Based Test}

\label{sec:testing-procedure}

The data consist of independent and identically distributed observations
$\left\{ (Y_{i},\boldsymbol{X}_{i}^{\text{\textsc{t}}})^{\text{\textsc{t}}}\right\} _{i=1}^{n}$, drawn
from the $\mathbb{R}^{1+k}$-valued random vector $(Y,\boldsymbol{X}^{\text{\textsc{t}}})^{\text{\textsc{t}}}$
with joint probability measure $P$. Let $F(y\mid\boldsymbol{x})=P(Y\leq y\mid\boldsymbol{X}=\boldsymbol{x})$
denote the conditional CDF of $Y$ given $\boldsymbol{X}=\boldsymbol{x}$,
and let $P_{\boldsymbol{X}}$ denote the marginal probability measure
of $\boldsymbol{X}$.

Consider the parametric family of conditional CDFs 
\[
\mathcal{F}=\left\{ (y,\boldsymbol{x})\mapsto F_{\boldsymbol{\theta}}(y\mid\boldsymbol{x}):\boldsymbol{\theta}\in\Theta\right\} ,
\]
where $\Theta\subset\mathbb{R}^{p}$ is the parameter space. The hypothesis
of interest is 
\[
H_{0}:F\in\mathcal{F}.
\]

We impose the following basic conditions. 
\begin{description}
\item [{A0:}] ~ 
\begin{enumerate}
\item[(i)] The map $(y,\boldsymbol{x},\boldsymbol{\theta})\mapsto F_{\boldsymbol{\theta}}(y\mid\boldsymbol{x})$
is jointly measurable. 
\item[(ii)] For every $(\boldsymbol{x},\boldsymbol{\theta})$, the map $y\mapsto F_{\boldsymbol{\theta}}(y\mid\boldsymbol{x})$
is continuous. 
\end{enumerate}
\end{description}
For any $\boldsymbol{\theta}\in\Theta$, define $U(\boldsymbol{\theta})=F_{\boldsymbol{\theta}}(Y\mid\boldsymbol{X})$.
Under $H_{0}$, $U_{0}\coloneqq U(\boldsymbol{\theta}_{0})$
is uniformly distributed on $[0,1]$ conditionally on $\boldsymbol{X}$.
Equivalently, $U_{0}$ is uniformly distributed on $[0,1]$ and independent
of $\boldsymbol{X}$. The proposed procedure exploits this characterization
by grouping the transformed observations and the explanatory variables
into a contingency table and applying a Pearson-type statistic.

We construct a contingency table with $LJ$ cells and require $J(L-1)>p$,
so that the limiting number of degrees of freedom under the composite
null hypothesis is positive. Let 
\[
\mathcal{U}=\left\{ I_{\ell}=(\vartheta_{\ell-1},\vartheta_{\ell}]:0=\vartheta_{0}<\vartheta_{1}<\cdots<\vartheta_{L}=1,\quad\ell=1,\ldots,L\right\} 
\]
be a partition of $[0,1]$, with the convention that the first cell
contains zero if necessary. Define $v_{\ell}=\vartheta_{\ell}-\vartheta_{\ell-1}$,
and $\boldsymbol{v}=(v_{1},\ldots,v_{L})^{\text{\textsc{t}}}$. Let $\mathcal{A}=\{A_{1},\ldots,A_{J}\}$ be a measurable partition
of $\mathbb{R}^{k}$ satisfying $\bigcup_{j=1}^{J}A_{j}=\mathbb{R}^{k}$, ${A_{j}\cap A_{j'}=\varnothing}$ for $j\neq j'$. Define
the indicator vectors $\boldsymbol{1}_{\mathcal{U}}(u)=\left(1_{I_{1}}(u),\ldots,1_{I_{L}}(u)\right)^{\text{\textsc{t}}}$
and $\boldsymbol{1}_{\mathcal{A}}(\boldsymbol{x})=\left(1_{A_{1}}(\boldsymbol{x}),\ldots,1_{A_{J}}(\boldsymbol{x})\right)^{\text{\textsc{t}}}$,
where $1_{C}$ denotes the indicator function of the set $C$. For each $\boldsymbol{\theta}\in\Theta$, the $(\ell,j)$th cell of
the contingency table is 
\[
D_{\ell j}(\boldsymbol{\theta})=\left\{ (y,\boldsymbol{x})\in\mathbb{R}^{1+k}:F_{\boldsymbol{\theta}}(y\mid\boldsymbol{x})\in I_{\ell},\quad\boldsymbol{x}\in A_{j}\right\} .
\]

Define 
\[
\boldsymbol{\Phi}_{\boldsymbol{\theta}}^{F}(\boldsymbol{x})=\mathbb{E}\left[\left.\boldsymbol{1}_{\mathcal{U}}\{U(\boldsymbol{\theta})\}\right|\boldsymbol{X}=\boldsymbol{x}\right],
\]
whose $\ell$th component is 
\[
\Phi_{\boldsymbol{\theta}\ell}^{F}(\boldsymbol{x})=P\left\{ F_{\boldsymbol{\theta}}(Y\mid\boldsymbol{x})\in I_{\ell}\mid\boldsymbol{X}=\boldsymbol{x}\right\} .
\]
The corresponding $LJ\times1$ cell-probability vector is 
\[
\boldsymbol{\pi}(\boldsymbol{\theta})=\mathbb{E}\left[\boldsymbol{\Phi}_{\boldsymbol{\theta}}^{F}(\boldsymbol{X})\otimes\boldsymbol{1}_{\mathcal{A}}(\boldsymbol{X})\right].
\]

Let 
\[
\boldsymbol{\Pi}=\left\{ \boldsymbol{\pi}\in[0,1]^{LJ}:\sum_{\ell=1}^{L}\sum_{j=1}^{J}\pi_{\ell j}=1\right\} 
\]
denote the multinomial parameter space, where $\pi_{\ell j}$ is the
probability of the $(\ell,j)$th cell.

The $LJ\times1$ vector of cell counts is 
\[
\widehat{\boldsymbol{O}}(\boldsymbol{\theta})=\sum_{i=1}^{n}\boldsymbol{1}_{\mathcal{U}}\left\{ U_{i}(\boldsymbol{\theta})\right\} \otimes\boldsymbol{1}_{\mathcal{A}}(\boldsymbol{X}_{i}),\qquad U_i(\boldsymbol{\theta})=F_{\boldsymbol{\theta}}(Y_{i}\mid\boldsymbol{X}_{i}).
\]
Its $(\ell,j)$th component, $\widehat{O}_{\ell j}(\boldsymbol{\theta})$,
is the number of observations in $D_{\ell j}(\boldsymbol{\theta})$.
Since the observations are independent and identically distributed,
$\widehat{\boldsymbol{O}}(\boldsymbol{\theta})\sim\operatorname{Multinomial}\left(n,\boldsymbol{\pi}(\boldsymbol{\theta})\right)$.

Under $H_{0}$, at the true parameter value $\boldsymbol{\theta}_{0}$,
$\Phi_{\boldsymbol{\theta}_{0}\ell}^{F}(\boldsymbol{x})=v_{\ell}$,
$\ell=1,\ldots,L$, for $P_{\boldsymbol{X}}$-almost every $\boldsymbol{x}$.
Consequently, $\boldsymbol{\pi}(\boldsymbol{\theta}_{0})=\boldsymbol{v}\otimes\boldsymbol{q}\eqqcolon\boldsymbol{\pi}_{0}$,
$\boldsymbol{q}=(q_{1},\ldots,q_{J})^{\text{\textsc{t}}}$, $q_{j}=P_{\boldsymbol{X}}(A_{j})$.
Because the cutpoints defining $\mathcal{U}$ are strictly ordered,
$v_{\ell}>0$ for every $\ell$. We also assume that $q_{j}>0$ for
every $j$. Hence every component of $\boldsymbol{\pi}_{0}=\boldsymbol{v}\otimes\boldsymbol{q}$
is strictly positive.

For the fixed partition $(\mathcal{U},\mathcal{A})$, the test exploits
the necessary implication of $H_{0}$, $\boldsymbol{\pi}(\boldsymbol{\theta}_{0})=\boldsymbol{v}\otimes\boldsymbol{q}$.
It is therefore designed to detect alternatives satisfying $H_{1}^{*}:\boldsymbol{\pi}(\boldsymbol{\theta})\neq\boldsymbol{v}\otimes\boldsymbol{q}$
for every $\boldsymbol{\theta}\in\Theta$. The resulting test is local to the chosen partition and is not omnibus. Omnibus consistency would require a sequence of increasingly fine partitions, which is beyond the scope of this paper.

\subsection{The statistic with known parameters}

Suppose initially that $\boldsymbol{\theta}_{0}$ is known. Let $\widehat{\boldsymbol{p}}(\boldsymbol{\theta})=n^{-1}\widehat{\boldsymbol{O}}(\boldsymbol{\theta})$ denote the vector of relative cell frequencies. Then $\widehat{\boldsymbol{p}}(\boldsymbol{\theta}_{0})$
is the unrestricted maximum likelihood estimator (MLE) of the multinomial
cell-probability vector. The multinomial log-likelihood for a generic
cell-probability vector $\boldsymbol{\pi}$ is 
\[
\widehat{\ell}(\boldsymbol{\pi})\propto\sum_{\ell=1}^{L}\sum_{j=1}^{J}\widehat{O}_{\ell j}(\boldsymbol{\theta}_{0})\log\pi_{\ell j},\qquad\boldsymbol{\pi}\in\boldsymbol{\Pi}.
\]

Thus, the restricted MLE of $\boldsymbol{\pi}_{0}=\boldsymbol{v}\otimes\boldsymbol{q}$
is $\widehat{\boldsymbol{\pi}}_{0}=\boldsymbol{v}\otimes\widehat{\boldsymbol{q}}$,
where $\widehat{\boldsymbol{q}}=n^{-1}\sum_{i=1}^{n}\boldsymbol{1}_{\mathcal{A}}(\boldsymbol{X}_{i})=(\widehat{q}_{1},\ldots,\widehat{q}_{J})^{\text{\textsc{t}}}$
and $\widehat{q}_{j}=n^{-1}\sum_{i=1}^{n}1_{A_{j}}(\boldsymbol{X}_{i})$.
Moreover, since $\widehat{\boldsymbol{\pi}}_{0}=\boldsymbol{v}\otimes\widehat{\boldsymbol{q}}$
and $v_{\ell}>0$ for every $\ell$, all components of $\widehat{\boldsymbol{\pi}}_{0}$
are strictly positive whenever every covariate cell is nonempty. For
fixed $J$ and $q_{j}>0$, $j=1,\ldots,J$, 
\[
P\left(\min_{1\leq j\leq J}\widehat{q}_{j}=0\right)\leq\sum_{j=1}^{J}P(\widehat{q}_{j}=0)=\sum_{j=1}^{J}(1-q_{j})^{n}=o(1).
\]
Therefore, all components of $\widehat{\boldsymbol{\pi}}_{0}$ are
strictly positive with probability approaching one.

For $\boldsymbol{\pi}$ in the relative interior of $\boldsymbol{\Pi}$,
define the multinomial score residual 
\[
\widehat{\boldsymbol{\Psi}}(\boldsymbol{\pi})=\operatorname{diag}^{-1}(\boldsymbol{\pi})\left[\widehat{\boldsymbol{p}}(\boldsymbol{\theta}_{0})-\boldsymbol{\pi}\right].
\]
Since the components of $\widehat{\boldsymbol{p}}(\boldsymbol{\theta}_{0})$
and $\boldsymbol{\pi}$ both sum to one, $\boldsymbol{\pi}^{\text{\textsc{t}}}\widehat{\boldsymbol{\Psi}}(\boldsymbol{\pi})=0$.
Let 
\[
\mathcal{I}^{-1}(\boldsymbol{\pi})=\operatorname{diag}(\boldsymbol{\pi})-\boldsymbol{\pi}\boldsymbol{\pi}^{\text{\textsc{t}}}
\]
denote the inverse Fisher information matrix on the multinomial simplex
\citep{birch1964new,pollard1979general}. The LM statistic for $H_{0}$ is therefore 
\begin{align*}
\widehat{LM}_{0}&=n\widehat{\boldsymbol{\Psi}}^{\text{\textsc{t}}}(\widehat{\boldsymbol{\pi}}_{0})\mathcal{I}^{-1}(\widehat{\boldsymbol{\pi}}_{0})\widehat{\boldsymbol{\Psi}}(\widehat{\boldsymbol{\pi}}_{0}) \\
&=n\sum_{\ell=1}^{L}\sum_{j=1}^{J}\frac{\left[\widehat{p}_{\ell j}(\boldsymbol{\theta}_{0})-v_{\ell}\widehat{q}_{j}\right]^{2}}{v_{\ell}\widehat{q}_{j}}\allowbreak\eqqcolon\widehat{X}_{0}^{2}.
\end{align*}
Thus, the LM statistic algebraically coincides with the Pearson statistic
$\widehat{X}_{0}^{2}$. It is the weighted squared distance between
the observed relative cell frequencies, $\widehat{p}_{\ell j}(\boldsymbol{\theta}_{0})$,
and their expected counterparts under $H_{0}$, $v_{\ell}\widehat{q}_{j}$,
with weights $(v_{\ell}\widehat{q}_{j})^{-1}$. Hence, the familiar
Pearson statistic admits, as in the classical case, an exact LM interpretation for our conditional-distribution specification problem.

The corresponding likelihood-ratio statistic is 
\[
\widehat{G}_{0}^{2}:=2\sum_{\ell=1}^{L}\sum_{j=1}^{J}\widehat{O}_{\ell j}(\boldsymbol{\theta}_{0})\log\left[\frac{\widehat{O}_{\ell j}(\boldsymbol{\theta}_{0})}{nv_{\ell}\widehat{q}_{j}}\right]=2n\sum_{\ell=1}^{L}\sum_{j=1}^{J}\widehat{p}_{\ell j}(\boldsymbol{\theta}_{0})\log\left[\frac{\widehat{p}_{\ell j}(\boldsymbol{\theta}_{0})}{v_{\ell}\widehat{q}_{j}}\right],
\]
where a cell with zero observed count contributes zero.

Define $\widehat{\boldsymbol{\xi}}_{0}=\sqrt{n}\left[\widehat{\boldsymbol{p}}(\boldsymbol{\theta}_{0})-\widehat{\boldsymbol{\pi}}_{0}\right]$, $\boldsymbol{\Lambda}_{0}=\operatorname{diag}^{1/2}(\boldsymbol{\pi}_{0})=\operatorname{diag}^{1/2}(\boldsymbol{v})\otimes\operatorname{diag}^{1/2}(\boldsymbol{q})$, and $\boldsymbol{\Omega}=\left(\boldsymbol{I}_{L}-\sqrt{\boldsymbol{v}}\sqrt{\boldsymbol{v}}^{\text{\textsc{t}}}\right)\otimes\boldsymbol{I}_{J}$,
where $\sqrt{\boldsymbol{v}}=(\sqrt{v_{1}},\ldots,\sqrt{v_{L}})^{\text{\textsc{t}}}$. Under $H_{0}$, $\operatorname{Var}(\widehat{\boldsymbol{\xi}}_{0})=\left[\operatorname{diag}(\boldsymbol{v})-\boldsymbol{v}\boldsymbol{v}^{\text{\textsc{t}}}\right]\otimes\operatorname{diag}(\boldsymbol{q})=\boldsymbol{\Lambda}_{0}\boldsymbol{\Omega}\boldsymbol{\Lambda}_{0}$.

By the law of large numbers and the central limit theorem,
\begin{equation}
\widehat{\boldsymbol{\Lambda}}_{0}\sqrt{n}\,\widehat{\boldsymbol{\Psi}}(\widehat{\boldsymbol{\pi}}_{0})=\widehat{\boldsymbol{\Lambda}}_{0}^{-1}\widehat{\boldsymbol{\xi}}_{0}\rightarrow_{d}N_{LJ}(\boldsymbol{0},\boldsymbol{\Omega}),\label{eq:1}
\end{equation}
where $\widehat{\boldsymbol{\Lambda}}_{0}=\operatorname{diag}^{1/2}(\widehat{\boldsymbol{\pi}}_{0})$. The matrix $\boldsymbol{\Omega}$ is symmetric and idempotent, with
$\operatorname{rank}(\boldsymbol{\Omega})=\operatorname{tr}(\boldsymbol{\Omega})=J(L-1)$.
Given $\boldsymbol{\theta}_{0}$, the unrestricted multinomial model
contains $LJ-1$ free parameters, whereas the restriction $\boldsymbol{\pi}_{0}=\boldsymbol{v}\otimes\boldsymbol{q}$
leaves $J-1$ free parameters. The number of restrictions is therefore
$(LJ-1)-(J-1)=J(L-1)$. Consequently, 
\[
\widehat{X}_{0}^{2}=\widehat{\boldsymbol{\xi}}_{0}^{\text{\textsc{t}}}\widehat{\boldsymbol{\Lambda}}_{0}^{-2}\widehat{\boldsymbol{\xi}}_{0}\rightarrow_{d}\chi_{J(L-1)}^{2},\qquad\widehat{G}_{0}^{2}=\widehat{X}_{0}^{2}+o_{P}\left(1\right)\rightarrow_{d}\chi_{J(L-1)}^{2}.
\]

For later comparison with the Wald-type procedure in Section~\ref{sec:wald-tests}, notice
that the Pearson statistic can equivalently be written as 
\[
\widehat{W}_{0}=n\left[\widehat{\boldsymbol{p}}(\boldsymbol{\theta}_{0})-\widehat{\boldsymbol{\pi}}_{0}\right]^{\text{\textsc{t}}}\left(\widehat{\boldsymbol{\Lambda}}_{0}\boldsymbol{\Omega}\widehat{\boldsymbol{\Lambda}}_{0}\right)^{-}\left[\widehat{\boldsymbol{p}}(\boldsymbol{\theta}_{0})-\widehat{\boldsymbol{\pi}}_{0}\right],
\]
where a superscript ``$^-$'' denotes a generalized inverse.

\subsection{Composite hypotheses}

When the parameter is unknown, the transformed observations $U_{i}(\boldsymbol{\theta}_{0})$
cannot be computed. We therefore first analyze an infeasible Pearson test in which the true parameter is assumed known when constructing the
contingency table but unknown in the multinomial restrictions. Although
infeasible, this benchmark reveals the asymptotic effect of parameter
estimation and serves as the basis for the feasible one-step procedure
developed in Section~\ref{sec:feasible-test}.

For each $\boldsymbol{\theta}\in\Theta$, let $P_{\boldsymbol{\theta}}$
denote the (semiparametric) joint probability measure on $\mathbb{R}^{1+k}$
induced by the conditional distribution $F_{\boldsymbol{\theta}}(\cdot\mid\cdot)$
and the marginal distribution $P_{\boldsymbol{X}}$. Thus, for every
measurable set $C\subset\mathbb{R}^{1+k}$,
\[
P_{\boldsymbol{\theta}}(C)=\int_{\mathbb{R}^{k}}\int_{\mathbb{R}}1_{C}(y,\boldsymbol{x})\,F_{\boldsymbol{\theta}}(dy\mid\boldsymbol{x})\,P_{\boldsymbol{X}}(d\boldsymbol{x}).
\]

For $\boldsymbol{\theta}_{1},\boldsymbol{\theta}_{2}\in\Theta$, define
the $L\times1$ vector $\boldsymbol{\Phi}_{\boldsymbol{\theta}_{1},\boldsymbol{\theta}_{2}}(\boldsymbol{x})$,
whose $\ell$th component is 
\[
\Phi_{\boldsymbol{\theta}_{1},\boldsymbol{\theta}_{2},\ell}(\boldsymbol{x})=P_{\boldsymbol{\theta}_{2}}\left\{ F_{\boldsymbol{\theta}_{1}}\left(Y\mid\boldsymbol{x}\right)\in I_{\ell}\mid\boldsymbol{X}=\boldsymbol{x}\right\} .
\]
Thus, $\Phi_{\boldsymbol{\theta}_{1},\boldsymbol{\theta}_{2},\ell}(\boldsymbol{x})$
is the conditional probability, under the model $F_{\boldsymbol{\theta}_{2}}$,
that $U(\boldsymbol{\theta}_{1})\in I_{\ell}$ given $\boldsymbol{X}=\boldsymbol{x}$.
The corresponding unconditional cell-probability vector is $\boldsymbol{\pi}_{\boldsymbol{\theta}_{1}}(\boldsymbol{\theta}_{2})=\mathbb{E}\left[\boldsymbol{\Phi}_{\boldsymbol{\theta}_{1},\boldsymbol{\theta}_{2}}(\boldsymbol{X})\otimes\boldsymbol{1}_{\mathcal{A}}(\boldsymbol{X})\right]$. The conditional expectation of the relative cell-frequency vector, given $\left\{ \boldsymbol{X}_{i}\right\} _{i=1}^{n}$, is
\[
\widehat{\boldsymbol{\pi}}_{\boldsymbol{\theta}_{1}}(\boldsymbol{\theta}_{2})=n^{-1}\sum_{i=1}^{n}\boldsymbol{\Phi}_{\boldsymbol{\theta}_{1},\boldsymbol{\theta}_{2}}(\boldsymbol{X}_{i})\otimes\boldsymbol{1}_{\mathcal{A}}(\boldsymbol{X}_{i}).
\]
Conditional on the observed covariates, the vector $\widehat{\boldsymbol{\pi}}_{\boldsymbol{\theta}_{1}}(\boldsymbol{\theta}_{2})$
gives the model-implied expected relative cell frequencies under $F_{\boldsymbol{\theta}_{2}}$.
Replacing the multinomial probabilities by these conditional expectations
yields the conditional grouped log-likelihood
\[
\widehat{\ell}_{\boldsymbol{\theta}_{1}}(\boldsymbol{\theta}_{2})\propto\sum_{\ell,j}\widehat{O}_{\ell j}(\boldsymbol{\theta}_{1})\log\widehat{\pi}_{\boldsymbol{\theta}_{1},\ell j}(\boldsymbol{\theta}_{2}).
\]

The grouped maximum likelihood estimator is $\widehat{\boldsymbol{\theta}}=\arg\max_{\boldsymbol{\theta}\in\Theta}\widehat{\ell}_{\boldsymbol{\theta}_{0}}(\boldsymbol{\theta})$,
or equivalently, $\widehat{\boldsymbol{\theta}}=\arg\min_{\boldsymbol{\theta}\in\Theta}\widehat{G}_{\boldsymbol{\theta}_{0}}^{2}(\boldsymbol{\theta})$,
where, for $\boldsymbol{\theta}_{1},\boldsymbol{\theta}_{2}\in\Theta$,
\[
\widehat{G}_{\boldsymbol{\theta}_{1}}^{2}(\boldsymbol{\theta}_{2})=2n\sum_{\ell=1}^{L}\sum_{j=1}^{J}\widehat{p}_{\ell j}(\boldsymbol{\theta}_{1})\log\left[\frac{\widehat{p}_{\ell j}(\boldsymbol{\theta}_{1})}{\widehat{\pi}_{\boldsymbol{\theta}_{1},\ell j}(\boldsymbol{\theta}_{2})}\right].
\]
The corresponding Pearson (minimum distance) criterion is 
\[
\widehat{X}_{\boldsymbol{\theta}_{1}}^{2}(\boldsymbol{\theta}_{2})=n\sum_{\ell=1}^{L}\sum_{j=1}^{J}\frac{\left[\widehat{p}_{\ell j}(\boldsymbol{\theta}_{1})-\widehat{\pi}_{\boldsymbol{\theta}_{1},\ell j}(\boldsymbol{\theta}_{2})\right]^{2}}{\widehat{\pi}_{\boldsymbol{\theta}_{1},\ell j}(\boldsymbol{\theta}_{2})}.
\]
The infeasible likelihood-ratio and Pearson statistics are $\widehat{G}_{\boldsymbol{\theta}_{0}}^{2}(\widehat{\boldsymbol{\theta}})$
and $\widehat{X}_{\boldsymbol{\theta}_{0}}^{2}(\widehat{\boldsymbol{\theta}})$,
respectively. They are infeasible because $\boldsymbol{\theta}_{0}$
is used to construct the ``observed'' cells.

\citet{birch1964new} provides minimal regularity conditions
for the limiting distribution of the Pearson statistic in the classical
multinomial case. We adapt those conditions to our setting,
following \citet{pollard1979general}. Henceforth, write $\boldsymbol{\pi}(\boldsymbol{\theta})=\boldsymbol{\pi}_{\boldsymbol{\theta}_{0}}(\boldsymbol{\theta})$,
$\boldsymbol{\Phi}_{\boldsymbol{\theta}}(\boldsymbol{x})=\boldsymbol{\Phi}_{\boldsymbol{\theta}_{0},\boldsymbol{\theta}}(\boldsymbol{x})$.
Then $\boldsymbol{\pi}(\boldsymbol{\theta}_{0})=\boldsymbol{\pi}_{0}=\boldsymbol{v}\otimes\boldsymbol{q}$, while $\boldsymbol{\Phi}_{\boldsymbol{\theta},\boldsymbol{\theta}}(\boldsymbol{X})=\boldsymbol{v}$,
$P_{\boldsymbol{X}}$-a.s., for every $\boldsymbol{\theta}\in\Theta$.

Let $\mathcal{N}_{\epsilon}=\left\{ \boldsymbol{\theta}\in\Theta:\|\boldsymbol{\theta}-\boldsymbol{\theta}_{0}\|\leq\epsilon\right\} $
be the neighborhood of $\boldsymbol{\theta}_{0}$ with radius $\epsilon>0$,
and define $\mathcal{M}=\left\{ \boldsymbol{x}\mapsto\boldsymbol{\Phi}_{\boldsymbol{\theta}}(\boldsymbol{x}):\boldsymbol{\theta}\in\Theta\right\} $
and $\mathcal{M}_{\epsilon}=\left\{ \boldsymbol{x}\mapsto\boldsymbol{\Phi}_{\boldsymbol{\theta}}(\boldsymbol{x}):\boldsymbol{\theta}\in\mathcal{N}_{\epsilon}\right\} $.
For a vector of functions $\boldsymbol{x}\mapsto\boldsymbol{f}(\boldsymbol{x})$,
the $L^{2}(Q)$ norm is defined as $\left\Vert \boldsymbol{f}\right\Vert _{Q,2}=\left[\int\left\Vert \boldsymbol{f}(\boldsymbol{x})\right\Vert ^{2}Q(d\boldsymbol{x})\right]^{1/2}$.
Unlike the classical setting studied by \citet{pollard1979general}, our argument requires conditions on the complexity of $\mathcal{M}$. 
\begin{description}
\item [{A1:}] ~ 
\begin{enumerate}
\item[(i)] The true parameter $\boldsymbol{\theta}_{0}$ is an interior point
of $\Theta$. 
\item[(ii)] For every $\epsilon>0$, $\inf_{\boldsymbol{\theta}\in\left.\Theta\right\backslash \mathcal{N}_{\epsilon}}\left\Vert \boldsymbol{\pi}(\boldsymbol{\theta})-\boldsymbol{\pi}_{0}\right\Vert >0$. 
\item[(iii)] Every component of $\boldsymbol{\pi}_{0}$ is strictly positive and
$J(L-1)>p$. 
\item[(iv)] The map $\boldsymbol{\theta}\mapsto\boldsymbol{\pi}(\boldsymbol{\theta})$
is differentiable at $\boldsymbol{\theta}_{0}$, and $\boldsymbol{\theta}\mapsto\boldsymbol{\Phi}_{\boldsymbol{\theta}}$
is $L^2(P_{\boldsymbol{X}})$-continuous at $\boldsymbol{\theta}_{0}$.
In particular, there exists an $LJ\times p$ matrix $\dot{\boldsymbol{\pi}}_{0}$
such that 
\begin{equation}
\left\Vert \boldsymbol{\pi}(\boldsymbol{\theta})-\boldsymbol{\pi}(\boldsymbol{\theta}_{0})-\dot{\boldsymbol{\pi}}_{0}(\boldsymbol{\theta}-\boldsymbol{\theta}_{0})\right\Vert  =  o\left(\|\boldsymbol{\theta}-\boldsymbol{\theta}_{0}\|\right)\quad\text{ as }\boldsymbol{\theta}\to\boldsymbol{\theta}_{0}.\label{eq:22}
\end{equation}
\item[(v)] The $LJ\times p$ matrix $\dot{\boldsymbol{\pi}}_{0}$ has full column
rank $p$. 
\item[(vi)] $\mathcal{M}$ is $P_{\boldsymbol{X}}$-Glivenko--Cantelli and $\mathcal{M}_{\epsilon}$
is $P_{\boldsymbol{X}}$-Donsker for some $\epsilon>0$. 
\end{enumerate}
\end{description}
The first five conditions are counterparts of the regularity conditions
used by \citet{birch1964new} and \citet{pollard1979general}
for multinomial goodness-of-fit statistics with estimated parameters.
Assumption A1(i) is standard. Assumption A1(ii)
is the global identification condition used to establish consistency
of the grouped estimator, whereas A1(iv)--A1(v)
provide the local smoothness and rank conditions required for the
asymptotic linearization. Assumption A1(iii)
guarantees that the limiting Pearson statistic is well defined and
has a positive number of degrees of freedom. Assumption A1(iv)
contains two distinct local requirements. First, differentiability
of $\boldsymbol{\theta}\mapsto\boldsymbol{\pi}(\boldsymbol{\theta})$
at $\boldsymbol{\theta}_{0}$ provides the first-order expansion used
in the quadratic approximation. Second, $L^{2}(P_{\boldsymbol{X}})$-continuity
of $\boldsymbol{\Phi}_{\boldsymbol{\theta}}$ at $\boldsymbol{\theta}_{0}$
is used together with the local Donsker condition in A1(vi)
to obtain stochastic equicontinuity at the estimated parameter. Assumption
A1(v) is the corresponding local rank/identification
condition. A convenient sufficient condition for both requirements
in A1(iv) is $L^{2}(P_{\boldsymbol{X}})$-differentiability
of $\boldsymbol{\theta}\mapsto\boldsymbol{\Phi}_{\boldsymbol{\theta}}$
at $\boldsymbol{\theta}_{0}$. In particular, suppose that there exists
an $L\times p$ matrix-valued function $\dot{\boldsymbol{\Phi}}_{0}$,
with components in $L^{2}(P_{\boldsymbol{X}})$, such that 
\[
\left\Vert \boldsymbol{\Phi}_{\boldsymbol{\theta}}-\boldsymbol{\Phi}_{\boldsymbol{\theta}_{0}}-\dot{\boldsymbol{\Phi}}_{0}(\boldsymbol{\theta}-\boldsymbol{\theta}_{0})\right\Vert _{P_{\boldsymbol{X}},2}=o\!\left(\|\boldsymbol{\theta}-\boldsymbol{\theta}_{0}\|\right).
\]
This condition implies $L^{2}(P_{\boldsymbol{X}})$-continuity immediately.
Moreover, by the Cauchy--Schwarz inequality, $\boldsymbol{\pi}(\boldsymbol{\theta})$
is differentiable at $\boldsymbol{\theta}_{0}$ with $\dot{\boldsymbol{\pi}}_{0}=\mathbb{E}\!\left[\dot{\boldsymbol{\Phi}}_{0}(\boldsymbol{X})\otimes\boldsymbol{1}_{\mathcal{A}}(\boldsymbol{X})\right]$.
Thus, A1(iv) is weaker than imposing $L^{2}(P_{\boldsymbol{X}})$-differentiability
directly on $\boldsymbol{\Phi}_{\boldsymbol{\theta}}$. Assumption
A1(vi) contains two conceptually different complexity
requirements. The global Glivenko-Cantelli property of $\mathcal{M}$
provides the uniform law of large numbers used for consistency of
the grouped estimator. The local Donsker property of $\mathcal{M}_{\epsilon}$
is only needed for the local asymptotic analysis. Since the indicators
associated with the fixed partition $\mathcal{A}$ are bounded and
finite in number, the corresponding Glivenko-Cantelli and Donsker
properties are preserved after multiplication by $\boldsymbol{1}_{\mathcal{A}}$.
The next remark discusses primitive conditions for A1(vi). 
\begin{rem}
\label{rem:primitive-a1}
Primitive sufficient conditions for A1(vi) are standard for finite-dimensional
parametric classes. For the global Glivenko--Cantelli requirement,
it is sufficient that $\Theta$ be compact and that the map $\boldsymbol{\theta}\mapsto\boldsymbol{\Phi}_{\boldsymbol{\theta}}(\boldsymbol{x})$
be continuous on $\Theta$ for $P_{\boldsymbol{X}}$-almost every
$\boldsymbol{x}$. The usual uniform law of large
numbers therefore applies; see, e.g., \citet{NeweyMcFadden1994}.
For the local Donsker requirement, a convenient sufficient condition
is that, for some $\epsilon>0$, there exists $G\in L^{2}(P_{\boldsymbol{X}})$
such that 
\[
\left\Vert \boldsymbol{\Phi}_{\boldsymbol{\theta}_{1}}(\boldsymbol{x})-\boldsymbol{\Phi}_{\boldsymbol{\theta}_{2}}(\boldsymbol{x})\right\Vert \leq G(\boldsymbol{x})\|\boldsymbol{\theta}_{1}-\boldsymbol{\theta}_{2}\|,\qquad\boldsymbol{\theta}_{1},\boldsymbol{\theta}_{2}\in\mathcal{N}_{\epsilon}.
\]
Since $\mathcal{N}_{\epsilon}$ is a bounded subset of $\mathbb{R}^{p}$,
this Lipschitz condition yields polynomial covering numbers and hence
$\mathcal{M}_{\epsilon}$ is $P_{\boldsymbol{X}}$-Donsker. A stronger
but often convenient primitive condition is differentiability of $\boldsymbol{\theta}\mapsto\boldsymbol{\Phi}_{\boldsymbol{\theta}}(\boldsymbol{x})$
throughout $\mathcal{N}_{\epsilon}$, together with $\sup_{\boldsymbol{\theta}\in\mathcal{N}_{\epsilon}}\left\Vert \dot{\boldsymbol{\Phi}}_{\boldsymbol{\theta}}(\boldsymbol{x})\right\Vert \leq G(\boldsymbol{x})$,
$G\in L^{2}(P_{\boldsymbol{X}})$. The preceding Lipschitz condition then follows from the mean value inequality.
\end{rem}

\begin{thm}
\label{thm:fixed-partition}
Suppose that $H_{0}$, A0, and A1 hold. Then 
\[
\widehat{X}_{\boldsymbol{\theta}_{0}}^{2}(\widehat{\boldsymbol{\theta}})\rightarrow_{d}\chi_{J(L-1)-p}^{2}.
\]
The same limiting distribution holds for $\widehat{G}_{\boldsymbol{\theta}_{0}}^{2}(\widehat{\boldsymbol{\theta}})$. 
\end{thm}

The proof in the Appendix adapts Pollard's \citeyearpar{pollard1979general} strategy for marginal CDFs to our conditional setting. Under the null hypothesis,
$U(\boldsymbol{\theta}_{0})$ is uniformly distributed on $[0,1]$
and independent of $\boldsymbol{X}$, so that the population cell-probability
vector is $\boldsymbol{\pi}_{0}=\boldsymbol{v}\otimes\boldsymbol{q}$.
The first step is to establish consistency of $\widehat{\boldsymbol{\theta}}$.
By the Glivenko--Cantelli condition in A1(vi),
$\sup_{\boldsymbol{\theta}\in\Theta}\left\Vert \widehat{\boldsymbol{\pi}}(\boldsymbol{\theta})-\boldsymbol{\pi}(\boldsymbol{\theta})\right\Vert =o_{P}(1)$,
while, by the central limit theorem, $\left\Vert \widehat{\boldsymbol{p}}(\boldsymbol{\theta}_{0})-\boldsymbol{\pi}_{0}\right\Vert =O_{P}(n^{-1/2})$.
The criterion-minimization argument therefore implies $\left\Vert \boldsymbol{\pi}(\widehat{\boldsymbol{\theta}})-\boldsymbol{\pi}_{0}\right\Vert =o_{P}(1)$,
and the global identification condition A1(ii)
yields $\widehat{\boldsymbol{\theta}}-\boldsymbol{\theta}_{0}=o_{P}(1)$.
Once consistency has localized $\widehat{\boldsymbol{\theta}}$ to
$\mathcal{N}_{\epsilon}$, the local Donsker condition in A1(vi),
together with the $L^{2}(P_{\boldsymbol{X}})$-continuity in A1(iv),
yields the stochastic equicontinuity required for the local expansion.
The differentiability of $\boldsymbol{\pi}(\boldsymbol{\theta})$
at $\boldsymbol{\theta}_{0}$ and the full-rank condition A1(v) then
strengthen consistency to $\widehat{\boldsymbol{\theta}}-\boldsymbol{\theta}_{0}=O_{P}(n^{-1/2})$.
It is therefore sufficient to consider values of $\boldsymbol{\theta}$
in an $O_{P}(n^{-1/2})$ neighborhood of $\boldsymbol{\theta}_{0}$.
Uniformly over such a neighborhood, $\widehat{G}_{\boldsymbol{\theta}_{0}}^{2}(\boldsymbol{\theta})$
and $\widehat{X}_{\boldsymbol{\theta}_{0}}^{2}(\boldsymbol{\theta})$
are asymptotically equivalent to 
\[
\widehat{Q}_{0}(\boldsymbol{\theta})=\left\Vert \boldsymbol{\Lambda}_{0}^{-1}\widehat{\boldsymbol{\xi}}_{0}-\boldsymbol{\Lambda}_{0}^{-1}\dot{\boldsymbol{\pi}}_{0}\sqrt{n}(\boldsymbol{\theta}-\boldsymbol{\theta}_{0})\right\Vert ^{2}.
\]
Define
\begin{equation}
\boldsymbol{A}_{0}=\boldsymbol{\Lambda}_{0}^{-1}\cdot\dot{\boldsymbol{\pi}}_{0},\qquad
\boldsymbol{R}_{0}=\boldsymbol{A}_{0}\cdot(\boldsymbol{A}_{0}^{\text{\textsc{t}}}\cdot\boldsymbol{A}_{0})^{-1}\cdot\boldsymbol{A}_{0}^{\text{\textsc{t}}},\qquad
\boldsymbol{H}_{0}=\boldsymbol{I}_{LJ}-\boldsymbol{R}_{0}.
\label{eq:fixed-projection-matrices}
\end{equation}
Then $\boldsymbol{R}_{0}$ is the orthogonal projection
onto the column space of $\boldsymbol{A}_{0}$, and $\boldsymbol{H}_{0}$
projects onto its orthogonal complement. Hence the minimum of the
quadratic approximation is $\left\Vert \boldsymbol{H}_{0}\boldsymbol{\Lambda}_{0}^{-1}\widehat{\boldsymbol{\xi}}_{0}\right\Vert ^{2}$.
The limiting chi-squared distribution then follows from the limiting
distribution of $\boldsymbol{\Lambda}_{0}^{-1}\widehat{\boldsymbol{\xi}}_{0}$,
the rank conditions, and the continuous mapping theorem.

\section{Data-Dependent Grouping}

\label{sec:data-dependent-grouping}

Since, under the null hypothesis, the Rosenblatt transform $F(Y\mid\boldsymbol{X})$ is uniformly
distributed on $[0,1]$ independently of $\boldsymbol{X}$, it is natural to
choose equal-probability cells in $\mathcal{U}$ by setting $\vartheta_{\ell}=\ell/L$
and $v_{\ell}=1/L$ for $\ell=1,\ldots,L$. Unequal-width cells may
also be used. In particular, narrower cells near zero and one may
increase power against alternatives involving tail misspecification,
as in the classical marginal-CDF case; see, for example, \citet{kallenberg1985number}.
\begin{rem}
\label{rem:fixed-u-partition}
A data-dependent partition of the transformed observations generally depends on the same uniform variables whose cell frequencies form the statistic. Consequently, conditional uniformity does not imply that the random cell lengths equal the corresponding conditional cell probabilities, and the usual multinomial/Pearson argument need not apply. For this reason, the partition $\mathcal{U}$ of $[0,1]$ is kept fixed, whereas the partition of the covariate space may be data-dependent.
\end{rem}

Once $\mathcal{U}$ has been selected, the observations of $\boldsymbol{X}$
may be grouped using a data-dependent partition. The classical Pearson
statistic remains asymptotically valid under suitable data-dependent
partitions; see, for example, \citet{Watson1957}, \citet{10.1214/aoms/Chernoff},
and \citet{moore1971chi}. We establish an analogous result for the
proposed conditional-distribution specification test. The essential
question is whether evaluation of the relevant grouped empirical process at the random
partition $\widehat{\boldsymbol{\Gamma}}$, rather than at its probability
limit $\boldsymbol{\Gamma}$, affects the first-order asymptotic distribution.

Let $\mathcal{C}\subseteq\mathcal{B}(\mathbb{R}^{k})$ be a class
of measurable subsets of $\mathbb{R}^{k}$. Equip $\mathcal{C}$ with
the semimetric 
\[
d_{P_{\boldsymbol{X}}}(C_{1},C_{2})=\left[P_{\boldsymbol{X}}(C_{1}\mathbin{\triangle}C_{2})\right]^{1/2},\qquad C_{1},C_{2}\in\mathcal{C},
\]
where $\triangle$ denotes symmetric difference, which is the $L^{2}(P_{\boldsymbol{X}})$ distance between the corresponding
indicator functions. The class of ordered partitions of $\mathbb{R}^{k}$ into $J$ cells
drawn from $\mathcal{C}$ is 
\[
\mathcal{G}=\left\{ \boldsymbol{\gamma}=(\gamma_{1},\ldots,\gamma_{J})^{\text{\textsc{t}}}\in\mathcal{C}^{J}:\bigcup_{j=1}^{J}\gamma_{j}=\mathbb{R}^{k},\quad\gamma_{j}\cap\gamma_{j'}=\varnothing\text{ for }j\neq j'\right\} .
\]
We equip $\mathcal{G}$ with the product semimetric 
\[
d_{P_{\boldsymbol{X}}}(\boldsymbol{\gamma},\widetilde{\boldsymbol{\gamma}})=\left\{ \sum_{j=1}^{J}P_{\boldsymbol{X}}(\gamma_{j}\mathbin{\triangle}\widetilde{\gamma}_{j})\right\} ^{1/2},
\]
which induces the product topology because $J$ is fixed, and with
the associated Borel $\sigma$-field.

For $\boldsymbol{\gamma}\in\mathcal{G}$, $\boldsymbol{\theta}\in\Theta$,
$\ell=1,\ldots,L$, and $j=1,\ldots,J$, define 
\[
D_{\ell j}(\boldsymbol{\gamma},\boldsymbol{\theta})=\left\{ (y,\boldsymbol{x})\in\mathbb{R}^{1+k}:F_{\boldsymbol{\theta}}(y\mid\boldsymbol{x})\in I_{\ell},\ \boldsymbol{x}\in\gamma_{j}\right\} .
\]
Also define $\boldsymbol{1}_{\boldsymbol{\gamma}}(\boldsymbol{x})=\left(1_{\gamma_{1}}(\boldsymbol{x}),\ldots,1_{\gamma_{J}}(\boldsymbol{x})\right)^{\text{\textsc{t}}}$,
and $\boldsymbol{\pi}(\boldsymbol{\gamma},\boldsymbol{\theta})=\mathbb{E}\left[\boldsymbol{\Phi}_{\boldsymbol{\theta}}(\boldsymbol{X})\otimes\boldsymbol{1}_{\boldsymbol{\gamma}}(\boldsymbol{X})\right]$,
where, for $\ell=1,\dots,L$ and $j=1,\dots,J$,
\begin{equation*}
\pi_{\ell j}(\boldsymbol{\gamma},\boldsymbol{\theta})=\mathbb{E}\left[\Phi_{\boldsymbol{\theta}\ell}(\boldsymbol{X})1_{\gamma_{j}}(\boldsymbol{X})\right]=\int_{\gamma_{j}}\Phi_{\boldsymbol{\theta}\ell}(\boldsymbol{x})P_{\boldsymbol{X}}(d\boldsymbol{x}),
\end{equation*}
or, equivalently, $\pi_{\ell j}(\boldsymbol{\gamma},\boldsymbol{\theta})=P_{\boldsymbol{\theta}}\left\{ D_{\ell j}(\boldsymbol{\gamma},\boldsymbol{\theta}_{0})\right\}$.

In contrast with the fixed-partition analysis, the random-partition
argument requires a first-order expansion that is uniform over the
cells in $\mathcal{C}$. Accordingly, as the analog of A1(iv),
assume that $\boldsymbol{\theta}\mapsto\boldsymbol{\Phi}_{\boldsymbol{\theta}}$
is $L^{2}(P_{\boldsymbol{X}})$-differentiable at $\boldsymbol{\theta}_{0}$,
which implies that there exists an $L\times p$ matrix-valued function
$\boldsymbol{x}\mapsto\dot{\boldsymbol{\Phi}}_{0}\left(\boldsymbol{x}\right)$,
such that 
\begin{equation}
\boldsymbol{\Phi}_{\boldsymbol{\theta}}\left(\boldsymbol{x}\right)=\boldsymbol{\Phi}_{\boldsymbol{\theta}_{0}}\left(\boldsymbol{x}\right)+\dot{\boldsymbol{\Phi}}_{0}\left(\boldsymbol{x}\right)(\boldsymbol{\theta}-\boldsymbol{\theta}_{0})+\boldsymbol{r}\left(\boldsymbol{x}\right),\label{eq:23}
\end{equation}
where $\left\Vert \boldsymbol{r}\right\Vert _{P_{\boldsymbol{X}},2}=o\left(\left\Vert \boldsymbol{\theta}-\boldsymbol{\theta}_{0}\right\Vert \right)$
as $\boldsymbol{\theta}\rightarrow\boldsymbol{\theta}_{0}$. Therefore,
there is an $LJ\times p$ matrix-valued function $\boldsymbol{\gamma}\mapsto\dot{\boldsymbol{\pi}}_{0}\left(\boldsymbol{\gamma}\right)=\mathbb{E}\left[\dot{\boldsymbol{\Phi}}_{0}\left(\boldsymbol{X}\right)\otimes\boldsymbol{1}_{\boldsymbol{\gamma}}\left(\boldsymbol{X}\right)\right]$
such that, using \eqref{eq:23},
\begin{eqnarray*}
\sup_{\boldsymbol{\gamma}\in\mathcal{G}}\left\Vert \begin{aligned} & \boldsymbol{\pi}(\boldsymbol{\gamma},\boldsymbol{\theta})-\boldsymbol{\pi}(\boldsymbol{\gamma},\boldsymbol{\theta}_{0})-\dot{\boldsymbol{\pi}}_{0}(\boldsymbol{\gamma})(\boldsymbol{\theta}-\boldsymbol{\theta}_{0})\\[-2pt]\end{aligned}
\right\Vert  & = & \sup_{\boldsymbol{\gamma}\in\mathcal{G}}\left\Vert \mathbb{E}\left[\boldsymbol{r}(\boldsymbol{X})\otimes\boldsymbol{1}_{\boldsymbol{\gamma}}(\boldsymbol{X})\right]\right\Vert \\
 & \leq & \left\Vert \boldsymbol{r}\right\Vert _{P_{\boldsymbol{X}},2}=o\left(\left\Vert \boldsymbol{\theta}-\boldsymbol{\theta}_{0}\right\Vert \right),
\end{eqnarray*}
where the inequality follows from Cauchy--Schwarz and $\|\boldsymbol{1}_{\boldsymbol{\gamma}}(\boldsymbol{X})\|=1$
because the cells form a partition. Thus, $\dot{\boldsymbol{\pi}}_{0}(\boldsymbol{\gamma})$
is the derivative of $\boldsymbol{\theta}\mapsto\boldsymbol{\pi}(\boldsymbol{\gamma},\boldsymbol{\theta})$
at $\boldsymbol{\theta}_{0}$, uniformly in $\boldsymbol{\gamma}\in\mathcal{G}$.
This is the uniform analog of \eqref{eq:22}.

A data-dependent partition $\widehat{\boldsymbol{\Gamma}}=(\widehat{\Gamma}_{1},\ldots,\widehat{\Gamma}_{J})^{\text{\textsc{t}}}$
is a measurable map into $\mathcal{G}$. We assume $\widehat{\boldsymbol{\Gamma}}\rightarrow_{P}\boldsymbol{\Gamma}=(\Gamma_{1},\ldots,\Gamma_{J})^{\text{\textsc{t}}}\in\mathcal{G}$,
where convergence is in the product topology. Equivalently, $P_{\boldsymbol{X}}\left(\widehat{\Gamma}_{j}\mathbin{\triangle}\Gamma_{j}\right)\rightarrow_{P}0$,
$j=1,\ldots,J$. At the limiting partition, the relevant derivative
matrix is $\dot{\boldsymbol{\pi}}_{0}(\boldsymbol{\Gamma})$, which
will be assumed to have full column rank.

For fixed $\boldsymbol{\gamma}\in\mathcal{G}$, define the observed
count and relative-frequency vectors by 
\[
\widehat{\boldsymbol{O}}(\boldsymbol{\gamma},\boldsymbol{\theta})=\sum_{i=1}^{n}\boldsymbol{1}_{\mathcal{U}}\left(U_{i}(\boldsymbol{\theta})\right)\otimes\boldsymbol{1}_{\boldsymbol{\gamma}}(\boldsymbol{X}_{i}),\qquad
\widehat{\boldsymbol{p}}(\boldsymbol{\gamma},\boldsymbol{\theta})=\frac{1}{n}\widehat{\boldsymbol{O}}(\boldsymbol{\gamma},\boldsymbol{\theta}).
\]
For $\boldsymbol{\theta}_{1},\boldsymbol{\theta}_{2}\in\Theta$, define
\[
\widehat{\boldsymbol{\pi}}_{\boldsymbol{\theta}_{1}}(\boldsymbol{\gamma},\boldsymbol{\theta}_{2})=\frac{1}{n}\sum_{i=1}^{n}\boldsymbol{\Phi}_{\boldsymbol{\theta}_{1},\boldsymbol{\theta}_{2}}(\boldsymbol{X}_{i})\otimes\boldsymbol{1}_{\boldsymbol{\gamma}}(\boldsymbol{X}_{i}).
\]
Conditional on the observed covariates, this vector is the model-implied
expected relative cell-frequency vector under $F_{\boldsymbol{\theta}_{2}}$
for the partition $\boldsymbol{\gamma}$.

For $\boldsymbol{\gamma}\in\mathcal{G}$, let $\widehat{\boldsymbol{\theta}}(\boldsymbol{\gamma})$
denote a minimizer of 
\[
\widehat{G}_{\boldsymbol{\theta}_{0}}^{2}(\boldsymbol{\gamma},\boldsymbol{\theta})=2n\sum_{\ell=1}^{L}\sum_{j=1}^{J}\widehat{p}_{\ell j}(\boldsymbol{\gamma},\boldsymbol{\theta}_{0})\log\left[\frac{\widehat{p}_{\ell j}(\boldsymbol{\gamma},\boldsymbol{\theta}_{0})}{\widehat{\pi}_{\boldsymbol{\theta}_{0},\ell j}(\boldsymbol{\gamma},\boldsymbol{\theta})}\right],
\]
over $\boldsymbol{\theta}\in\Theta$. The estimator based on the random
partition is $\widehat{\boldsymbol{\theta}}=\widehat{\boldsymbol{\theta}}(\widehat{\boldsymbol{\Gamma}})$. For generic $\boldsymbol{\theta}_{1},\boldsymbol{\theta}_{2}\in\Theta$,
define the Pearson statistic 
\[
\widehat{X}_{\boldsymbol{\theta}_{1}}^{2}(\boldsymbol{\gamma},\boldsymbol{\theta}_{2})=n\sum_{\ell=1}^{L}\sum_{j=1}^{J}\frac{\left[\widehat{p}_{\ell j}(\boldsymbol{\gamma},\boldsymbol{\theta}_{1})-\widehat{\pi}_{\boldsymbol{\theta}_{1},\ell j}(\boldsymbol{\gamma},\boldsymbol{\theta}_{2})\right]^{2}}{\widehat{\pi}_{\boldsymbol{\theta}_{1},\ell j}(\boldsymbol{\gamma},\boldsymbol{\theta}_{2})}.
\]
The infeasible statistic based on the estimated partition is therefore
$\widehat{X}_{\boldsymbol{\theta}_{0}}^{2}(\widehat{\boldsymbol{\Gamma}},\widehat{\boldsymbol{\theta}})$.

Define 
\begin{equation}
\widehat{\boldsymbol{\xi}}_{\boldsymbol{\theta}_{1}}(\boldsymbol{\gamma},\boldsymbol{\theta}_{2})=\sqrt{n}\left[\widehat{\boldsymbol{p}}(\boldsymbol{\gamma},\boldsymbol{\theta}_{1})-\widehat{\boldsymbol{\pi}}_{\boldsymbol{\theta}_{1}}(\boldsymbol{\gamma},\boldsymbol{\theta}_{2})\right],
\label{eq:frequency-residual-definition}
\end{equation}
and $\widehat{\boldsymbol{\xi}}_{0}(\boldsymbol{\gamma}):=\widehat{\boldsymbol{\xi}}_{\boldsymbol{\theta}_{0}}(\boldsymbol{\gamma},\boldsymbol{\theta}_{0})$. For every $\boldsymbol{\gamma}\in\mathcal{G}$ and $\boldsymbol{\theta}\in\Theta$,
\[
\widehat{\boldsymbol{\pi}}_{\boldsymbol{\theta}}(\boldsymbol{\gamma},\boldsymbol{\theta})=\boldsymbol{v}\otimes\widehat{\boldsymbol{q}}(\boldsymbol{\gamma})=:\widehat{\boldsymbol{\pi}}_{0}(\boldsymbol{\gamma}),
\]
 where $\widehat{\boldsymbol{q}}(\boldsymbol{\gamma})=n^{-1}\sum_{i=1}^{n}\boldsymbol{1}_{\boldsymbol{\gamma}}(\boldsymbol{X}_{i})$
estimates $\boldsymbol{q}(\boldsymbol{\gamma})=\mathbb{E}\left[\boldsymbol{1}_{\boldsymbol{\gamma}}(\boldsymbol{X})\right]$.
Thus, $q_{j}(\boldsymbol{\gamma})=P_{\boldsymbol{X}}(\gamma_{j})$,
$j=1,\ldots,J$. The corresponding population probability vector and weighting matrices are
\begin{equation}
\boldsymbol{\pi}_{0}(\boldsymbol{\gamma})=\boldsymbol{v}\otimes\boldsymbol{q}(\boldsymbol{\gamma}),
\qquad
\boldsymbol{\Lambda}_{0}(\boldsymbol{\gamma})=
\operatorname{diag}^{1/2}\!\left[\boldsymbol{\pi}_{0}(\boldsymbol{\gamma})\right].
\label{eq:population-probability-weighting}
\end{equation}
Let $\widehat{\boldsymbol{\Lambda}}_{0}(\boldsymbol{\gamma})=\operatorname{diag}^{1/2}\left[\widehat{\boldsymbol{\pi}}_{0}(\boldsymbol{\gamma})\right]$.
Then $\widehat{X}_{\boldsymbol{\theta}_{0}}^{2}(\boldsymbol{\gamma},\boldsymbol{\theta}_{0})=\left\Vert \widehat{\boldsymbol{\Lambda}}_{0}^{-1}(\boldsymbol{\gamma})\widehat{\boldsymbol{\xi}}_{0}(\boldsymbol{\gamma})\right\Vert ^{2}$.
Moreover,
\begin{equation}
\widehat{\boldsymbol{\xi}}_{0}(\boldsymbol{\gamma})=\allowbreak\frac{1}{\sqrt{n}}\sum_{i=1}^{n}\boldsymbol{w}(Y_{i},\boldsymbol{X}_{i})\allowbreak\otimes\boldsymbol{1}_{\boldsymbol{\gamma}}(\boldsymbol{X}_{i})
\label{eq:marked-process}
\end{equation}
is a marked empirical process indexed by $\boldsymbol{\gamma}\in\mathcal{G}$,
with marks $\boldsymbol{w}(y,\boldsymbol{x})=\boldsymbol{1}_{\mathcal{U}}\left[F_{\boldsymbol{\theta}_{0}}(y\mid\boldsymbol{x})\right]-\boldsymbol{v}$. 

To obtain a limit theorem valid at $\widehat{\boldsymbol{\Gamma}}$,
the pointwise central limit theorem used in Section~\ref{sec:testing-procedure} must
be replaced by a uniform analog. Define 
\begin{equation}
\mathcal{Z}=\left\{ (y,\boldsymbol{x})\mapsto\left[\boldsymbol{1}_{\mathcal{U}}\left(F_{\boldsymbol{\theta}_{0}}(y\mid\boldsymbol{x})\right)-\boldsymbol{v}\right]\otimes\boldsymbol{1}_{\boldsymbol{\gamma}}(\boldsymbol{x}):\boldsymbol{\gamma}\in\mathcal{G}\right\} .
\label{eq:marked-class}
\end{equation}
The process $\widehat{\boldsymbol{\xi}}_{0}(\cdot)$ may be viewed
as a random element of $\ell^{\infty}(\mathcal{G};\mathbb{R}^{LJ})$, the space of bounded functions from $\mathcal{G}$ into $\mathbb{R}^{LJ}$ equipped with the supremum norm. Assume that the indicator class $\left\{ \boldsymbol{x}\mapsto1_{C}(\boldsymbol{x}):C\in\mathcal{C}\right\} $
is $P_{\boldsymbol{X}}$-Donsker. Since the marks $1_{I_{\ell}}\left[F_{\boldsymbol{\theta}_{0}}(Y\mid\boldsymbol{X})\right]-v_{\ell}$,
$\ell=1,\ldots,L$, are fixed and bounded, multiplication by them preserves
Donskerity. Because $J$ and $L$ are fixed, the finite vector class
$\mathcal{Z}$ is therefore $P$-Donsker. Consequently, $\widehat{\boldsymbol{\xi}}_{0}(\cdot)\rightarrow_{d}\boldsymbol{\xi}_{\infty}(\cdot)$ in $\ell^{\infty}(\mathcal{G};\mathbb{R}^{LJ})$, where $\boldsymbol{\xi}_{\infty}(\cdot)$ is a tight, mean-zero Gaussian
process with the covariance kernel 
\[
\mathbb{E}\left[\boldsymbol{\xi}_{\infty}(\boldsymbol{\gamma})\boldsymbol{\xi}_{\infty}^{\text{\textsc{t}}}(\boldsymbol{\gamma}^{*})\right]=\left[\operatorname{diag}(\boldsymbol{v})-\boldsymbol{v}\boldsymbol{v}^{\text{\textsc{t}}}\right]\otimes\boldsymbol{S}(\boldsymbol{\gamma},\boldsymbol{\gamma}^{*}),
\]
for $\boldsymbol{\gamma},\boldsymbol{\gamma}^{*}\in\mathcal{G}$,
where $\boldsymbol{S}(\boldsymbol{\gamma},\boldsymbol{\gamma^{*}})=\mathbb{E}\left[\boldsymbol{1}_{\boldsymbol{\gamma}}(\boldsymbol{X})\boldsymbol{1}_{\boldsymbol{\gamma}^{*}}^{\text{\textsc{t}}}(\boldsymbol{X})\right]$,
whose $(j,m)$th component is $S_{jm}(\boldsymbol{\gamma},\boldsymbol{\gamma}^{*})=P_{\boldsymbol{X}}\left(\gamma_{j}\cap\gamma_{m}^{*}\right)$.

The following assumption adapts the fixed-cell conditions to a random
partition. 
\begin{description}
\item [{A2:}] ~ 
\begin{enumerate}
\item[(i)] The true value $\boldsymbol{\theta}_{0}$ is an interior point of
$\Theta$. 
\item[(ii)] The sequence $\{\widehat{\boldsymbol{\Gamma}}\}$ consists of measurable
random elements of $\mathcal{G}$ satisfying $\widehat{\boldsymbol{\Gamma}}\rightarrow_{P}\boldsymbol{\Gamma}\in\mathcal{G}$. 
\item[(iii)] Every component of $\boldsymbol{q}(\boldsymbol{\Gamma})$ is strictly
positive. 
\item[(iv)] The map $\boldsymbol{\theta}\mapsto\boldsymbol{\Phi}_{\boldsymbol{\theta}}$
is $L^{2}(P_{\boldsymbol{X}})$-differentiable at $\boldsymbol{\theta}_{0}$
in the sense of \eqref{eq:23}. 
\item[(v)] The matrix $\dot{\boldsymbol{\pi}}_{0}(\boldsymbol{\Gamma})$ has
column rank $p$, and $J(L-1)>p$. 
\item[(vi)] The class $\mathcal{M}$ is $P_{\boldsymbol{X}}$-Glivenko--Cantelli
and $\mathcal{M}_{\epsilon}$ has a bounded uniform entropy integral
(BUEI) for some $\epsilon>0$. 
\item[(vii)] $\mathcal{C}$ is a VC class of sets. 
\item[(viii)] The estimator $\widehat{\boldsymbol{\theta}}=\widehat{\boldsymbol{\theta}}(\widehat{\boldsymbol{\Gamma}})$
is consistent for $\boldsymbol{\theta}_{0}$ and satisfies 
\[
\widehat{G}_{\boldsymbol{\theta}_{0}}^{2}(\widehat{\boldsymbol{\Gamma}},\widehat{\boldsymbol{\theta}})\leq\inf_{\boldsymbol{\theta}\in\Theta}\widehat{G}_{\boldsymbol{\theta}_{0}}^{2}(\widehat{\boldsymbol{\Gamma}},\boldsymbol{\theta})+o_{P}(1).
\]
\end{enumerate}
\end{description}
We impose consistency directly in A2(viii). It could be established
as in the fixed-partition case, replacing pointwise convergence arguments
with their uniform counterparts over the partition class. Because
the estimator is infeasible and serves only as an intermediate device,
we omit the additional conditions needed for this purpose and focus
on the effect of data-dependent grouping.

The remaining conditions are the random-partition counterparts of
those in A1. Assumption A2(ii)
specifies convergence of the estimated cells, while A2(iv)
provides a uniform derivative representation for the cell probabilities.
Assumptions A2(vi) and A2(vii),
together with the relevant measurability and boundedness conditions,
imply that the product class $\left\{ \boldsymbol{x}\mapsto\boldsymbol{\Phi}_{\boldsymbol{\theta}}(\boldsymbol{x})\otimes\boldsymbol{1}_{\boldsymbol{\gamma}}(\boldsymbol{x}):\boldsymbol{\theta}\in\mathcal{N}_{\epsilon},\ \boldsymbol{\gamma}\in\mathcal{G}\right\} $
is $P_{\boldsymbol{X}}$-Donsker. It follows that 
\[
\sup_{\boldsymbol{\theta}\in\mathcal{N}_{\epsilon},\,\boldsymbol{\gamma}\in\mathcal{G}}\left\Vert \widehat{\boldsymbol{\pi}}_{\boldsymbol{\theta}_{0}}(\boldsymbol{\gamma},\boldsymbol{\theta})-\boldsymbol{\pi}(\boldsymbol{\gamma},\boldsymbol{\theta})\right\Vert =O_{P}(n^{-1/2}),
\]
which is used in the asymptotic analysis below. 
\begin{thm}
\label{thm:random-partition} Suppose that $H_{0}$, A0,
and A2 hold. Then 
\[
\widehat{X}_{\boldsymbol{\theta}_{0}}^{2}(\widehat{\boldsymbol{\Gamma}},\widehat{\boldsymbol{\theta}})\rightarrow_{d}\chi_{J(L-1)-p}^{2}.
\]
The same limiting distribution holds for the $\widehat{G}_{\boldsymbol{\theta}_{0}}^{2}(\widehat{\boldsymbol{\Gamma}},\widehat{\boldsymbol{\theta}})$.
\end{thm}

The proof follows the strategy of Theorem~\ref{thm:fixed-partition}, but the relevant approximations must now hold uniformly over the partition. The conditions in A2 imply $\widehat{\boldsymbol{\theta}}-\boldsymbol{\theta}_{0}=O_{P}(n^{-1/2})$.
Thus, as in the fixed-partition case, it suffices to consider $\boldsymbol{\theta}$
in an $O_{P}(n^{-1/2})$ neighborhood of $\boldsymbol{\theta}_{0}$.
Uniformly over such values of $\boldsymbol{\theta}$ and over $\boldsymbol{\gamma}$
in a shrinking neighborhood of $\boldsymbol{\Gamma}$, the Pearson
and likelihood-ratio criteria are asymptotically equivalent to 
\[
\widehat{Q}_{0}(\boldsymbol{\gamma},\boldsymbol{\theta})=\left\Vert \boldsymbol{\Lambda}_{0}^{-1}(\boldsymbol{\gamma})\widehat{\boldsymbol{\xi}}_{0}(\boldsymbol{\gamma})-\boldsymbol{\Lambda}_{0}^{-1}(\boldsymbol{\gamma})\dot{\boldsymbol{\pi}}_{0}(\boldsymbol{\gamma})\sqrt{n}\,(\boldsymbol{\theta}-\boldsymbol{\theta}_{0})\right\Vert ^{2}.
\]
This is the random-partition analog of the quadratic approximation
used in the proof of Theorem~\ref{thm:fixed-partition}. Define $\boldsymbol{A}_{0}(\boldsymbol{\gamma})=\boldsymbol{\Lambda}_{0}^{-1}(\boldsymbol{\gamma})\cdot\dot{\boldsymbol{\pi}}_{0}(\boldsymbol{\gamma})$ and
\begin{equation}
\boldsymbol{R}_{0}(\boldsymbol{\gamma})=\boldsymbol{A}_{0}(\boldsymbol{\gamma})\cdot
\left[\boldsymbol{A}_{0}^{\text{\textsc{t}}}(\boldsymbol{\gamma})\cdot\boldsymbol{A}_{0}(\boldsymbol{\gamma})\right]^{-1}
\cdot\boldsymbol{A}_{0}^{\text{\textsc{t}}}(\boldsymbol{\gamma}), \qquad
\boldsymbol{H}_{0}(\boldsymbol{\gamma})=\boldsymbol{I}_{LJ}-\boldsymbol{R}_{0}(\boldsymbol{\gamma}).
\label{eq:projection-matrices}
\end{equation}
The corresponding minimized residual is
$\boldsymbol{H}_{0}(\widehat{\boldsymbol{\Gamma}})\cdot\boldsymbol{\Lambda}_{0}^{-1}(\widehat{\boldsymbol{\Gamma}})\cdot\widehat{\boldsymbol{\xi}}_{0}(\widehat{\boldsymbol{\Gamma}})$.
Here $\boldsymbol{R}_{0}(\boldsymbol{\gamma})$ is the orthogonal
projection onto the column space of $\boldsymbol{A}_{0}(\boldsymbol{\gamma})$,
while $\boldsymbol{H}_{0}(\boldsymbol{\gamma})$ projects onto its
orthogonal complement. Uniformly over a shrinking neighborhood of $\boldsymbol{\Gamma}$,
the minimized statistic admits the representation 
\[
\widehat{X}_{\boldsymbol{\theta}_{0}}^{2}(\widehat{\boldsymbol{\Gamma}},\widehat{\boldsymbol{\theta}})=\left\Vert \boldsymbol{H}_{0}(\widehat{\boldsymbol{\Gamma}})\boldsymbol{\Lambda}_{0}^{-1}(\widehat{\boldsymbol{\Gamma}})\widehat{\boldsymbol{\xi}}_{0}(\widehat{\boldsymbol{\Gamma}})\right\Vert ^{2}+o_{P}(1).
\]
Weak convergence of the marked process, asymptotic equicontinuity
with respect to the partition semimetric, and $\widehat{\boldsymbol{\Gamma}}\rightarrow_{P}\boldsymbol{\Gamma}$
then imply that evaluation at the estimated partition is asymptotically
equivalent to evaluation at $\boldsymbol{\Gamma}$. No asymptotic
independence between $\widehat{\boldsymbol{\Gamma}}$ and the empirical
process is required. 
\begin{rem}
\label{rem:random-partition-limit}
Assumption A2(ii), which requires the data-dependent
partition $\widehat{\boldsymbol{\Gamma}}$ to converge in probability
to a fixed partition $\boldsymbol{\Gamma}$, is a convenient sufficient
condition rather than an essential requirement. Together with the
Donsker conditions in A2(vi) and A2(vii),
it permits stochastic equicontinuity to reduce the asymptotic analysis
to evaluation at the fixed partition $\boldsymbol{\Gamma}$, so that
the randomness of the estimated partition has no first-order effect
on the limiting distribution. As discussed by \citet[Section 6]{pollard1979general}
in the classical grouped-data setting, convergence to a nonrandom
limiting partition is not necessary in principle; more general random-partition
schemes may lead to the same asymptotic distribution, although their
treatment requires a more general argument. 
\end{rem}

\section{The Feasible Test}

\label{sec:feasible-test}

The infeasible grouped-data estimator introduced in the previous section
depends on the unknown parameter $\boldsymbol{\theta}_{0}$ through
the construction of the observed cells. A feasible estimator that
is asymptotically equivalent to the infeasible estimator can be obtained
by a one-step scoring procedure.

Let $\widetilde{\boldsymbol{\theta}}$ be a preliminary estimator
satisfying $\sqrt{n}(\widetilde{\boldsymbol{\theta}}-\boldsymbol{\theta}_{0})=O_{P}(1)$. The
basic idea is to approximate $\widehat{X}_{\boldsymbol{\theta}_{0}}^{2}(\widehat{\boldsymbol{\Gamma}},\boldsymbol{\theta})$
by expanding the model-implied cell probabilities around $\widetilde{\boldsymbol{\theta}}$,
rather than around the unknown $\boldsymbol{\theta}_{0}$.

For $\boldsymbol{\theta}_{1},\boldsymbol{\theta}_{2}\in\Theta$, recall
that $\boldsymbol{\Phi}_{\boldsymbol{\theta}_{1},\boldsymbol{\theta}_{2}}(\boldsymbol{x})$
is the vector of conditional cell probabilities obtained by constructing
the Rosenblatt transform using $\boldsymbol{\theta}_{1}$ and evaluating
probabilities under the model indexed by $\boldsymbol{\theta}_{2}$.
Define the $L\times p$ derivative matrix 
\[
\dot{\boldsymbol{\Phi}}_{\boldsymbol{\theta}}(\boldsymbol{x})=\left.\frac{\partial}{\partial\boldsymbol{\eta}^{\text{\textsc{t}}}}\boldsymbol{\Phi}_{\boldsymbol{\theta},\boldsymbol{\eta}}(\boldsymbol{x})\right|_{\boldsymbol{\eta}=\boldsymbol{\theta}}.
\]
Although the asymptotic theory defines the derivative abstractly through
differentiability in $L^{2}(P_{\boldsymbol{X}})$, its computation
is straightforward in smooth parametric models. In fact, 
\[
{\displaystyle \Phi_{\boldsymbol{\theta}_{0},\boldsymbol{\theta},\ell}(\boldsymbol{x})=F_{\boldsymbol{\theta}}\left(F_{\boldsymbol{\theta}_{0}}^{-1}(\vartheta_{\ell}\mid\boldsymbol{x})\mid\boldsymbol{x}\right)-F_{\boldsymbol{\theta}}\left(F_{\boldsymbol{\theta}_{0}}^{-1}(\vartheta_{\ell-1}\mid\boldsymbol{x})\mid\boldsymbol{x}\right).}
\]
Hence, since the cell boundaries $F_{\boldsymbol{\theta}_{0}}^{-1}(\vartheta_{\ell}\mid\boldsymbol{x})$
are fixed when differentiating,
\[
{\displaystyle \dot{\Phi}_{0\ell}(\boldsymbol{x})=\dot{F}_{\boldsymbol{\theta}_{0}}\left(F_{\boldsymbol{\theta}_{0}}^{-1}(\vartheta_{\ell}\mid\boldsymbol{x})\mid\boldsymbol{x}\right)-\dot{F}_{\boldsymbol{\theta}_{0}}\left(F_{\boldsymbol{\theta}_{0}}^{-1}(\vartheta_{\ell-1}\mid\boldsymbol{x})\mid\boldsymbol{x}\right),}
\]
where $\dot{F}_{\boldsymbol{\theta}_{0}}(y\mid\boldsymbol{x})$ denotes
the derivative of $F_{\boldsymbol{\theta}}(y\mid\boldsymbol{x})$
with respect to $\boldsymbol{\theta}$, evaluated at $\boldsymbol{\theta}_{0}$.
Thus, computation of $\dot{\boldsymbol{\Phi}}_{0}$ requires only
the parameter derivative of the conditional CDF; no derivative of
the inverse CDF and no differentiation under an integral are required.

For $\boldsymbol{\gamma}\in\mathcal{G}$, define the sample and population
derivative matrices 
\[
\widehat{\dot{\boldsymbol{\pi}}}(\boldsymbol{\gamma},\boldsymbol{\theta})=\frac{1}{n}\sum_{i=1}^{n}\dot{\boldsymbol{\Phi}}_{\boldsymbol{\theta}}(\boldsymbol{X}_{i})\otimes\boldsymbol{1}_{\boldsymbol{\gamma}}(\boldsymbol{X}_{i}),\qquad\dot{\boldsymbol{\pi}}(\boldsymbol{\gamma},\boldsymbol{\theta})=\mathbb{E}\left[\dot{\boldsymbol{\Phi}}_{\boldsymbol{\theta}}(\boldsymbol{X})\otimes\boldsymbol{1}_{\boldsymbol{\gamma}}(\boldsymbol{X})\right],
\]
and let $\dot{\boldsymbol{\pi}}_{0}(\boldsymbol{\gamma}):=\dot{\boldsymbol{\pi}}(\boldsymbol{\gamma},\boldsymbol{\theta}_{0})$.
This agrees with the derivative matrix introduced in Section~\ref{sec:data-dependent-grouping};
hence $\dot{\boldsymbol{\pi}}_{0}(\boldsymbol{\Gamma})$ is the derivative
relevant at the limiting partition.

Recall that $\widehat{\boldsymbol{\Lambda}}_{0}(\boldsymbol{\gamma})=\operatorname{diag}^{1/2}\left[\widehat{\boldsymbol{\pi}}_{0}(\boldsymbol{\gamma})\right]$.
Using the arguments in the proofs of Theorems~\ref{thm:fixed-partition} and~\ref{thm:random-partition}, the Pearson
and likelihood-ratio criteria can be approximated locally by the quadratic
form 
\[
\widehat{Q}_{\boldsymbol{\theta}_{0}}(\boldsymbol{\gamma},\boldsymbol{\theta})=\left\Vert \widehat{\boldsymbol{\Lambda}}_{0}^{-1}(\boldsymbol{\gamma})\left[\widehat{\boldsymbol{\xi}}_{\boldsymbol{\theta}_{0}}(\boldsymbol{\gamma},\widetilde{\boldsymbol{\theta}})-\widehat{\dot{\boldsymbol{\pi}}}(\boldsymbol{\gamma},\widetilde{\boldsymbol{\theta}})\sqrt{n}(\boldsymbol{\theta}-\widetilde{\boldsymbol{\theta}})\right]\right\Vert ^{2}.
\]
This quadratic form is infeasible because the observed cells are constructed
using $\boldsymbol{\theta}_{0}$. Replacing $\boldsymbol{\theta}_{0}$
by $\widetilde{\boldsymbol{\theta}}$ gives the feasible criterion
\[
\widehat{Q}_{\widetilde{\boldsymbol{\theta}}}(\boldsymbol{\gamma},\boldsymbol{\theta})=\left\Vert \widehat{\boldsymbol{\Lambda}}_{0}^{-1}(\boldsymbol{\gamma})\left[\widehat{\boldsymbol{\xi}}_{\widetilde{\boldsymbol{\theta}}}(\boldsymbol{\gamma},\widetilde{\boldsymbol{\theta}})-\widehat{\dot{\boldsymbol{\pi}}}(\boldsymbol{\gamma},\widetilde{\boldsymbol{\theta}})\sqrt{n}(\boldsymbol{\theta}-\widetilde{\boldsymbol{\theta}})\right]\right\Vert ^{2}.
\]
The minimizer of $\boldsymbol{\theta}\mapsto\widehat{Q}_{\widetilde{\boldsymbol{\theta}}}(\widehat{\boldsymbol{\Gamma}},\boldsymbol{\theta})$
is the one-step estimator 
\begin{equation*}
{\displaystyle \widehat{\boldsymbol{\theta}}^{*}=\widetilde{\boldsymbol{\theta}}+\frac{1}{\sqrt{n}}\widehat{\boldsymbol{J}}^{-1}(\widehat{\boldsymbol{\Gamma}},\widetilde{\boldsymbol{\theta}})\widehat{\dot{\boldsymbol{\pi}}}^{\text{\textsc{t}}}(\widehat{\boldsymbol{\Gamma}},\widetilde{\boldsymbol{\theta}})\widehat{\boldsymbol{\Lambda}}_{0}^{-2}(\widehat{\boldsymbol{\Gamma}})\widehat{\boldsymbol{\xi}}_{\widetilde{\boldsymbol{\theta}}}(\widehat{\boldsymbol{\Gamma}},\widetilde{\boldsymbol{\theta}}),}
\end{equation*}
where
\[
\widehat{\boldsymbol{J}}(\boldsymbol{\gamma},\boldsymbol{\theta})=\widehat{\dot{\boldsymbol{\pi}}}^{\text{\textsc{t}}}(\boldsymbol{\gamma},\boldsymbol{\theta})\widehat{\boldsymbol{\Lambda}}_{0}^{-2}(\boldsymbol{\gamma})\widehat{\dot{\boldsymbol{\pi}}}(\boldsymbol{\gamma},\boldsymbol{\theta}).
\]
Under the regularity conditions below, 
\begin{equation}
\widehat{\boldsymbol{J}}(\widehat{\boldsymbol{\Gamma}},\widetilde{\boldsymbol{\theta}})\to_{P}\boldsymbol{J}_{0}(\boldsymbol{\Gamma}),\;\text{ with }\;\boldsymbol{J}_{0}(\boldsymbol{\Gamma})=\dot{\boldsymbol{\pi}}_{0}^{\text{\textsc{t}}}(\boldsymbol{\Gamma})\boldsymbol{\Lambda}_{0}^{-2}(\boldsymbol{\Gamma})\dot{\boldsymbol{\pi}}_{0}(\boldsymbol{\Gamma}).\label{eq:36}
\end{equation}

Define the estimated projection onto the standardized derivative space
by 
\[
\widehat{\boldsymbol{R}}(\boldsymbol{\gamma},\boldsymbol{\theta})=\widehat{\boldsymbol{\Lambda}}_{0}^{-1}(\boldsymbol{\gamma})\widehat{\dot{\boldsymbol{\pi}}}(\boldsymbol{\gamma},\boldsymbol{\theta})\widehat{\boldsymbol{J}}^{-1}(\boldsymbol{\gamma},\boldsymbol{\theta})\widehat{\dot{\boldsymbol{\pi}}}^{\text{\textsc{t}}}(\boldsymbol{\gamma},\boldsymbol{\theta})\widehat{\boldsymbol{\Lambda}}_{0}^{-1}(\boldsymbol{\gamma}),
\]
and let $\widehat{\boldsymbol{H}}(\boldsymbol{\gamma},\boldsymbol{\theta})=\boldsymbol{I}_{LJ}-\widehat{\boldsymbol{R}}(\boldsymbol{\gamma},\boldsymbol{\theta})$. Thus, $\widehat{\boldsymbol{H}}(\boldsymbol{\gamma},\boldsymbol{\theta})$
is the orthogonal projection onto the orthogonal complement of the
column space of $\widehat{\boldsymbol{\Lambda}}_{0}^{-1}(\boldsymbol{\gamma})\widehat{\dot{\boldsymbol{\pi}}}(\boldsymbol{\gamma},\boldsymbol{\theta})$.
At the one-step estimator, 
\[
\widehat{Q}_{\widetilde{\boldsymbol{\theta}}}(\widehat{\boldsymbol{\Gamma}},\widehat{\boldsymbol{\theta}}^{*})=\left\Vert \widehat{\boldsymbol{H}}(\widehat{\boldsymbol{\Gamma}},\widetilde{\boldsymbol{\theta}})\widehat{\boldsymbol{\Lambda}}_{0}^{-1}(\widehat{\boldsymbol{\Gamma}})\widehat{\boldsymbol{\xi}}_{\widetilde{\boldsymbol{\theta}}}(\widehat{\boldsymbol{\Gamma}},\widetilde{\boldsymbol{\theta}})\right\Vert ^{2}.
\]

The feasible Pearson statistic is $\widehat{X}_{\widehat{\boldsymbol{\theta}}^{*}}^{2}(\widehat{\boldsymbol{\Gamma}},\widehat{\boldsymbol{\theta}}^{*})$.
Its observed cells and model-implied expected probabilities are both
evaluated at $\widehat{\boldsymbol{\theta}}^{*}$. Since $\widehat{\boldsymbol{\pi}}_{\boldsymbol{\theta}}(\boldsymbol{\gamma},\boldsymbol{\theta})=\boldsymbol{v}\otimes\widehat{\boldsymbol{q}}(\boldsymbol{\gamma})$,
\[
\widehat{X}_{\widehat{\boldsymbol{\theta}}^{*}}^{2}(\widehat{\boldsymbol{\Gamma}},\widehat{\boldsymbol{\theta}}^{*})=\sum_{\ell=1}^{L}\sum_{j=1}^{J}\frac{\left[\widehat{O}_{\ell j}(\widehat{\boldsymbol{\Gamma}},\widehat{\boldsymbol{\theta}}^{*})-nv_{\ell}\widehat{q}_{j}(\widehat{\boldsymbol{\Gamma}})\right]^{2}}{nv_{\ell}\widehat{q}_{j}(\widehat{\boldsymbol{\Gamma}})}.
\]
Its asymptotic distribution is established by showing that 
\[
\widehat{X}_{\widehat{\boldsymbol{\theta}}^{*}}^{2}(\widehat{\boldsymbol{\Gamma}},\widehat{\boldsymbol{\theta}}^{*})=\widehat{X}_{\boldsymbol{\theta}_{0}}^{2}(\widehat{\boldsymbol{\Gamma}},\widehat{\boldsymbol{\theta}}^{*})+o_{P}(1).
\]

\begin{description}
\item [{A3:}] ~ 
\begin{enumerate}
\item[(i)] The preliminary estimator satisfies $\sqrt{n}(\widetilde{\boldsymbol{\theta}}-\boldsymbol{\theta}_{0})=O_{P}(1)$. 
\item[(ii)] The map $\boldsymbol{\theta}\mapsto F_{\boldsymbol{\theta}}$, with
$F_{\boldsymbol{\theta}}\in\mathcal{F}$, is $L^{2}(P)$-continuous
at $\boldsymbol{\theta}_{0}$; that is, $\left\Vert F_{\boldsymbol{\theta}}-F_{\boldsymbol{\theta}_{0}}\right\Vert _{P,2}\to0$ as $\boldsymbol{\theta}\to\boldsymbol{\theta}_{0}$. 
\item[(iii)] The map $(\boldsymbol{\alpha},\boldsymbol{\beta})\mapsto\boldsymbol{\Phi}_{\boldsymbol{\alpha},\boldsymbol{\beta}}$
is continuously $L^{2}(P_{\boldsymbol{X}})$-differentiable in a neighborhood
of $(\boldsymbol{\theta}_{0},\boldsymbol{\theta}_{0})$, with partial
derivatives $\dot{\boldsymbol{\Phi}}_{1,\boldsymbol{\alpha},\boldsymbol{\beta}}$
and $\dot{\boldsymbol{\Phi}}_{2,\boldsymbol{\alpha},\boldsymbol{\beta}}$. 
\item[(iv)] The derivative class $\dot{\mathcal{M}}=\left\{ \boldsymbol{x}\mapsto\dot{\boldsymbol{\Phi}}_{\boldsymbol{\theta}}(\boldsymbol{x}):\boldsymbol{\theta}\in\mathcal{N}_{\epsilon}\right\} $
is $P_{\boldsymbol{X}}$-Glivenko--Cantelli and admits a $P_{\boldsymbol{X}}$-integrable
envelope. 
\item[(v)] For each partition boundary $\vartheta_{\ell}$, $P\left\{ F_{\boldsymbol{\theta}_{0}}(Y\mid\boldsymbol{X})=\vartheta_{\ell}\right\} =0$,
$\ell=0,\ldots,L$. 
\item[(vi)] For each $\ell=0,\ldots,L$, the class $\left\{ (y,\boldsymbol{x})\mapsto1\left\{ F_{\boldsymbol{\theta}}(y\mid\boldsymbol{x})\leq\vartheta_{\ell}\right\} :\boldsymbol{\theta}\in\mathcal{N}_{\epsilon}\right\} $
is $P$-Donsker. 
\item[(vii)] The matrix $\boldsymbol{J}_{0}(\boldsymbol{\Gamma})$ is positive
definite; see \eqref{eq:36}.
\end{enumerate}
\end{description}
By A3(iv), A2(vii), and the product Glivenko--Cantelli preservation
property, 
\[
\sup_{\boldsymbol{\theta}\in\mathcal{N}_{\epsilon},\,
\boldsymbol{\gamma}\in\mathcal{G}
}\left\Vert \widehat{\dot{\boldsymbol{\pi}}}(\boldsymbol{\gamma},\boldsymbol{\theta})-\dot{\boldsymbol{\pi}}(\boldsymbol{\gamma},\boldsymbol{\theta})\right\Vert =o_{P}(1).
\]
Together with $\widehat{\boldsymbol{\Gamma}}\to_{P}\boldsymbol{\Gamma}$
and $\widetilde{\boldsymbol{\theta}}\to_{P}\boldsymbol{\theta}_{0}$,
this yields $\widehat{\dot{\boldsymbol{\pi}}}(\widehat{\boldsymbol{\Gamma}},\widetilde{\boldsymbol{\theta}})\to_{P}\dot{\boldsymbol{\pi}}_{0}(\boldsymbol{\Gamma})$.

Assumptions A3(ii)--A3(vi), together with the $\sqrt{n}$-consistency
of $\widetilde{\boldsymbol{\theta}}$, imply 
\[
\widehat{\boldsymbol{\xi}}_{\widetilde{\boldsymbol{\theta}}}(\widehat{\boldsymbol{\Gamma}},\widetilde{\boldsymbol{\theta}})=\widehat{\boldsymbol{\xi}}_{\boldsymbol{\theta}_{0}}(\widehat{\boldsymbol{\Gamma}},\widetilde{\boldsymbol{\theta}})+o_{P}(1).
\]
They also imply that replacing $\boldsymbol{\theta}_{0}$ by $\widehat{\boldsymbol{\theta}}^{*}$
in the transformation changes the Pearson statistic by $o_{P}(1)$.
Therefore, 
\[
\widehat{X}_{\widehat{\boldsymbol{\theta}}^{*}}^{2}(\widehat{\boldsymbol{\Gamma}},\widehat{\boldsymbol{\theta}}^{*})=\widehat{Q}_{\widetilde{\boldsymbol{\theta}}}(\widehat{\boldsymbol{\Gamma}},\widehat{\boldsymbol{\theta}}^{*})+o_{P}(1).
\]

\begin{thm}
\label{thm:feasible} Suppose that $H_{0}$, A0, A2,
and A3 hold. Then $\widehat{X}_{\widehat{\boldsymbol{\theta}}^{*}}^{2}(\widehat{\boldsymbol{\Gamma}},\widehat{\boldsymbol{\theta}}^{*})\rightarrow_{d}\chi_{J(L-1)-p}^{2}$. 
\end{thm}

\section{The Wald Test}

\label{sec:wald-tests}

An alternative feasible procedure can be constructed using an estimator
of $\boldsymbol{\theta}_{0}$ based on the ungrouped observations.
\citet{10.1214/aoms/Chernoff} showed that substituting such an estimator
into the Pearson statistic generally produces a limiting distribution
that is a weighted sum of independent chi-squared random variables.
\citet{nikulin1973chi} and, independently, \citet{rao1974chi} proposed
a Wald-type correction based on the asymptotic covariance matrix of
the difference between observed and model-implied cell frequencies. \citet{heckman1984chi} and \citet{andrews1988chi} extended this approach to specification
tests for conditional distributions using general $\sqrt{n}$-consistent
estimators.

The Wald statistic is a quadratic form in the frequency residuals,
weighted by a consistent estimator of a generalized inverse of their
asymptotic covariance matrix. Under suitable regularity conditions,
it converges to a chi-squared distribution with degrees of freedom
equal to the rank of the limiting covariance matrix; see
\citet{andrews1988chi,andrews1988theory}. In important cases this rank
can be characterized a priori. It can be interpreted as a GMM-type
$J$-statistic based on the $LJ$ overidentifying conditions, of which
$J(L-1)$ are linearly independent because of the adding-up restrictions.
For the conditional-CDF problem and the partitions considered here,
use of the conditional maximum likelihood estimator based on the ungrouped
observations yields a limiting covariance matrix whose rank is known
and equals $J(L-1)$, provided that parameter estimation introduces no
additional degeneracy on the $J(L-1)$-dimensional multinomial support.
Thus the Wald statistic has the standard $\chi^2_{J(L-1)}$ null distribution.

Suppose that, for each $\boldsymbol{x}$, $F_{\boldsymbol{\theta}}(\cdot\mid\boldsymbol{x})$
admits a density $f_{\boldsymbol{\theta}}(\cdot\mid\boldsymbol{x})$
with respect to a $\sigma$-finite dominating measure $\mu(\cdot\mid\boldsymbol{x})$.
Let 
\[
\widehat{\boldsymbol{\theta}}^{\dagger}\in\arg\max_{\boldsymbol{\theta}\in\Theta}\sum_{i=1}^{n}\log f_{\boldsymbol{\theta}}(Y_{i}\mid\boldsymbol{X}_{i})
\]
denote the conditional maximum likelihood estimator based on the ungrouped
observations. Define the conditional score and Fisher information
matrix by 
\[
\mathbb{I}(\boldsymbol{\theta})=\mathbb{E}\!\left[\boldsymbol{s}_{\boldsymbol{\theta}}\boldsymbol{s}_{\boldsymbol{\theta}}^{\text{\textsc{t}}}\right], \qquad
\boldsymbol{s}_{\boldsymbol{\theta}}(y,\boldsymbol{x})=\frac{\partial}{\partial\boldsymbol{\theta}}\log f_{\boldsymbol{\theta}}(y\mid\boldsymbol{x}).
\]
The efficient influence function associated with the conditional MLE
is 
\[
\boldsymbol{\ell}_{\boldsymbol{\theta}}(y,\boldsymbol{x})=\mathbb{I}^{-1}(\boldsymbol{\theta})\boldsymbol{s}_{\boldsymbol{\theta}}(y,\boldsymbol{x}).
\]

The following conditions are used to derive the asymptotic distribution
of the frequency residuals evaluated at $\widehat{\boldsymbol{\theta}}^{\dagger}$. 
\begin{description}
\item [{A4:}] ~ 
\begin{enumerate}
\item[(i)] The conditional density model is differentiable in quadratic mean
at $\boldsymbol{\theta}_{0}$, $\mathbb{I}(\boldsymbol{\theta}_{0})$
is nonsingular, and the conditional MLE admits the asymptotic linear
representation 
\[
\sqrt{n}\left(\widehat{\boldsymbol{\theta}}^{\dagger}-\boldsymbol{\theta}_{0}\right)=\frac{1}{\sqrt{n}}\sum_{i=1}^{n}\boldsymbol{\ell}_{\boldsymbol{\theta}_{0}}(Y_{i},\boldsymbol{X}_{i})+o_{P}(1).
\]
 
\item[(ii)] The matrix $\mathbb{I}(\boldsymbol{\theta}_{0})-\boldsymbol{J}_{0}(\boldsymbol{\Gamma})$
is positive definite, and 
$\widehat{\mathbb{I}}(\widehat{\boldsymbol{\theta}}^{\dagger})\to_{P}\mathbb{I}(\boldsymbol{\theta}_{0})$. 
\end{enumerate}
\end{description}
Assumption A4(i) is the standard asymptotic linearity condition for
the conditional MLE. Assumption A4(ii) requires the Fisher information
in the ungrouped observations to strictly exceed the information contained
in the grouped cell probabilities. This condition is stronger than
is needed merely to define a Wald statistic: its role is to ensure
that parameter estimation introduces no additional degeneracy on the
$J(L-1)$-dimensional multinomial support. The convergence in A4(ii) holds, for example, under local
$P$-a.s.\ continuity of the score in $\boldsymbol{\theta}$,
together with a local $L^{2}(P)$ envelope.

To describe the limiting covariance matrix, define, for $\boldsymbol{\gamma}\in\mathcal{G}$,
$\boldsymbol{B}_{0}(\boldsymbol{\gamma})=\boldsymbol{\Lambda}_{0}^{-1}(\boldsymbol{\gamma})\dot{\boldsymbol{\pi}}_{0}(\boldsymbol{\gamma})$,
and $\boldsymbol{S}_{0}(\boldsymbol{\gamma})=\boldsymbol{B}_{0}(\boldsymbol{\gamma})\mathbb{I}^{-1}(\boldsymbol{\theta}_{0})\boldsymbol{B}_{0}^{\text{\textsc{t}}}(\boldsymbol{\gamma})$.
Recall that $\boldsymbol{\Omega}=\left(\boldsymbol{I}_{L}-\sqrt{\boldsymbol{v}}\sqrt{\boldsymbol{v}}^{\text{\textsc{t}}}\right)\otimes\boldsymbol{I}_{J}$,
and set $\boldsymbol{\Sigma}_{0}(\boldsymbol{\gamma})=\boldsymbol{\Omega}-\boldsymbol{S}_{0}(\boldsymbol{\gamma})$.
The matrix $\boldsymbol{S}_{0}(\boldsymbol{\gamma})$ is the reduction
in the standardized covariance matrix caused by estimating $\boldsymbol{\theta}_{0}$
with the efficient conditional MLE. Unlike the projection matrix $\boldsymbol{R}_{0}(\boldsymbol{\gamma})$
appearing in Theorems~\ref{thm:fixed-partition}--\ref{thm:feasible}, $\boldsymbol{S}_{0}(\boldsymbol{\gamma})$
is not generally idempotent.

The adding-up restrictions imply that $\boldsymbol{S}_{0}(\boldsymbol{\gamma})$
acts only on the range of $\boldsymbol{\Omega}$. To see this, let
$\boldsymbol{\Upsilon}=\sqrt{\boldsymbol{v}}\sqrt{\boldsymbol{v}}^{\text{\textsc{t}}}\otimes\boldsymbol{I}_{J}$, so that $\boldsymbol{\Omega}=\boldsymbol{I}_{LJ}-\boldsymbol{\Upsilon}$.
Since $\sqrt{\boldsymbol{v}}^{\text{\textsc{t}}}\operatorname{diag}^{-1/2}(\boldsymbol{v})=\boldsymbol{\iota}_{L}^{\text{\textsc{t}}}$, where $\boldsymbol{\iota}_{L}$ denotes the $L\times 1$ vector of ones, and the components of $\boldsymbol{\Phi}_{\boldsymbol{\theta}_{0},\boldsymbol{\theta}}(\boldsymbol{x})$
sum to one for every $\boldsymbol{\theta}$, $\boldsymbol{\iota}_{L}^{\text{\textsc{t}}}\dot{\boldsymbol{\Phi}}_{0}(\boldsymbol{x})=\boldsymbol{0}^{\text{\textsc{t}}}$.
Consequently, 
\begin{equation}
\boldsymbol{\Upsilon}\boldsymbol{B}_{0}(\boldsymbol{\gamma})=\boldsymbol{0},\qquad\boldsymbol{\gamma}\in\mathcal{G},\label{eq:wald-orthogonality}
\end{equation}
and therefore $\boldsymbol{\Upsilon}\boldsymbol{S}_{0}(\boldsymbol{\gamma})=\boldsymbol{S}_{0}(\boldsymbol{\gamma})\boldsymbol{\Upsilon}=\boldsymbol{0}$.

Under A4(ii), we use the following generalized inverse of the limiting
covariance matrix: 
\begin{equation}
\boldsymbol{\Sigma}_{0}^{-}(\boldsymbol{\gamma})=\left[\boldsymbol{I}_{LJ}-\boldsymbol{S}_{0}(\boldsymbol{\gamma})\right]^{-1}.\label{eq:wald-generalized-inverse}
\end{equation}
Its validity and the corresponding rank result are established in
the proof of Theorem~\ref{thm:wald}.

We now construct sample analogs. Let 
\[
\widehat{\mathbb{I}}(\boldsymbol{\theta})=\frac{1}{n}\sum_{i=1}^{n}\boldsymbol{s}_{\boldsymbol{\theta}}(Y_{i},\boldsymbol{X}_{i})\boldsymbol{s}_{\boldsymbol{\theta}}^{\text{\textsc{t}}}(Y_{i},\boldsymbol{X}_{i}).
\]
Using the derivative notation of Section~\ref{sec:feasible-test}, define $\widehat{\boldsymbol{B}}(\boldsymbol{\gamma},\boldsymbol{\theta})=\widehat{\boldsymbol{\Lambda}}_{0}^{-1}(\boldsymbol{\gamma})\widehat{\dot{\boldsymbol{\pi}}}(\boldsymbol{\gamma},\boldsymbol{\theta})$,
and $\widehat{\boldsymbol{S}}(\boldsymbol{\gamma},\boldsymbol{\theta})=\allowbreak\widehat{\boldsymbol{B}}(\boldsymbol{\gamma},\boldsymbol{\theta})\widehat{\mathbb{I}}^{-1}(\boldsymbol{\theta})\widehat{\boldsymbol{B}}^{\text{\textsc{t}}}(\boldsymbol{\gamma},\boldsymbol{\theta})$.
For brevity, write $\widehat{\boldsymbol{S}}(\widehat{\boldsymbol{\Gamma}})=\widehat{\boldsymbol{S}}(\widehat{\boldsymbol{\Gamma}},\widehat{\boldsymbol{\theta}}^{\dagger})$
and $\allowbreak\widehat{\boldsymbol{\Sigma}}(\widehat{\boldsymbol{\Gamma}})=\boldsymbol{\Omega}-\widehat{\boldsymbol{S}}(\widehat{\boldsymbol{\Gamma}})$.
Under the consistency conditions used below, $\widehat{\boldsymbol{S}}(\widehat{\boldsymbol{\Gamma}})\to_{P}\boldsymbol{S}_{0}(\boldsymbol{\Gamma})$ and $\widehat{\boldsymbol{\Sigma}}(\widehat{\boldsymbol{\Gamma}})\to_{P}\boldsymbol{\Sigma}_{0}(\boldsymbol{\Gamma})$.

The standardized frequency residual at the conditional MLE
is $\widehat{\boldsymbol{\Lambda}}_{0}^{-1}(\widehat{\boldsymbol{\Gamma}})\widehat{\boldsymbol{\xi}}_{\widehat{\boldsymbol{\theta}}^{\dagger}}(\widehat{\boldsymbol{\Gamma}},\widehat{\boldsymbol{\theta}}^{\dagger})$.
Under $H_{0}$, A2(i)--A2(vii), A3(ii)--A3(vi), and A4(i), the expansion
established for the feasible transformation in Section~\ref{sec:feasible-test}, applied
to $\widehat{\boldsymbol{\theta}}^{\dagger}=\boldsymbol{\theta}_{0}+O_{P}(n^{-1/2})$,
together with the asymptotic linear representation in A4(i), yields
\begin{equation*}
\widehat{\boldsymbol{\Lambda}}_{0}^{-1}(\widehat{\boldsymbol{\Gamma}})\widehat{\boldsymbol{\xi}}_{\widehat{\boldsymbol{\theta}}^{\dagger}}(\widehat{\boldsymbol{\Gamma}},\widehat{\boldsymbol{\theta}}^{\dagger})\rightarrow_{d}N_{LJ}\left(\boldsymbol{0},\boldsymbol{\Sigma}_{0}(\boldsymbol{\Gamma})\right).
\end{equation*}
Notice that A2(viii), which concerns the infeasible minimum likelihood-ratio
estimator, and A3(i), which concerns the preliminary estimator used
by the one-step procedure, are not needed for the Wald statistic.

Under A4(ii), take $\widehat{\boldsymbol{\Sigma}}^{-}(\widehat{\boldsymbol{\Gamma}})=\left[\boldsymbol{I}_{LJ}-\widehat{\boldsymbol{S}}(\widehat{\boldsymbol{\Gamma}})\right]^{-1}$,
which is consistent for \eqref{eq:wald-generalized-inverse}. Define
\[
\widehat{W}(\widehat{\boldsymbol{\Gamma}})=\widehat{\boldsymbol{\xi}}_{\widehat{\boldsymbol{\theta}}^{\dagger}}^{\text{\textsc{t}}}(\widehat{\boldsymbol{\Gamma}},\widehat{\boldsymbol{\theta}}^{\dagger})\widehat{\boldsymbol{\Lambda}}_{0}^{-1}(\widehat{\boldsymbol{\Gamma}})\widehat{\boldsymbol{\Sigma}}^{-}(\widehat{\boldsymbol{\Gamma}})\widehat{\boldsymbol{\Lambda}}_{0}^{-1}(\widehat{\boldsymbol{\Gamma}})\widehat{\boldsymbol{\xi}}_{\widehat{\boldsymbol{\theta}}^{\dagger}}(\widehat{\boldsymbol{\Gamma}},\widehat{\boldsymbol{\theta}}^{\dagger}).
\]
Equivalently, under A4(ii), with probability tending to one, 
\[
\widehat{W}(\widehat{\boldsymbol{\Gamma}})=\left\Vert \left[\boldsymbol{I}_{LJ}-\widehat{\boldsymbol{S}}(\widehat{\boldsymbol{\Gamma}})\right]^{-1/2}\widehat{\boldsymbol{\Lambda}}_{0}^{-1}(\widehat{\boldsymbol{\Gamma}})\widehat{\boldsymbol{\xi}}_{\widehat{\boldsymbol{\theta}}^{\dagger}}(\widehat{\boldsymbol{\Gamma}},\widehat{\boldsymbol{\theta}}^{\dagger})\right\Vert ^{2}.
\]

\begin{thm}
\label{thm:wald} Suppose that $H_{0}$, A0, A2(i)--A2(vii),
A3(ii)--A3(vi), and A4 hold. Then $\widehat{W}(\widehat{\boldsymbol{\Gamma}})\rightarrow_{d}\chi_{J(L-1)}^{2}$.
\end{thm}

\section{Power}
\label{sec:power}

Section~\ref{sec:testing-procedure} noted that, for fixed $L$ and $J$, the test is local
to the chosen partition and is therefore not omnibus. We first formalize the corresponding
consistency result under fixed alternatives and then characterize local asymptotic power.

For a fixed alternative, define the population cell-probability vector obtained by evaluating
the Rosenblatt transform at $\boldsymbol{\theta}$ as
\[
\boldsymbol{\pi}^{A}(\boldsymbol{\gamma},\boldsymbol{\theta})
=
\left(
P\left\{
F_{\boldsymbol{\theta}}(Y\mid\boldsymbol X)\in I_\ell,
\ \boldsymbol X\in\gamma_j
\right\}:
\ell=1,\ldots,L,\ j=1,\ldots,J
\right)^{\text{\textsc{t}}}.
\]
Let $\boldsymbol{\theta}^{*}$ denote the probability limit of the feasible estimator under
the alternative and let $\boldsymbol{\Gamma}$ denote the probability limit of the estimated
partition. Since the cell probabilities implied by correct specification are
$\boldsymbol v\otimes\boldsymbol q(\boldsymbol\gamma)$, define
\[
\boldsymbol{\Delta}(\boldsymbol\gamma)
=
\boldsymbol{\pi}^{A}(\boldsymbol\gamma,\boldsymbol\theta^{*})
-
\boldsymbol v\otimes\boldsymbol q(\boldsymbol\gamma).
\]
Recall from (\ref{eq:population-probability-weighting}) that
$\boldsymbol\Lambda_0(\boldsymbol\gamma)
=\operatorname{diag}^{1/2}\{\boldsymbol v\otimes\boldsymbol q(\boldsymbol\gamma)\}$.
Suppose that, under the fixed alternative,
$\widehat{\boldsymbol\theta}^{*}\to_P\boldsymbol\theta^{*}$ and
$\widehat{\boldsymbol\Gamma}\to_P\boldsymbol\Gamma$, and that the corresponding
sample cell frequencies converge to their population counterparts. If every component of
$\boldsymbol v\otimes\boldsymbol q(\boldsymbol\Gamma)$ is strictly positive and
$\boldsymbol\Delta(\boldsymbol\Gamma)\neq\boldsymbol 0$, then
\begin{equation*}
 n^{-1}\widehat X^2_{\widehat{\boldsymbol\theta}^{*}}
 \left(\widehat{\boldsymbol\Gamma},\widehat{\boldsymbol\theta}^{*}\right)
 \to_P
 \boldsymbol\Delta^{\text{\textsc{t}}}(\boldsymbol\Gamma)
 \boldsymbol\Lambda_0^{-2}(\boldsymbol\Gamma)
 \boldsymbol\Delta(\boldsymbol\Gamma)>0.
 \label{eq:fixed-power}
\end{equation*}
Consequently, the statistic diverges at rate $n$, and the test is consistent against such
fixed alternatives. This is the fixed-partition consistency notion already implicit in
$H_1^{*}$: alternatives that leave the selected cell probabilities unchanged are not detected.

A more informative characterization is obtained under alternatives contiguous to the null.
In the remainder of this section, $P$ denotes the joint probability measure under the null
model $F_{\boldsymbol\theta_0}$ and the marginal distribution $P_{\boldsymbol X}$. Let $P_n$
denote the joint probability measure induced by the conditional distribution
$F_n(\cdot\mid\cdot)$ and the same marginal distribution $P_{\boldsymbol X}$. Thus, for every
Borel set $A\subset\mathbb R^{1+k}$,
\[
P_n(A)=\int_{\mathbb R^k}\int_{\mathbb R}
1_A(y,\boldsymbol x)F_n(dy\mid\boldsymbol x)P_{\boldsymbol X}(d\boldsymbol x).
\]
Following \citet{delgado2008distribution}, consider the following contiguous alternatives.

\begin{description}
\item [{A5:}] Under $P_n$, the marginal distribution of $\boldsymbol X$ remains
$P_{\boldsymbol X}$, while the conditional distribution of $Y$ given
$\boldsymbol X=\boldsymbol x$ satisfies
\[
\frac{F_n(dy\mid\boldsymbol x)}
     {F_{\boldsymbol\theta_0}(dy\mid\boldsymbol x)}
=1+\frac{t_n(y,\boldsymbol x)}{\sqrt n},
\]
where
\[
\int t_n(y,\boldsymbol x)F_{\boldsymbol\theta_0}(dy\mid\boldsymbol x)=0
\]
for $P_{\boldsymbol X}$-almost every $\boldsymbol x$, and $t_n\to t$ in $L^2(P)$.
The regularity conditions used below continue to hold along the sequence $\{P_n\}$.
\end{description}

The relevant object for the test is the first-order effect of A5 on the $LJ$ cell
probabilities. For
$\boldsymbol\gamma=(\gamma_1,\ldots,\gamma_J)^{\text{\textsc{t}}}\in\mathcal G$, define 
\[
\boldsymbol{\delta}^{1}(\boldsymbol{\gamma})
=\left(\boldsymbol{\delta}^{1}_{1}(\boldsymbol{\gamma})^{\text{\textsc{t}}},
\ldots,\boldsymbol{\delta}^{1}_{L}(\boldsymbol{\gamma})^{\text{\textsc{t}}}\right)^{\text{\textsc{t}}},
\qquad
\boldsymbol{\delta}^{1}_{\ell}(\boldsymbol{\gamma})
=\left(\delta_{\ell}^{1}(\gamma_1),\ldots,\delta_{\ell}^{1}(\gamma_J)\right)^{\text{\textsc{t}}},
\]
where
\begin{equation}
\delta_\ell^1(\gamma)
=
\int 1_\gamma(\boldsymbol x)
\int 1_{I_\ell}\!\left(F_{\boldsymbol\theta_0}(y\mid\boldsymbol x)\right)
 t(y,\boldsymbol x)F_{\boldsymbol\theta_0}(dy\mid\boldsymbol x)
 P_{\boldsymbol X}(d\boldsymbol x).
\label{eq:cell-drift}
\end{equation}
Defining $\delta_{\ell,n}^1(\gamma)$ analogously with $t_n$ in place of $t$, A5 gives
\[
\pi_{\ell,n}(\gamma)
=
\pi_{\ell,0}(\gamma)+n^{-1/2}\delta_{\ell,n}^1(\gamma).
\]
Since $t_n\to t$ in $L^2(P)$, the Cauchy--Schwarz inequality yields
$\delta_{\ell,n}^1(\gamma)\to\delta_\ell^1(\gamma)$, and therefore
\begin{equation}
\boldsymbol\pi_n(\boldsymbol\gamma)
=
\boldsymbol\pi_0(\boldsymbol\gamma)
+n^{-1/2}\boldsymbol\delta^1(\boldsymbol\gamma)
+o(n^{-1/2}).
\label{eq:pi-local-expansion}
\end{equation}
Moreover, the normalization in A5 and $t_n\to t$ in $L^2(P)$ imply
\[
\int t(y,\boldsymbol X)F_{\boldsymbol\theta_0}(dy\mid\boldsymbol X)=0
\qquad P_{\boldsymbol X}\text{-a.s.}
\]
Since $\{I_\ell\}_{\ell=1}^L$ is a partition of $[0,1]$,
\begin{equation}
\sum_{\ell=1}^L\delta_\ell^1(\gamma)
=
\int 1_\gamma(\boldsymbol x)
\left[\int t(y,\boldsymbol x)F_{\boldsymbol\theta_0}(dy\mid\boldsymbol x)\right]
P_{\boldsymbol X}(d\boldsymbol x)=0.
\label{eq:local-adding-up}
\end{equation}
Thus, within each covariate cell, the local alternative redistributes probability mass across
the $L$ transformed-data cells while leaving the marginal probability of the covariate cell
unchanged.

The vector $\boldsymbol\delta^1(\boldsymbol\gamma)$ is the direct first-order effect of the
local alternative on the cell probabilities. The alternative also affects the statistic
indirectly through parameter estimation; see \cite{delgado2008distribution} and references therein. Under $\{P_n\}$, the same first-order
minimization argument used under $H_0$ gives
\begin{equation*}
\sqrt n\{\widehat{\boldsymbol\theta}(\boldsymbol\gamma)-\boldsymbol\theta_0\}
=
\boldsymbol J_0^{-1}(\boldsymbol\gamma)
\dot{\boldsymbol\pi}_0^{\text{\textsc{t}}}(\boldsymbol\gamma)
\boldsymbol\Lambda_0^{-2}(\boldsymbol\gamma)
\widehat{\boldsymbol\xi}_0(\boldsymbol\gamma)
+o_{P_n}(1).
\label{eq:local-estimator-representation}
\end{equation*}
Accordingly, the deterministic local displacement of the grouped estimator is
\begin{equation}
\boldsymbol b(\boldsymbol\gamma)
=
\boldsymbol J_0^{-1}(\boldsymbol\gamma)
\dot{\boldsymbol\pi}_0^{\text{\textsc{t}}}(\boldsymbol\gamma)
\boldsymbol\Lambda_0^{-2}(\boldsymbol\gamma)
\boldsymbol\delta^1(\boldsymbol\gamma),
\label{eq:local-b}
\end{equation}
and the corresponding first-order displacement of the fitted cell probabilities is
\begin{equation}
\boldsymbol\delta^2(\boldsymbol\gamma)
=
\dot{\boldsymbol\pi}_0(\boldsymbol\gamma)\boldsymbol b(\boldsymbol\gamma).
\label{eq:local-estimation-drift}
\end{equation}
Hence the deterministic drift remaining in the cell residuals is
\begin{equation}
\boldsymbol\delta(\boldsymbol\gamma)
=
\boldsymbol\delta^1(\boldsymbol\gamma)-\boldsymbol\delta^2(\boldsymbol\gamma).
\label{eq:net-local-drift}
\end{equation}
Using $\boldsymbol J_0(\boldsymbol\gamma)$ from (\ref{eq:36}) and the projection matrices
$\boldsymbol R_0(\boldsymbol\gamma)$ and $\boldsymbol H_0(\boldsymbol\gamma)$ defined in
(\ref{eq:projection-matrices}), equations (\ref{eq:local-b})--(\ref{eq:net-local-drift}) imply
\begin{equation*}
\boldsymbol\Lambda_0^{-1}(\boldsymbol\gamma)
\boldsymbol\delta(\boldsymbol\gamma)
=
\boldsymbol H_0(\boldsymbol\gamma)
\boldsymbol\Lambda_0^{-1}(\boldsymbol\gamma)
\boldsymbol\delta^1(\boldsymbol\gamma).
\label{eq:projected-local-drift}
\end{equation*}
Thus parameter estimation removes precisely the component of the standardized direct drift
lying in the standardized parametric derivative space.

\begin{thm}
\label{thm:local-power}
Suppose that A0, A2, A3, and A5 hold. Then, under the sequence of contiguous alternatives
$\{P_n\}$ in A5,
\begin{equation*}
\widehat X^2_{\widehat{\boldsymbol\theta}^{*}}
\left(\widehat{\boldsymbol\Gamma},\widehat{\boldsymbol\theta}^{*}\right)
\to_d
\chi^2_{J(L-1)-p}\{\lambda(\boldsymbol\Gamma)\},
\label{eq:local-power-limit}
\end{equation*}
where
\begin{align}
\lambda(\boldsymbol\Gamma)
&=
\left\|
\boldsymbol\Lambda_0^{-1}(\boldsymbol\Gamma)
\boldsymbol\delta(\boldsymbol\Gamma)
\right\|^2
=
\left\|
\boldsymbol H_0(\boldsymbol\Gamma)
\boldsymbol\Lambda_0^{-1}(\boldsymbol\Gamma)
\boldsymbol\delta^1(\boldsymbol\Gamma)
\right\|^2.
\label{eq:lambda-projected}
\end{align}
\end{thm}

Theorem~\ref{thm:local-power} shows the two effects of parameter estimation.
The direct cell-probability drift is $\boldsymbol\delta^1(\boldsymbol\Gamma)$, while
$\boldsymbol\delta^2(\boldsymbol\Gamma)$ is the part reproduced by the local displacement
of the estimator. Only their difference contributes to the noncentrality parameter. In
particular, if
$\boldsymbol\Lambda_0^{-1}(\boldsymbol\Gamma)
\boldsymbol\delta^1(\boldsymbol\Gamma)
\in
\operatorname{col}\{\boldsymbol A_0(\boldsymbol\Gamma)\},$
then $\lambda(\boldsymbol\Gamma)=0$. After grouping, the local departure is
indistinguishable to first order from a local change in the parameter of the null model.
For a given local departure $t$, different partitions generally produce different vectors
$\boldsymbol\delta^1(\boldsymbol\Gamma)$ and hence different noncentrality parameters.
This gives a local-power interpretation to the choice of partition.

\medskip
\noindent\textbf{Example: local conditional heteroskedasticity.}
\medskip

The conditional-normality example in \citet{delgado2008distribution} provides a simple
illustration of the two components of the local drift. Suppose that under the null
\[
Y\mid\boldsymbol X=\boldsymbol x
\sim N\!\left(\mu(\boldsymbol x),\sigma^2\right),
\]
and consider the contiguous heteroskedastic alternative
\begin{equation*}
\sigma_n^2(\boldsymbol x)
=
\sigma^2\left\{1+\frac{g(\boldsymbol x)}{\sqrt n}\right\}.
\label{eq:local-hetero}
\end{equation*}
The corresponding first-order term in A5 is
\[
t(y,\boldsymbol x)
=
\frac{g(\boldsymbol x)}{2}
\left[
\left\{\frac{y-\mu(\boldsymbol x)}{\sigma}\right\}^{2}-1
\right].
\]
Let $z_\ell=\Phi^{-1}(\vartheta_\ell)$, where $\Phi$ and $\phi$
denote the standard normal distribution and density functions.
Substitution into~\eqref{eq:cell-drift}, using
$\int(z^2-1)\phi(z)\,dz=-z\phi(z)$, gives
\begin{equation}
\delta_\ell^1(\gamma_j)
=
\frac{1}{2}
\left[
z_{\ell-1}\phi(z_{\ell-1})
-z_\ell\phi(z_\ell)
\right]
\int_{\gamma_j}g(\boldsymbol x)P_{\boldsymbol X}(d\boldsymbol x).
\label{eq:hetero-cell-drift}
\end{equation}
Thus, the partition of $[0,1]$ determines the first factor in
(\ref{eq:hetero-cell-drift}), whereas the covariate partition determines
the integral of $g(\boldsymbol{x})$ over each cell. A partition that averages
$g(\boldsymbol x)$ nearly to zero within its cells therefore has little
local power against this alternative, while a partition separating
regions with different values of $g(\boldsymbol x)$ can produce larger
cell-probability drifts.

Parameter estimation also affects this drift through $\boldsymbol{\delta}^{2}(\boldsymbol{\Gamma})$ in~\eqref{eq:local-estimation-drift}. In particular, when $\sigma^{2}$ is estimated, write
\[
g(\boldsymbol X)
=
\mathbb{E}[g(\boldsymbol X)] +
\{g(\boldsymbol X)-\mathbb{E}[g(\boldsymbol X)]\}.
\]
The component generated by $\mathbb{E}[g(\boldsymbol X)]$ corresponds to a local change in the common variance parameter and is therefore absorbed by parameter estimation. The centered component cannot, in general, be
reproduced by changing a common variance parameter. Thus, the local power in Theorem~\ref{thm:local-power} is determined by the part of the direct drift $\boldsymbol{\delta}^{1}(\boldsymbol{\Gamma})$ that remains
after subtracting the estimator-induced drift $\boldsymbol{\delta}^{2}(\boldsymbol{\Gamma})$.

The same arguments also characterize the power of the Wald statistic in
Section~\ref{sec:wald-tests}. Under a fixed alternative, let
$\boldsymbol\theta_A^{\dagger}$ denote the probability limit of the conditional
maximum likelihood estimator and define
\[
\boldsymbol\Delta_W(\boldsymbol\Gamma)
=\boldsymbol\pi^A(\boldsymbol\Gamma,\boldsymbol\theta_A^{\dagger})
-\boldsymbol v\otimes\boldsymbol q(\boldsymbol\Gamma).
\]
Provided that the estimated Wald weighting matrix converges to a generalized
inverse that yields a strictly positive quadratic form whenever $\boldsymbol\Delta_W(\boldsymbol\Gamma)\neq\boldsymbol 0$, $\widehat W(\widehat{\boldsymbol\Gamma})$ diverges at rate $n$ for any such alternative. Thus the Wald
test has the same directional-consistency interpretation as the Pearson test:
with fixed $L$ and $J$, it is consistent against fixed alternatives detected by
the selected cell probabilities, but it is not omnibus.

For the contiguous alternatives in A5, let $\boldsymbol b_{\dagger}
=\mathbb E\!\left[\boldsymbol\ell_{\boldsymbol\theta_0}(Y,\boldsymbol X)t(Y,\boldsymbol X)\right]$ denote the deterministic local displacement of the conditional maximum likelihood estimator. The corresponding first-order drift of the Wald frequency
residuals is
\[
\boldsymbol\delta_W(\boldsymbol\Gamma)
=\boldsymbol\delta^1(\boldsymbol\Gamma)
-\dot{\boldsymbol\pi}_0(\boldsymbol\Gamma)\boldsymbol b_{\dagger}.
\]
Hence, under A4(ii) and the regularity conditions used above,
\[
\widehat W(\widehat{\boldsymbol\Gamma})\ \to_d
\chi^2_{J(L-1)}\!\left\{\lambda_W(\boldsymbol\Gamma)\right\},
\qquad
\lambda_W(\boldsymbol\Gamma)
=\boldsymbol\delta_W^{\text{\textsc{t}}}(\boldsymbol\Gamma)
 \boldsymbol\Lambda_0^{-1}(\boldsymbol\Gamma)
 \boldsymbol\Sigma_0^{-}(\boldsymbol\Gamma)
 \boldsymbol\Lambda_0^{-1}(\boldsymbol\Gamma)
 \boldsymbol\delta_W(\boldsymbol\Gamma).
\]
Thus the local-power mechanism is the same as for the Pearson procedure. The
alternative generates a direct drift in the cell probabilities, while parameter
estimation absorbs the component reproduced by a local change in the fitted
parametric model. The difference is that, for the Wald statistic, this component
is determined by the conditional maximum likelihood estimator computed from the
ungrouped observations rather than by the grouped estimator.

\section{Monte Carlo}
\label{sec:monte-carlo}

\subsection{Partitioning algorithm}\label{subsec:algorithms}

\citet{gessaman1970} has proposed a simple deterministic rule to obtain $\widehat{\Gamma}_j$ containing approximately the same number of points $\{\bs{X}_i\}_{i=1}^n$. Let an integer $T\geq 2$ such that $n\geq T^k$ be given, and let $\bs{X} = (X_1,\dots,X_k)^{\textsc{t}}$. Start by splitting $\{\bs{X}_i\}_{i=1}^n$ into $T$ sets having the same number of points, except for perhaps the boundary sets, using hyperplanes perpendicular to the axis of $X_1$. If $k = 1$, the process terminates here. Otherwise, proceed recursively, next partitioning each of the obtained $T$ cylindrical sets. The procedure yields $\widehat{\boldsymbol\Gamma}=(\widehat\Gamma_1,\ldots,\widehat\Gamma_J)^{\textsc{t}}$ with $J=T^k$ so that $n \geq T^k$ is required to make sure that the cells are non-empty. It can be shown that $\max_{j=1,\dots,J}\widehat{N}_j - \min_{j=1,\dots,J}\widehat{N}_j \leq 1$, where $\widehat{N}_j = \sum_{i=1}^n\unit_{\widehat{\Gamma}_j}(\bs{X}_i)$ is the number of points in $\widehat{\Gamma}_j$, $j=1,\dots,J$. One can expect cells with approximately the same number of points to improve finite sample behavior. However, even with $T=L=2$, the total number of cells, $LJ=2^{k+1}$, increases rapidly with $k$ and demands an alternative approach for relatively large values of $k$.

We propose a Random Tree Partition (RTP) resulting in significantly fewer than $T^k$ cells while preserving the possibility of controlling the number of points per cell. Let $T \geq 2$ and $r\geq 1$ be integers. Let 
\begin{equation*}
    \mathcal{I} := (\underbrace{1,\ldots,1}_{r\text{ times}},\underbrace{2,\ldots,2}_{r\text{ times}},\ldots,\underbrace{k,\ldots,k}_{r\text{ times}})^{\textsc t}
\end{equation*}
be a $kr\times 1$ vector. Start with a tree containing a single node (the root) associated with $C := \mbb{R}^k$.
\begin{enumerate}
\item[1.] Select $j\in\mathcal{I}$ uniformly at random.
\item[2.] Using $T-1$ hyperplanes perpendicular to the $X_j$ axis, partition $C$ into $T$ cells $\widehat{C}_1,\dots,\widehat{C}_T$ such that each cell contains approximately the same number of points $\{\bs{X}_i\}_{i=1}^n$. Split the current node associated with $C$ by adding $T$ child nodes associated with $\widehat{C}_1,\dots,\widehat{C}_T$.
\item[3.] Remove one instance of $j$ from $\mathcal{I}$.
\item[4.] If $\mathcal{I}$ has no elements left, the procedure terminates here. Otherwise, select a terminal node associated with $\widehat{C}$ that contains the largest number of points and repeat Step 1 with $C := \widehat{C}$.
\end{enumerate}

Once the algorithm terminates, denote its collection of terminal cells by $\widehat{\boldsymbol\Gamma}=(\widehat\Gamma_1,\ldots,\widehat\Gamma_J)^{\textsc{t}}$. The underlying class $\mathcal{C}$ consists of axis-aligned hyperrectangles in $\mathbb{R}^k$ and therefore has VC dimension at most $2k$. Hence, $\mathcal{C}$ is a VC class and satisfies A2(vii).

Each $j\in\mathcal{I}$ induces a single split, resulting in $kr$ splits in total. As we start with a single-node tree and each split increases the number of terminal nodes by $T-1$, the final tree has $J=1+kr(T-1)$ terminal nodes. To be able to perform all $kr$ splits and avoid empty terminal nodes, we henceforth require $n\geq 1+kr(T-1)$. It can be shown that
\begin{equation*}
\max_{j=1,\dots,J}\widehat{N}_j - \min_{j=1,\dots,J}\widehat{N}_j \leq  1+ (T-1) \min_{j=1,\dots,J}\widehat{N}_j \quad \Longleftrightarrow \quad \frac{\max_{j=1,\dots,J}\widehat{N}_j}{\min_{j=1,\dots,J}\widehat{N}_j} \leq T + \frac{1}{\min_{j=1,\dots,J}\widehat{N}_j}.
\end{equation*}
That is, the number of points in any cell is at most a little more than $T$ times the number of points in any other cell. For simplicity and to avoid high values of $J$ with large $k$, we consider binary trees with median splits so that $T=2$, and the number of cells equals $J=1+kr$.

We emphasize that, for $n\geq J=1+kr(T-1)$, the algorithm guarantees that each $\widehat{\Gamma}_j$, $j=1,\dots,J$, is nonempty and hence that $\widehat{q}_j>0$. Since $v_\ell>0$, $\ell=1,\dots,L$, this also guarantees strictly positive model-implied probabilities for all $JL$ cells. Consequently, the $X^2$ and $G^2$ statistics are always well defined when using an RTP, while the corresponding cell-probability denominators in the Wald statistic are strictly positive. Notice, however, that the RTP only controls the marginal number of observations falling in each $\widehat{\Gamma}_j$ and cannot prevent some of the $JL$ joint cells from having zero observed counts. Such empty cells are admissible for $G^2$, with their contribution defined to be zero, but may become frequent when $JL$ is large relative to $n$.

\subsection{Simulation design}

We consider the null hypothesis
\begin{equation*}
  H_0 : Y \mid \bs{X} \sim \mcc{N}\left(\mu(\bs{X}), \sigma_0^2\right),
\end{equation*}
where the regression model is multivariate linear, i.e., $\mu(\bs{X}) = \beta_{00} + \bs{X}^{\textsc{t}}\bs{\beta}_0$, with $\bs{\beta}_0 = (\beta_{10},\dots,\beta_{k0})^{\textsc{t}}$. We set $\beta_{00}=\dots=\beta_{k0}=\sigma_0^2=1$. Therefore, $\bt_0=(\beta_{00},\bs{\beta}_0^{\textsc{t}},\sigma_0^2)^{\textsc{t}}$, $\bt=(\beta_0,\beta_1,\dots,\beta_k,\sigma^2)^{\textsc{t}}$, and $p=k+2$. The data consist of $\{Y_i, \bs{X}_i\}_{i=1}^n$, where $\{\bs{X}_i\}_{i=1}^n$ are i.i.d.\ observations of $\bs{X}$, a $k\times 1$ vector of independent random variables uniformly distributed on $[0,1]$, and
\begin{equation*}
  Y_i = \mu(\bs{X}_i) + \delta(\bs{X}_i) + \sigma(\bs{X}_i)\cdot\varepsilon_i,  
\end{equation*}
where $\{\varepsilon_i\}_{i=1}^n$ are i.i.d.\ with mean zero and variance one,
\begin{equation*}
  \delta(\bs{x}) = a\cdot \sum_{i=1}^k\ln(x_i), \qquad \text{and} \qquad \sigma(\bs{x}) = \exp\left(-b\cdot\sum_{i=1}^kx_i\right)\frac{\sqrt{(2b)^ke^{2bk}}}{\sqrt{\left(e^{2b}-1\right)^k}},
\end{equation*}
where $\sigma(\cdot)$ is such that $\Var\UB*{\sigma(\bs{X})\varepsilon_i} = \E\UB*{\sigma^2(\bs{X})} = 1$ and $a,b\geq 0$, with $\sigma(\bs{x})=1$ when $b=0$. Under $H_0$, $\{\varepsilon_i\}_{i=1}^n$ are i.i.d.\ standard normal and $a=b=0$. Let $\mu_r = \E\UB*{\varepsilon_i^r}$. We consider the following alternatives.
\begin{enumerate}
\item From linear regression specification: $a=1$, $b=0$, and $\{\varepsilon_i\}_{i=1}^n$ are i.i.d.\ $\mcc{N}(0,1)$.
\item From conditional homoskedasticity: $a=0$, $b=\frac{2}{3}$, and $\{\varepsilon_i\}_{i=1}^n$ are i.i.d.\ $\mcc{N}(0,1)$.
\item From conditional symmetry: $a=b=0$, and $\{\varepsilon_i\}_{i=1}^n$ are i.i.d.\ and follow a skewed generalized $t$ (SGT) distribution proposed by \citet{theodossiou1998}, with parameters $(\lambda,p,q)$. This allows us to generate asymmetric distributions (with $\mu_3>0$) but without excess kurtosis ($\mu_4=3$). We report results of mild ($\text{SGT}_{1/3}$) and strong ($\text{SGT}_{2/3}$) asymmetry using the following parameter values.
  \begin{itemize}
  \item[(i)] $(\lambda,p,q) = (0.2537, 2.9769, 3.8400)$ with $\mu_1 = 0$, $\mu_2=1$, $\mu_3=1/3$, $\mu_4=3$.
  \item[(ii)] $(\lambda,p,q) = (0.9988, 3.2732, 8.6073)$ with $\mu_1 = 0$, $\mu_2=1$, $\mu_3=2/3$, $\mu_4=3$.
  \end{itemize}
\item From conditional mesokurtosis: $a=b=0$, and standardized $\{\varepsilon_i\}_{i=1}^n$ are i.i.d.\ and follow a $t$ distribution. We provide two cases, $t_5$ and $t_{2.1}$, with the latter having heavier tails and, unlike $t_5$, infinite kurtosis.
  \begin{itemize}
  \item[(i)] $\sqrt{5/3}\cdot\varepsilon_i \sim t_5$,  $i=1,\dots,n$, with $\mu_1=0$, $\mu_2=1$, $\mu_3=0$, $\mu_4 = 9$.
  \item[(ii)] $\sqrt{21}\cdot \varepsilon_i \sim t_{2.1}$, $i=1,\dots,n$, with $\mu_1=0$, $\mu_2=1$, $\mu_3$ undefined, $\mu_4 = \infty$.
  \end{itemize}
\end{enumerate}

We use $\widehat{\bt}^{*}$ iterated until convergence as it significantly improves size accuracy. We have run simulations for tests based on $\widehat{X}^2_{\widehat{\bt}^{*}}(\widehat{\bs{\Gamma}},\widehat{\bt}^{*})$, $\widehat{G}^2_{\widehat{\bt}^{*}}(\widehat{\bs{\Gamma}},\widehat{\bt}^{*})$, and $\widehat{W}(\widehat{\bs{\Gamma}})$. We also compare these tests with the omnibus conditional Kolmogorov-Smirnov (KS) bootstrap test proposed by \citet{andrews1997conditional}, which is based on the difference between the sample joint distribution and its restricted version imposing the conditional CDF model specification under $H_0$. Results are based on 4000 Monte Carlo iterations, and the KS bootstrap test is based on 2000 resamples. We report results for $n=50,100$, and $500$.

We first consider balanced partitions $\mathcal{U}$, i.e., partitions with $\vartheta_{\ell} = \ell/L$, $\ell=1,\dots,L$, which do not favor, in principle, any alternative. Next we consider unbalanced partitions, with small cells in the tails. This has been proposed in the classical literature to favor alternatives with heavy-tailed marginal distributions \citep[see, e.g.,][]{kallenberg1985number}. The power of the $X^2$ and $W$ tests, particularly $X^2$, improves for any of the considered alternatives, not only leptokurtic ones. In all the tables we consider $k=1, 10$ to assess the effect of the curse of dimensionality on the different tests and employ the RTP algorithm with $T=2$ to partition $\mbb{R}^k$.

Balanced partitions are considered with $L=3,6,12,24$, where $L=3,6$ are combined with $r=1,2,5,10$, and $L=12,24$ are used only with $r=1$.

Additionally, we consider $J=1$, i.e., without partitioning $\mbb{R}^k$ at all, with $L=2,3$. These cases cannot be used with the $X^2$ statistic because $J(L-1)<p$ and hence $\rank\{\dot{\bs{\pi}}_{0}(\bs{\Gamma})\}<p$. The rejection rates are not provided when $J=1+kr>n$.

Although $G^2$ remains well defined in the presence of zero observed cell counts, sparse joint cells may adversely affect its finite sample performance. The results for $G^2$, however, are almost identical to those of $X^2$. We therefore do not report the results for $G^2$.

\subsection{Results}

\begin{table}[!hb]
  \centering
  \begin{footnotesize}
    \setlength{\tabcolsep}{9pt}
    \renewcommand{\arraystretch}{0.6}
    \begin{tabular}{cccc|ccc|cccHHR|ccc}
      \midrule \midrule \\  [-1.8ex]
      \multirow{2}{*}[-0.4em]{$L$} & \multirow{2}{*}[-0.4em]{$r$} & \multirow{2}{*}[-0.4em]{$J$} & \multirow{2}{*}[-0.4em]{$k$} & \multicolumn{3}{c}{$n = 50$} & \multicolumn{3}{c}{$n = 100$} & \multicolumn{3}{@{}c@{}}{} & \multicolumn{3}{c}{$n = 500$} \\ \cmidrule(lr){5-7} \cmidrule(l){8-10} \cmidrule(lr){14-16}
                                   &  &  &  &  0.01 & 0.05 & 0.10 & 0.01 & 0.05 & 0.10 & 0.01 & 0.05 & 0.1 & 0.01 & 0.05 & 0.10 \\ 
      \midrule \multicolumn{16}{c}{$W$ test using $\widehat{\bt}^{\dagger}$} \\ \midrule 
2 & -- & 1 & 1 & 0.01 & 0.03 & 0.10 & 0.01 & 0.06 & 0.13 & -- & -- & -- & 0.01 & 0.04 & 0.08 \\
3 & -- & 1 & 1 & 0.01 & 0.05 & 0.11 & 0.01 & 0.05 & 0.11 & -- & -- & -- & 0.01 & 0.06 & 0.10 \\
3 & 1 & 2 & 1 & 0.01 & 0.05 & 0.10 & 0.01 & 0.05 & 0.10 & -- & -- & -- & 0.01 & 0.05 & 0.09 \\
3 & 2 & 3 & 1 & 0.01 & 0.04 & 0.10 & 0.01 & 0.05 & 0.10 & -- & -- & -- & 0.01 & 0.05 & 0.10 \\
3 & 5 & 6 & 1 & 0.01 & 0.05 & 0.09 & 0.01 & 0.05 & 0.10 & -- & -- & -- & 0.01 & 0.04 & 0.10 \\
3 & 10 & 11 & 1 & 0.01 & 0.04 & 0.09 & 0.01 & 0.05 & 0.10 & -- & -- & -- & 0.01 & 0.05 & 0.10 \\
6 & 1 & 2 & 1 & 0.01 & 0.05 & 0.09 & 0.01 & 0.05 & 0.09 & -- & -- & -- & 0.01 & 0.04 & 0.09 \\
6 & 2 & 3 & 1 & 0.01 & 0.05 & 0.09 & 0.01 & 0.05 & 0.10 & -- & -- & -- & 0.01 & 0.05 & 0.09 \\
6 & 5 & 6 & 1 & 0.00 & 0.04 & 0.09 & 0.01 & 0.05 & 0.10 & -- & -- & -- & 0.01 & 0.06 & 0.11 \\
6 & 10 & 11 & 1 & 0.00 & 0.04 & 0.08 & 0.01 & 0.04 & 0.09 & -- & -- & -- & 0.01 & 0.05 & 0.10 \\
12 & 1 & 2 & 1 & 0.01 & 0.05 & 0.10 & 0.01 & 0.05 & 0.10 & -- & -- & -- & 0.01 & 0.05 & 0.10 \\
24 & 1 & 2 & 1 & 0.01 & 0.05 & 0.09 & 0.01 & 0.04 & 0.10 & -- & -- & -- & 0.01 & 0.05 & 0.10 \\
      \midrule
2 & -- & 1 & 10 & 0.01 & 0.03 & 0.10 & 0.01 & 0.07 & 0.14 & -- & -- & -- & 0.01 & 0.04 & 0.09 \\
3 & -- & 1 & 10 & 0.01 & 0.05 & 0.11 & 0.01 & 0.05 & 0.10 & -- & -- & -- & 0.01 & 0.05 & 0.09 \\
3 & 1 & 11 & 10 & 0.01 & 0.06 & 0.14 & 0.01 & 0.06 & 0.12 & -- & -- & -- & 0.01 & 0.05 & 0.10 \\
3 & 2 & 21 & 10 & 0.00 & 0.05 & 0.12 & 0.01 & 0.05 & 0.11 & -- & -- & -- & 0.01 & 0.06 & 0.11 \\
3 & 5 & 51 & 10 & -- & -- & -- & 0.00 & 0.02 & 0.07 & -- & -- & -- & 0.01 & 0.05 & 0.10 \\
3 & 10 & 101 & 10 & -- & -- & -- & -- & -- & -- & -- & -- & -- & 0.01 & 0.04 & 0.09 \\
6 & 1 & 11 & 10 & 0.01 & 0.04 & 0.10 & 0.01 & 0.06 & 0.12 & -- & -- & -- & 0.01 & 0.05 & 0.10 \\
6 & 2 & 21 & 10 & 0.00 & 0.03 & 0.09 & 0.01 & 0.04 & 0.10 & -- & -- & -- & 0.01 & 0.04 & 0.09 \\
6 & 5 & 51 & 10 & -- & -- & -- & 0.00 & 0.02 & 0.06 & -- & -- & -- & 0.01 & 0.05 & 0.10 \\
6 & 10 & 101 & 10 & -- & -- & -- & -- & -- & -- & -- & -- & -- & 0.01 & 0.04 & 0.08 \\
12 & 1 & 11 & 10 & 0.01 & 0.05 & 0.11 & 0.01 & 0.05 & 0.10 & -- & -- & -- & 0.01 & 0.05 & 0.10 \\
24 & 1 & 11 & 10 & 0.01 & 0.05 & 0.10 & 0.01 & 0.04 & 0.09 & -- & -- & -- & 0.01 & 0.05 & 0.10 \\
      \midrule \multicolumn{16}{c}{$X^2$ test using $\widehat{\bt}^{*}$} \\ \midrule
3 & 1 & 2 & 1 & 0.01 & 0.05 & 0.15 & 0.01 & 0.06 & 0.11 & -- & -- & -- & 0.01 & 0.06 & 0.11 \\
3 & 2 & 3 & 1 & 0.01 & 0.05 & 0.11 & 0.01 & 0.05 & 0.11 & -- & -- & -- & 0.01 & 0.05 & 0.10 \\
3 & 5 & 6 & 1 & 0.01 & 0.05 & 0.10 & 0.01 & 0.05 & 0.11 & -- & -- & -- & 0.01 & 0.04 & 0.09 \\
3 & 10 & 11 & 1 & 0.00 & 0.04 & 0.10 & 0.01 & 0.04 & 0.09 & -- & -- & -- & 0.01 & 0.05 & 0.10 \\
6 & 1 & 2 & 1 & 0.01 & 0.05 & 0.09 & 0.01 & 0.04 & 0.10 & -- & -- & -- & 0.01 & 0.05 & 0.09 \\
6 & 2 & 3 & 1 & 0.01 & 0.05 & 0.09 & 0.01 & 0.05 & 0.10 & -- & -- & -- & 0.01 & 0.05 & 0.10 \\
6 & 5 & 6 & 1 & 0.01 & 0.04 & 0.09 & 0.01 & 0.05 & 0.09 & -- & -- & -- & 0.01 & 0.05 & 0.10 \\
6 & 10 & 11 & 1 & 0.01 & 0.04 & 0.08 & 0.01 & 0.05 & 0.09 & -- & -- & -- & 0.01 & 0.05 & 0.10 \\
12 & 1 & 2 & 1 & 0.01 & 0.05 & 0.09 & 0.01 & 0.04 & 0.09 & -- & -- & -- & 0.01 & 0.05 & 0.10 \\
24 & 1 & 2 & 1 & 0.01 & 0.06 & 0.10 & 0.01 & 0.06 & 0.10 & -- & -- & -- & 0.01 & 0.05 & 0.09 \\
      \midrule      
3 & 1 & 11 & 10 & 0.23 & 0.30 & 0.36 & 0.06 & 0.13 & 0.21 & -- & -- & -- & 0.01 & 0.06 & 0.13 \\
3 & 2 & 21 & 10 & 0.00 & 0.04 & 0.11 & 0.01 & 0.04 & 0.09 & -- & -- & -- & 0.01 & 0.05 & 0.10 \\
3 & 5 & 51 & 10 & -- & -- & -- & 0.00 & 0.02 & 0.07 & -- & -- & -- & 0.01 & 0.03 & 0.08 \\
3 & 10 & 101 & 10 & -- & -- & -- & -- & -- & -- & -- & -- & -- & 0.00 & 0.04 & 0.08 \\
6 & 1 & 11 & 10 & 0.08 & 0.15 & 0.23 & 0.04 & 0.08 & 0.15 & -- & -- & -- & 0.01 & 0.06 & 0.11 \\
6 & 2 & 21 & 10 & 0.00 & 0.01 & 0.06 & 0.01 & 0.03 & 0.07 & -- & -- & -- & 0.01 & 0.04 & 0.09 \\
6 & 5 & 51 & 10 & -- & -- & -- & 0.00 & 0.01 & 0.05 & -- & -- & -- & 0.01 & 0.04 & 0.08 \\
6 & 10 & 101 & 10 & -- & -- & -- & -- & -- & -- & -- & -- & -- & 0.00 & 0.03 & 0.07 \\
12 & 1 & 11 & 10 & 0.06 & 0.11 & 0.16 & 0.03 & 0.07 & 0.12 & -- & -- & -- & 0.01 & 0.04 & 0.09 \\
24 & 1 & 11 & 10 & 0.05 & 0.08 & 0.12 & 0.02 & 0.06 & 0.11 & -- & -- & -- & 0.01 & 0.05 & 0.10 \\
      \midrule
      \multicolumn{16}{c}{Kolmogorov-Smirnov test using $\widehat{\bt}^{\dagger}$} \\ \midrule 
      \multirow{2}{*}{--} & \multirow{2}{*}{--} & \multirow{2}{*}{--} & 1 & 0.01 & 0.05 & 0.09 & 0.01 & 0.05 & 0.10 & 0.01 & 0.06 & 0.11 & 0.01 & 0.05 & 0.10 \\
                                   & & & 10 & 0.01 & 0.05 & 0.10 & 0.01 & 0.05 & 0.10 & 0.01 & 0.05 & 0.09 & 0.01 & 0.05 & 0.11 \\ 
      \bottomrule
    \end{tabular}
  \end{footnotesize}
  \caption{Size accuracy using balanced partitions ($k = 1, 10$).}
  \label{tab:sizeG}
\end{table}

\begin{sidewaystable}
  \centering
  \begin{scriptsize}
    \centering
    \setlength{\tabcolsep}{7.5pt}
    \renewcommand{\arraystretch}{0.9}
    \begin{tabular}{cccc|ccRc|ccRc|ccRc|ccRc|ccRc|ccRc}
      \midrule \midrule \\  [-1.8ex]
      \multirow{2}{*}[-0.4em]{$L$} & \multirow{2}{*}[-0.4em]{$r$} & \multirow{2}{*}[-0.4em]{$J$} & \multirow{2}{*}[-0.4em]{$k$} & \multicolumn{4}{c}{$a = 1$} & \multicolumn{4}{c}{$b = 2/3$} & \multicolumn{4}{c}{$\text{SGT}_{1/3}$} & \multicolumn{4}{c}{$\text{SGT}_{2/3}$} & \multicolumn{4}{c}{$t_5$} & \multicolumn{4}{c}{$t_{2.1}$} \\ \cmidrule(lr){5-8} \cmidrule(lr){9-12} \cmidrule(lr){13-16} \cmidrule(lr){17-20} \cmidrule(lr){21-24} \cmidrule(lr){25-28}
                                   & & & & $50$ & $100$ & $200$ & $500$ & $50$ & $100$ & $200$ & $500$ & $50$ & $100$ & $200$ & $500$  & $50$ & $100$ & $200$ & $500$ &  $50$ & $100$ & $200$ & $500$  & $50$ & $100$ & $200$ & $500$ \\
      \midrule \multicolumn{28}{c}{$W$ test using $\widehat{\bt}^{\dagger}$} \\ \midrule
2 & -- & 1 & 1 & 0.04 & 0.09 & 0.09 & 0.13 & 0.04 & 0.07 & 0.07 & 0.05 & 0.06 & 0.13 & 0.13 & 0.33 & 0.15 & 0.39 & 0.39 & 0.95 & 0.07 & 0.13 & 0.13 & 0.10 & 0.27 & 0.41 & 0.41 & 0.45 \\
3 & -- & 1 & 1 & 0.06 & 0.08 & 0.08 & 0.21 & 0.06 & 0.07 & 0.07 & 0.12 & 0.09 & 0.16 & 0.16 & 0.66 & 0.28 & 0.54 & 0.54 & 1.00 & 0.16 & 0.29 & 0.29 & 0.81 & 0.66 & 0.91 & 0.91 & 1.00 \\
3 & 1 & 2 & 1 & 0.11 & 0.15 & 0.15 & 0.59 & 0.09 & 0.14 & 0.14 & 0.57 & 0.08 & 0.12 & 0.12 & 0.50 & 0.21 & 0.44 & 0.44 & 1.00 & 0.15 & 0.22 & 0.22 & 0.75 & 0.64 & 0.89 & 0.89 & 1.00 \\
3 & 2 & 3 & 1 & 0.13 & 0.26 & 0.26 & 0.91 & 0.07 & 0.12 & 0.12 & 0.51 & 0.07 & 0.10 & 0.10 & 0.45 & 0.17 & 0.36 & 0.36 & 0.99 & 0.13 & 0.21 & 0.21 & 0.70 & 0.60 & 0.87 & 0.87 & 1.00 \\
3 & 5 & 6 & 1 & 0.20 & 0.44 & 0.44 & 1.00 & 0.07 & 0.11 & 0.11 & 0.47 & 0.06 & 0.09 & 0.09 & 0.31 & 0.13 & 0.26 & 0.26 & 0.98 & 0.11 & 0.17 & 0.17 & 0.60 & 0.53 & 0.82 & 0.82 & 1.00 \\
3 & 10 & 11 & 1 & 0.14 & 0.35 & 0.35 & 0.99 & 0.06 & 0.09 & 0.09 & 0.36 & 0.05 & 0.07 & 0.07 & 0.24 & 0.09 & 0.18 & 0.18 & 0.92 & 0.08 & 0.13 & 0.13 & 0.48 & 0.42 & 0.74 & 0.74 & 1.00 \\
6 & 1 & 2 & 1 & 0.10 & 0.16 & 0.16 & 0.62 & 0.10 & 0.18 & 0.18 & 0.81 & 0.07 & 0.11 & 0.11 & 0.51 & 0.22 & 0.51 & 0.51 & 1.00 & 0.18 & 0.31 & 0.31 & 0.90 & 0.69 & 0.92 & 0.92 & 1.00 \\
6 & 2 & 3 & 1 & 0.13 & 0.23 & 0.23 & 0.89 & 0.09 & 0.16 & 0.16 & 0.78 & 0.07 & 0.10 & 0.10 & 0.42 & 0.16 & 0.41 & 0.41 & 1.00 & 0.16 & 0.27 & 0.27 & 0.86 & 0.64 & 0.91 & 0.91 & 1.00 \\
6 & 5 & 6 & 1 & 0.16 & 0.37 & 0.37 & 0.99 & 0.07 & 0.14 & 0.14 & 0.69 & 0.05 & 0.08 & 0.08 & 0.29 & 0.12 & 0.28 & 0.28 & 1.00 & 0.12 & 0.21 & 0.21 & 0.77 & 0.55 & 0.85 & 0.85 & 1.00 \\
6 & 10 & 11 & 1 & 0.12 & 0.29 & 0.29 & 0.99 & 0.05 & 0.10 & 0.10 & 0.54 & 0.04 & 0.07 & 0.07 & 0.21 & 0.08 & 0.19 & 0.19 & 0.98 & 0.10 & 0.16 & 0.16 & 0.65 & 0.48 & 0.79 & 0.79 & 1.00 \\
12 & 1 & 2 & 1 & 0.10 & 0.15 & 0.15 & 0.59 & 0.09 & 0.15 & 0.15 & 0.79 & 0.07 & 0.10 & 0.10 & 0.46 & 0.21 & 0.53 & 0.53 & 1.00 & 0.19 & 0.29 & 0.29 & 0.90 & 0.67 & 0.92 & 0.92 & 1.00 \\
24 & 1 & 2 & 1 & 0.09 & 0.14 & 0.14 & 0.53 & 0.09 & 0.12 & 0.12 & 0.70 & 0.06 & 0.09 & 0.09 & 0.35 & 0.18 & 0.42 & 0.42 & 1.00 & 0.19 & 0.30 & 0.30 & 0.89 & 0.65 & 0.91 & 0.91 & 1.00 \\
      \midrule
2 & -- & 1 & 10 & 0.06 & 0.24 & 0.24 & 0.86 & 0.03 & 0.08 & 0.08 & 0.08 & 0.04 & 0.12 & 0.12 & 0.35 & 0.09 & 0.33 & 0.33 & 0.94 & 0.05 & 0.11 & 0.11 & 0.09 & 0.13 & 0.30 & 0.30 & 0.40 \\
3 & -- & 1 & 10 & 0.08 & 0.22 & 0.22 & 0.93 & 0.20 & 0.59 & 0.59 & 1.00 & 0.07 & 0.13 & 0.13 & 0.61 & 0.14 & 0.39 & 0.39 & 1.00 & 0.10 & 0.21 & 0.21 & 0.78 & 0.40 & 0.81 & 0.81 & 1.00 \\
3 & 1 & 11 & 10 & 0.09 & 0.14 & 0.14 & 0.74 & 0.17 & 0.44 & 0.44 & 1.00 & 0.08 & 0.07 & 0.07 & 0.22 & 0.10 & 0.16 & 0.16 & 0.90 & 0.09 & 0.11 & 0.11 & 0.45 & 0.27 & 0.60 & 0.60 & 1.00 \\
3 & 2 & 21 & 10 & 0.07 & 0.12 & 0.12 & 0.70 & 0.10 & 0.32 & 0.32 & 1.00 & 0.05 & 0.06 & 0.06 & 0.16 & 0.07 & 0.12 & 0.12 & 0.78 & 0.06 & 0.09 & 0.09 & 0.33 & 0.21 & 0.52 & 0.52 & 1.00 \\
3 & 5 & 51 & 10 & -- & 0.06 & 0.06 & 0.59 & -- & 0.13 & 0.13 & 0.99 & -- & 0.02 & 0.02 & 0.11 & -- & 0.05 & 0.05 & 0.53 & -- & 0.03 & 0.03 & 0.21 & -- & 0.34 & 0.34 & 1.00 \\
3 & 10 & 101 & 10 & -- & -- & -- & 0.46 & -- & -- & -- & 0.97 & -- & -- & -- & 0.08 & -- & -- & -- & 0.35 & -- & -- & -- & 0.14 & -- & -- & -- & 0.99 \\
6 & 1 & 11 & 10 & 0.08 & 0.13 & 0.13 & 0.76 & 0.15 & 0.47 & 0.47 & 1.00 & 0.07 & 0.06 & 0.06 & 0.20 & 0.08 & 0.15 & 0.15 & 0.96 & 0.09 & 0.14 & 0.14 & 0.61 & 0.30 & 0.68 & 0.68 & 1.00 \\
6 & 2 & 21 & 10 & 0.06 & 0.11 & 0.11 & 0.69 & 0.10 & 0.35 & 0.35 & 1.00 & 0.04 & 0.06 & 0.06 & 0.14 & 0.05 & 0.11 & 0.11 & 0.86 & 0.06 & 0.10 & 0.10 & 0.48 & 0.21 & 0.58 & 0.58 & 1.00 \\
6 & 5 & 51 & 10 & -- & 0.05 & 0.05 & 0.56 & -- & 0.14 & 0.14 & 1.00 & -- & 0.02 & 0.02 & 0.10 & -- & 0.04 & 0.04 & 0.59 & -- & 0.04 & 0.04 & 0.31 & -- & 0.39 & 0.39 & 1.00 \\
6 & 10 & 101 & 10 & -- & -- & -- & 0.38 & -- & -- & -- & 0.98 & -- & -- & -- & 0.07 & -- & -- & -- & 0.35 & -- & -- & -- & 0.20 & -- & -- & -- & 1.00 \\
12 & 1 & 11 & 10 & 0.06 & 0.11 & 0.11 & 0.71 & 0.14 & 0.43 & 0.43 & 1.00 & 0.05 & 0.06 & 0.06 & 0.17 & 0.07 & 0.11 & 0.11 & 0.95 & 0.08 & 0.13 & 0.13 & 0.64 & 0.30 & 0.67 & 0.67 & 1.00 \\
24 & 1 & 11 & 10 & 0.06 & 0.09 & 0.09 & 0.62 & 0.13 & 0.37 & 0.37 & 1.00 & 0.05 & 0.06 & 0.06 & 0.13 & 0.05 & 0.09 & 0.09 & 0.87 & 0.08 & 0.13 & 0.13 & 0.60 & 0.29 & 0.66 & 0.66 & 1.00 \\           
      \midrule \multicolumn{28}{c}{$X^2$ test using $\widehat{\bt}^{*}$} \\ \midrule                                
3 & 1 & 2 & 1 & 0.05 & 0.07 & 0.07 & 0.12 & 0.10 & 0.20 & 0.20 & 0.68 & 0.04 & 0.06 & 0.06 & 0.06 & 0.04 & 0.06 & 0.06 & 0.05 & 0.08 & 0.09 & 0.09 & 0.08 & 0.20 & 0.33 & 0.33 & 0.47 \\
3 & 2 & 3 & 1 & 0.17 & 0.30 & 0.30 & 0.93 & 0.10 & 0.13 & 0.13 & 0.55 & 0.05 & 0.05 & 0.05 & 0.05 & 0.06 & 0.06 & 0.06 & 0.05 & 0.08 & 0.09 & 0.09 & 0.07 & 0.14 & 0.25 & 0.25 & 0.43 \\
3 & 5 & 6 & 1 & 0.20 & 0.43 & 0.43 & 1.00 & 0.08 & 0.11 & 0.11 & 0.44 & 0.06 & 0.05 & 0.05 & 0.05 & 0.05 & 0.05 & 0.05 & 0.06 & 0.06 & 0.07 & 0.07 & 0.06 & 0.09 & 0.19 & 0.19 & 0.39 \\
3 & 10 & 11 & 1 & 0.13 & 0.30 & 0.30 & 0.99 & 0.05 & 0.08 & 0.08 & 0.32 & 0.04 & 0.05 & 0.05 & 0.04 & 0.04 & 0.05 & 0.05 & 0.06 & 0.04 & 0.07 & 0.07 & 0.05 & 0.07 & 0.16 & 0.16 & 0.35 \\
6 & 1 & 2 & 1 & 0.05 & 0.07 & 0.07 & 0.13 & 0.10 & 0.17 & 0.17 & 0.81 & 0.06 & 0.07 & 0.07 & 0.20 & 0.12 & 0.18 & 0.18 & 0.67 & 0.06 & 0.05 & 0.05 & 0.08 & 0.11 & 0.20 & 0.20 & 0.61 \\
6 & 2 & 3 & 1 & 0.09 & 0.18 & 0.18 & 0.87 & 0.08 & 0.13 & 0.13 & 0.77 & 0.06 & 0.06 & 0.06 & 0.16 & 0.10 & 0.14 & 0.14 & 0.57 & 0.05 & 0.05 & 0.05 & 0.06 & 0.10 & 0.15 & 0.15 & 0.56 \\
6 & 5 & 6 & 1 & 0.14 & 0.33 & 0.33 & 1.00 & 0.08 & 0.12 & 0.12 & 0.66 & 0.05 & 0.06 & 0.06 & 0.11 & 0.08 & 0.11 & 0.11 & 0.40 & 0.05 & 0.04 & 0.04 & 0.06 & 0.08 & 0.13 & 0.13 & 0.47 \\
6 & 10 & 11 & 1 & 0.09 & 0.23 & 0.23 & 0.98 & 0.06 & 0.08 & 0.08 & 0.50 & 0.04 & 0.05 & 0.05 & 0.09 & 0.06 & 0.08 & 0.08 & 0.28 & 0.04 & 0.05 & 0.05 & 0.05 & 0.06 & 0.10 & 0.10 & 0.40 \\
12 & 1 & 2 & 1 & 0.05 & 0.06 & 0.06 & 0.13 & 0.08 & 0.14 & 0.14 & 0.78 & 0.07 & 0.08 & 0.08 & 0.31 & 0.21 & 0.47 & 0.47 & 1.00 & 0.05 & 0.06 & 0.06 & 0.12 & 0.11 & 0.16 & 0.16 & 0.72 \\
24 & 1 & 2 & 1 & 0.06 & 0.07 & 0.07 & 0.14 & 0.08 & 0.13 & 0.13 & 0.66 & 0.07 & 0.08 & 0.08 & 0.30 & 0.19 & 0.42 & 0.42 & 1.00 & 0.06 & 0.07 & 0.07 & 0.17 & 0.13 & 0.22 & 0.22 & 0.93 \\
      \midrule                          
3 & 1 & 11 & 10 & 0.33 & 0.14 & 0.14 & 0.09 & 0.32 & 0.35 & 0.35 & 0.89 & 0.28 & 0.13 & 0.13 & 0.07 & 0.29 & 0.13 & 0.13 & 0.08 & 0.29 & 0.17 & 0.17 & 0.08 & 0.34 & 0.36 & 0.36 & 0.39 \\
3 & 2 & 21 & 10 & 0.04 & 0.04 & 0.04 & 0.14 & 0.05 & 0.12 & 0.12 & 0.79 & 0.04 & 0.04 & 0.04 & 0.05 & 0.04 & 0.04 & 0.04 & 0.06 & 0.04 & 0.05 & 0.05 & 0.05 & 0.05 & 0.10 & 0.10 & 0.31 \\
3 & 5 & 51 & 10 & -- & 0.03 & 0.03 & 0.15 & -- & 0.05 & 0.05 & 0.60 & -- & 0.02 & 0.02 & 0.04 & -- & 0.02 & 0.02 & 0.04 & -- & 0.02 & 0.02 & 0.04 & -- & 0.04 & 0.04 & 0.23 \\
3 & 10 & 101 & 10 & -- & -- & -- & 0.14 & -- & -- & -- & 0.41 & -- & -- & -- & 0.03 & -- & -- & -- & 0.04 & -- & -- & -- & 0.03 & -- & -- & -- & 0.17 \\
6 & 1 & 11 & 10 & 0.17 & 0.08 & 0.08 & 0.17 & 0.16 & 0.14 & 0.14 & 0.78 & 0.15 & 0.08 & 0.08 & 0.08 & 0.15 & 0.08 & 0.08 & 0.29 & 0.16 & 0.09 & 0.09 & 0.06 & 0.21 & 0.24 & 0.24 & 0.41 \\
6 & 2 & 21 & 10 & 0.02 & 0.04 & 0.04 & 0.17 & 0.02 & 0.06 & 0.06 & 0.72 & 0.02 & 0.03 & 0.03 & 0.06 & 0.02 & 0.04 & 0.04 & 0.20 & 0.02 & 0.03 & 0.03 & 0.05 & 0.02 & 0.06 & 0.06 & 0.32 \\
6 & 5 & 51 & 10 & -- & 0.01 & 0.01 & 0.15 & -- & 0.02 & 0.02 & 0.54 & -- & 0.01 & 0.01 & 0.05 & -- & 0.01 & 0.01 & 0.12 & -- & 0.01 & 0.01 & 0.04 & -- & 0.02 & 0.02 & 0.24 \\
6 & 10 & 101 & 10 & -- & -- & -- & 0.13 & -- & -- & -- & 0.34 & -- & -- & -- & 0.04 & -- & -- & -- & 0.07 & -- & -- & -- & 0.03 & -- & -- & -- & 0.16 \\
12 & 1 & 11 & 10 & 0.11 & 0.07 & 0.07 & 0.23 & 0.11 & 0.10 & 0.10 & 0.79 & 0.10 & 0.07 & 0.07 & 0.10 & 0.10 & 0.07 & 0.07 & 0.52 & 0.10 & 0.07 & 0.07 & 0.06 & 0.14 & 0.13 & 0.13 & 0.23 \\
24 & 1 & 11 & 10 & 0.10 & 0.07 & 0.07 & 0.21 & 0.09 & 0.08 & 0.08 & 0.77 & 0.09 & 0.06 & 0.06 & 0.09 & 0.08 & 0.07 & 0.07 & 0.46 & 0.08 & 0.06 & 0.06 & 0.07 & 0.10 & 0.09 & 0.09 & 0.29 \\   
      \midrule
      \multicolumn{28}{c}{Kolmogorov-Smirnov test using $\widehat{\bt}^{\dagger}$} \\ \midrule 
      \multirow{2}{*}{--} & \multirow{2}{*}{--} & \multirow{2}{*}{--} & 1 & 0.09 & 0.20 & 0.20 & 0.90 & 0.07 & 0.07 & 0.07 & 0.13 & 0.13 & 0.17 & 0.17 & 0.63 & 0.30 & 0.57 & 0.57 & 1.00 & 0.15 & 0.23 & 0.23 & 0.83 & 0.58 & 0.89 & 0.89 & 1.00 \\
&  &  & 10 & 0.09 & 0.05 & 0.05 & 0.21 & 0.10 & 0.06 & 0.06 & 0.09 & 0.05 & 0.05 & 0.05 & 0.07 & 0.05 & 0.07 & 0.07 & 0.11 & 0.08 & 0.04 & 0.04 & 0.05 & 0.12 & 0.04 & 0.04 & 0.22 \\
      \bottomrule
    \end{tabular}
  \end{scriptsize}
  \caption{Power under deviations in mean, variance, skewness, and kurtosis using balanced partitions ($k=1,10$ and $\alpha=5\%$).}
  \label{tab:powerG}
\end{sidewaystable}

Table~\ref{tab:sizeG} reports the results under $H_0$. The $W$ and KS tests exhibit excellent size accuracy. The $X^2$ test is also very accurate for $k=1$, while for $k=10$ some size distortions arise when $n$ is small. These distortions diminish rapidly with the sample size and are negligible for $n=500$.

The rejection rates under $H_1$ are reported in Table~\ref{tab:powerG}. For $k=1$, $X^2$ and $W$ generally exhibit good power, with $W$ performing particularly well against asymmetric and leptokurtic alternatives. The $X^2$ test is more sensitive to the choice of partition, with larger values of $L$ tending to improve its power against asymmetry. For $k=10$, $W$ is considerably more robust to the curse of dimensionality than the KS test and generally than $X^2$. Both $X^2$ and $W$ perform particularly well under conditional heteroskedasticity, where the KS test has low power. Overall, $W$ performs strongly across the alternatives, whereas the performance of $X^2$ depends more on the partition choice.

The decline in power as the dimension of the covariate vector increases can be interpreted by analogy with the spectral properties of Cram\'{e}r-von Mises statistics. In the univariate case, their spectral decomposition assigns progressively smaller weights to higher-frequency components, while in higher dimensions the departure from the null may be dispersed over a larger collection of such components \citep{durbin1972components, krivyakova1977distribution}. Although this spectral argument does not apply directly to the supremum-based KS statistic, a similar dispersion of the departure may contribute to the pronounced loss of power observed in our higher-dimensional designs.

Next we present results for unbalanced partitions $\mathcal{U}$ with $\vartheta_{\ell}\neq \ell/L$ and small cells in the tails in the hope of closing the gap between $X^2$ and $W$. We report results for the following partitions $(\vartheta_0,\vartheta_1,\dots,\vartheta_L)$:
\begin{itemize}
\item[(i)] $6^*$ with (0, 0.01, 0.25, 0.5, 0.75, 0.99, 1),
\item[(ii)] $6^{**}$ with (0, 0.05, 0.1, 0.5, 0.9, 0.95, 1),
\item[(iii)] $8^{*}$ with (0, 0.01, 0.06, 0.16, 0.5, 0.84, 0.94, 0.99, 1).
\item[(iv)] $8^{**}$ with (0, 0.01, 0.10, 0.33, 0.5, 0.66, 0.90, 0.99, 1).
\end{itemize}
These unbalanced partitions are intended, in principle, to improve the power in the direction of heavy-tailed alternatives.

Table~\ref{tab:sizeS} reports size accuracy for the unbalanced partitions. For $k=1$, both the $X^2$ and $W$ tests have size accuracy comparable to that of the KS bootstrap test. For $k=10$, size distortions arise for the smaller sample sizes, particularly for $X^2$, but decrease substantially with $n$ and are small for $n=500$.

\begin{table}[!htb]
  \centering
    \setlength{\tabcolsep}{8pt}
    \renewcommand{\arraystretch}{0.85}
    \begin{tabular}{lccc|ccc|cccHHR|ccc}
      \midrule \midrule \\  [-1.8ex]
      \multirow{2}{*}[-0.4em]{$L$} & \multirow{2}{*}[-0.4em]{$r$} & \multirow{2}{*}[-0.4em]{$J$} & \multirow{2}{*}[-0.4em]{$k$} & \multicolumn{3}{c}{$n = 50$} & \multicolumn{3}{c}{$n = 100$} & \multicolumn{3}{@{}c@{}}{} & \multicolumn{3}{c}{$n = 500$} \\ \cmidrule(lr){5-7} \cmidrule(l){8-10} \cmidrule(lr){14-16}
                                   &  &  &  &  0.01 & 0.05 & 0.10 & 0.01 & 0.05 & 0.10 & 0.01 & 0.05 & 0.1 & 0.01 & 0.05 & 0.10 \\ 
      \midrule \multicolumn{16}{c}{$W$ test using $\widehat{\bt}^{\dagger}$} \\ \midrule
$6^*$ & 1 & 2 & 1 & 0.02 & 0.07 & 0.11 & 0.01 & 0.05 & 0.09 & -- & -- & -- & 0.01 & 0.05 & 0.10 \\
$6^{**}$ & 1 & 2 & 1 & 0.01 & 0.05 & 0.10 & 0.01 & 0.05 & 0.10 & -- & -- & -- & 0.01 & 0.04 & 0.10 \\
$8^*$ & 1 & 2 & 1 & 0.01 & 0.04 & 0.09 & 0.01 & 0.05 & 0.10 & -- & -- & -- & 0.01 & 0.05 & 0.10 \\
$8^{**}$ & 1 & 2 & 1 & 0.01 & 0.05 & 0.09 & 0.01 & 0.05 & 0.10 & -- & -- & -- & 0.01 & 0.05 & 0.10 \\                            
      \midrule
$6^*$ & 1 & 11 & 10 & 0.09 & 0.17 & 0.23 & 0.04 & 0.11 & 0.16 & -- & -- & -- & 0.02 & 0.07 & 0.11 \\
$6^{**}$ & 1 & 11 & 10 & 0.04 & 0.11 & 0.17 & 0.02 & 0.07 & 0.13 & -- & -- & -- & 0.01 & 0.06 & 0.10 \\
$8^*$ & 1 & 11 & 10 & 0.05 & 0.13 & 0.19 & 0.03 & 0.09 & 0.14 & -- & -- & -- & 0.01 & 0.06 & 0.10 \\
$8^{**}$ & 1 & 11 & 10 & 0.06 & 0.14 & 0.20 & 0.03 & 0.09 & 0.14 & -- & -- & -- & 0.02 & 0.06 & 0.11 \\
      \midrule \multicolumn{16}{c}{$X^2$ test using $\widehat{\bt}^{*}$} \\ \midrule
$6^*$ & 1 & 2 & 1 & 0.02 & 0.06 & 0.10 & 0.01 & 0.05 & 0.09 & -- & -- & -- & 0.01 & 0.05 & 0.10 \\
$6^{**}$ & 1 & 2 & 1 & 0.02 & 0.07 & 0.13 & 0.02 & 0.06 & 0.11 & -- & -- & -- & 0.01 & 0.06 & 0.11 \\
$8^*$ & 1 & 2 & 1 & 0.01 & 0.05 & 0.10 & 0.01 & 0.05 & 0.09 & -- & -- & -- & 0.01 & 0.05 & 0.11 \\
$8^{**}$ & 1 & 2 & 1 & 0.01 & 0.05 & 0.09 & 0.01 & 0.05 & 0.09 & -- & -- & -- & 0.01 & 0.05 & 0.10 \\                                    
      \midrule
$6^*$ & 1 & 11 & 10 & 0.32 & 0.40 & 0.45 & 0.17 & 0.29 & 0.36 & -- & -- & -- & 0.03 & 0.09 & 0.15 \\
$6^{**}$ & 1 & 11 & 10 & 0.28 & 0.38 & 0.46 & 0.12 & 0.23 & 0.32 & -- & -- & -- & 0.02 & 0.08 & 0.14 \\
$8^*$ & 1 & 11 & 10 & 0.29 & 0.37 & 0.42 & 0.16 & 0.28 & 0.37 & -- & -- & -- & 0.02 & 0.08 & 0.14 \\
$8^{**}$ & 1 & 11 & 10 & 0.26 & 0.34 & 0.39 & 0.14 & 0.26 & 0.34 & -- & -- & -- & 0.03 & 0.08 & 0.14 \\
      \midrule
      \multicolumn{16}{c}{Kolmogorov-Smirnov test using $\widehat{\bt}^{\dagger}$} \\ \midrule 
      \multirow{2}{*}{--} & \multirow{2}{*}{--} & \multirow{2}{*}{--} & 1 & 0.01 & 0.05 & 0.09 & 0.01 & 0.05 & 0.10 & 0.01 & 0.06 & 0.11 & 0.01 & 0.05 & 0.10 \\
                                   & & & 10 & 0.01 & 0.05 & 0.10 & 0.01 & 0.05 & 0.10 & 0.01 & 0.05 & 0.09 & 0.01 & 0.05 & 0.11 \\       
      \bottomrule      
    \end{tabular}
  \caption{Size accuracy using unbalanced partitions ($k = 1, 10$).}
  \label{tab:sizeS}
\end{table}

Table~\ref{tab:powerS} reports power for the unbalanced partitions, together with the best-performing balanced partition for each alternative. The results show that the choice of partition can have a substantial effect on power. In particular, unbalanced partitions considerably improve the performance of the $X^2$ test against several alternatives and substantially reduce the gap with $W$. The gains are especially pronounced for $k=10$, where both $X^2$ and $W$ are generally much more robust to the curse of dimensionality than the KS test. The results also illustrate the directional nature of the procedure: no single partition dominates across all alternatives, but suitable choices can produce large power gains in particular directions.

\begin{sidewaystable}[ph!]
  \centering
  \begin{footnotesize}
    \centering
    \renewcommand{\arraystretch}{0.9}
    \begin{tabular}{lccc|ccRc|ccRc|ccRc|ccRc|ccRc|ccRc}
      \midrule \midrule \\  [-1.8ex]
      \multirow{2}{*}[-0.4em]{$L$} & \multirow{2}{*}[-0.4em]{$r$} & \multirow{2}{*}[-0.4em]{$J$} & \multirow{2}{*}[-0.4em]{$k$} & \multicolumn{4}{c}{$a = 1$} & \multicolumn{4}{c}{$b = 2/3$} & \multicolumn{4}{c}{$\text{SGT}_{1/3}$} & \multicolumn{4}{c}{$\text{SGT}_{2/3}$} & \multicolumn{4}{c}{$t_5$} & \multicolumn{4}{c}{$t_{2.1}$} \\ \cmidrule(lr){5-8} \cmidrule(lr){9-12} \cmidrule(lr){13-16} \cmidrule(lr){17-20} \cmidrule(lr){21-24} \cmidrule(lr){25-28}
                                   & & & & $50$ & $100$ & $200$ & $500$ & $50$ & $100$ & $200$ & $500$ & $50$ & $100$ & $200$ & $500$  & $50$ & $100$ & $200$ & $500$ &  $50$ & $100$ & $200$ & $500$  & $50$ & $100$ & $200$ & $500$ \\
      \midrule \multicolumn{28}{c}{$W$ test using $\widehat{\bt}^{\dagger}$} \\ \midrule
$6^*$ & 1 & 2 & 1 & 0.10 & 0.16 & 0.16 & 0.65 & 0.14 & 0.21 & 0.21 & 0.84 & 0.07 & 0.09 & 0.09 & 0.45 & 0.17 & 0.38 & 0.38 & 1.00 & 0.21 & 0.30 & 0.30 & 0.84 & 0.67 & 0.91 & 0.91 & 1.00 \\
$6^{**}$ & 1 & 2 & 1 & 0.07 & 0.13 & 0.13 & 0.54 & 0.08 & 0.17 & 0.17 & 0.89 & 0.08 & 0.10 & 0.10 & 0.43 & 0.27 & 0.70 & 0.70 & 1.00 & 0.08 & 0.15 & 0.15 & 0.65 & 0.39 & 0.68 & 0.68 & 1.00 \\
$8^*$ & 1 & 2 & 1 & 0.11 & 0.16 & 0.16 & 0.67 & 0.12 & 0.20 & 0.20 & 0.91 & 0.06 & 0.09 & 0.09 & 0.42 & 0.20 & 0.63 & 0.63 & 1.00 & 0.20 & 0.30 & 0.30 & 0.86 & 0.62 & 0.87 & 0.87 & 1.00 \\
$8^{**}$ & 1 & 2 & 1 & 0.11 & 0.17 & 0.17 & 0.65 & 0.13 & 0.21 & 0.21 & 0.90 & 0.06 & 0.09 & 0.09 & 0.38 & 0.20 & 0.39 & 0.39 & 0.99 & 0.21 & 0.31 & 0.31 & 0.87 & 0.70 & 0.92 & 0.92 & 1.00 \\
      \multicolumn{3}{r}{Best balanced} & 1 & 0.20 & 0.44 & 0.44 & 1.00 & 0.10 & 0.18 & 0.18 & 0.81 & 0.09 & 0.16 & 0.16 & 0.66 & 0.28 & 0.54 & 0.54 & 1.00 & 0.19 & 0.31 & 0.31 & 0.90 & 0.69 & 0.92 & 0.92 & 1.00 \\ \midrule
$6^*$ & 1 & 11 & 10 & 0.24 & 0.26 & 0.26 & 0.77 & 0.49 & 0.71 & 0.71 & 1.00 & 0.16 & 0.10 & 0.10 & 0.14 & 0.17 & 0.18 & 0.18 & 0.87 & 0.33 & 0.35 & 0.35 & 0.69 & 0.57 & 0.80 & 0.80 & 1.00 \\
$6^{**}$ & 1 & 11 & 10 & 0.09 & 0.07 & 0.07 & 0.52 & 0.05 & 0.06 & 0.06 & 0.96 & 0.11 & 0.09 & 0.09 & 0.15 & 0.12 & 0.14 & 0.14 & 1.00 & 0.06 & 0.04 & 0.04 & 0.22 & 0.03 & 0.20 & 0.20 & 0.99 \\
$8^*$ & 1 & 11 & 10 & 0.15 & 0.16 & 0.16 & 0.72 & 0.29 & 0.49 & 0.49 & 1.00 & 0.10 & 0.07 & 0.07 & 0.14 & 0.13 & 0.13 & 0.13 & 0.99 & 0.21 & 0.21 & 0.21 & 0.63 & 0.39 & 0.69 & 0.69 & 1.00 \\
$8^{**}$ & 1 & 11 & 10 & 0.19 & 0.21 & 0.21 & 0.76 & 0.42 & 0.67 & 0.67 & 1.00 & 0.12 & 0.08 & 0.08 & 0.12 & 0.14 & 0.13 & 0.13 & 0.75 & 0.27 & 0.30 & 0.30 & 0.70 & 0.53 & 0.81 & 0.81 & 1.00 \\
      \multicolumn{3}{r}{Best balanced} & 10 & 0.09 & 0.24 & 0.24 & 0.93 & 0.20 & 0.59 & 0.59 & 1.00 & 0.08 & 0.13 & 0.13 & 0.61 & 0.14 & 0.39 & 0.39 & 1.00 & 0.10 & 0.21 & 0.21 & 0.78 & 0.40 & 0.81 & 0.81 & 1.00 \\
      \midrule \multicolumn{28}{c}{$X^2$ test using $\widehat{\bt}^{*}$} \\ \midrule
$6^*$ & 1 & 2 & 1 & 0.12 & 0.16 & 0.16 & 0.58 & 0.15 & 0.23 & 0.23 & 0.89 & 0.04 & 0.05 & 0.05 & 0.22 & 0.09 & 0.08 & 0.08 & 0.61 & 0.23 & 0.31 & 0.31 & 0.81 & 0.63 & 0.85 & 0.85 & 1.00 \\
$6^{**}$ & 1 & 2 & 1 & 0.07 & 0.08 & 0.08 & 0.28 & 0.10 & 0.19 & 0.19 & 0.92 & 0.10 & 0.12 & 0.12 & 0.42 & 0.37 & 0.79 & 0.79 & 1.00 & 0.05 & 0.04 & 0.04 & 0.11 & 0.06 & 0.12 & 0.12 & 0.78 \\
$8^*$ & 1 & 2 & 1 & 0.10 & 0.14 & 0.14 & 0.53 & 0.13 & 0.22 & 0.22 & 0.93 & 0.06 & 0.09 & 0.09 & 0.47 & 0.23 & 0.68 & 0.68 & 1.00 & 0.14 & 0.19 & 0.19 & 0.69 & 0.39 & 0.66 & 0.66 & 1.00 \\
$8^{**}$ & 1 & 2 & 1 & 0.11 & 0.15 & 0.15 & 0.54 & 0.15 & 0.23 & 0.23 & 0.91 & 0.06 & 0.08 & 0.08 & 0.40 & 0.18 & 0.38 & 0.38 & 0.99 & 0.19 & 0.25 & 0.25 & 0.74 & 0.55 & 0.78 & 0.78 & 1.00 \\
      \multicolumn{3}{r}{Best balanced} & 1 & 0.20 & 0.43 & 0.43 & 1.00 & 0.10 & 0.20 & 0.20 & 0.81 & 0.07 & 0.08 & 0.08 & 0.31 & 0.21 & 0.47 & 0.47 & 1.00 & 0.08 & 0.09 & 0.09 & 0.17 & 0.20 & 0.33 & 0.33 & 0.93 \\ \midrule
$6^*$ & 1 & 11 & 10 & 0.38 & 0.27 & 0.27 & 0.77 & 0.44 & 0.55 & 0.55 & 1.00 & 0.40 & 0.27 & 0.27 & 0.09 & 0.40 & 0.27 & 0.27 & 0.27 & 0.41 & 0.38 & 0.38 & 0.64 & 0.59 & 0.68 & 0.68 & 1.00 \\
$6^{**}$ & 1 & 11 & 10 & 0.35 & 0.16 & 0.16 & 0.51 & 0.41 & 0.26 & 0.26 & 0.87 & 0.41 & 0.24 & 0.24 & 0.20 & 0.40 & 0.27 & 0.27 & 0.85 & 0.40 & 0.24 & 0.24 & 0.06 & 0.53 & 0.27 & 0.27 & 0.15 \\
$8^*$ & 1 & 11 & 10 & 0.30 & 0.19 & 0.19 & 0.76 & 0.42 & 0.51 & 0.51 & 1.00 & 0.36 & 0.27 & 0.27 & 0.16 & 0.36 & 0.29 & 0.29 & 0.78 & 0.40 & 0.36 & 0.36 & 0.48 & 0.61 & 0.57 & 0.57 & 0.98 \\
$8^{**}$ & 1 & 11 & 10 & 0.30 & 0.22 & 0.22 & 0.78 & 0.41 & 0.54 & 0.54 & 1.00 & 0.34 & 0.25 & 0.25 & 0.13 & 0.34 & 0.24 & 0.24 & 0.59 & 0.37 & 0.38 & 0.38 & 0.55 & 0.59 & 0.62 & 0.62 & 0.99 \\
      \multicolumn{3}{r}{Best balanced} & 10 & 0.33 & 0.14 & 0.14 & 0.23 & 0.32 & 0.35 & 0.35 & 0.89 & 0.28 & 0.13 & 0.13 & 0.10 & 0.29 & 0.13 & 0.13 & 0.52 & 0.29 & 0.17 & 0.17 & 0.08 & 0.34 & 0.36 & 0.36 & 0.41 \\
      \midrule
      \multicolumn{28}{c}{Kolmogorov-Smirnov test using $\widehat{\bt}^{\dagger}$} \\ \midrule 
      \multirow{2}{*}{--} & \multirow{2}{*}{--} & \multirow{2}{*}{--} & 1 & 0.09 & 0.20 & 0.20 & 0.90 & 0.07 & 0.07 & 0.07 & 0.13 & 0.13 & 0.17 & 0.17 & 0.63 & 0.30 & 0.57 & 0.57 & 1.00 & 0.15 & 0.23 & 0.23 & 0.83 & 0.58 & 0.89 & 0.89 & 1.00 \\
&  &  & 10 & 0.09 & 0.05 & 0.05 & 0.21 & 0.10 & 0.06 & 0.06 & 0.09 & 0.05 & 0.05 & 0.05 & 0.07 & 0.05 & 0.07 & 0.07 & 0.11 & 0.08 & 0.04 & 0.04 & 0.05 & 0.12 & 0.04 & 0.04 & 0.22 \\      
      
      \bottomrule
    \end{tabular}
  \end{footnotesize}
  \caption{Power under deviations in mean, variance, skewness, and kurtosis using unbalanced partitions ($k=1,10$ and $\alpha=5\%$).}
  \label{tab:powerS}
\end{sidewaystable}

\section{Concluding Remarks}
\label{sec:final-remarks}

This paper develops a specification test for parametric conditional distribution models by exploiting the characterization provided by the Rosenblatt transform. Cross-classifying the transformed response and the covariates leads to a Pearson statistic with a chi-squared limiting distribution with known degrees of freedom, while accounting for parameter estimation. The same limiting distribution remains valid for the class of convergent data-dependent covariate partitions considered in the paper, allowing the partition to be adapted to features of the data or to alternatives of particular interest. The framework also accommodates estimation from the ungrouped observations through a corresponding Wald statistic.

For fixed $L$ and $J$, the Pearson procedure is directional rather than omnibus. It is
consistent against fixed alternatives that change the cell probabilities selected by the
partition, but no finite partition can detect every departure from the null. The local-power analysis shows that only the component of the local cell-probability drift that is not absorbed by parameter estimation contributes to the noncentrality parameter. In return, the test has a standard chi-squared null distribution and
does not require bootstrap critical values. The simulations indicate that this trade-off can
be favorable, especially in the higher-dimensional designs considered here.

The choice of partition should therefore be viewed as part of the design of the test. For the
partition of $[0,1]$, balanced cells provide a natural default, while unbalanced cells can be
used when departures in particular regions, such as the tails, are of special interest. For the
covariate space, the data-dependent procedures considered here are designed to produce
nonempty cells containing approximately equal numbers of observations, which helps to
control finite-sample behavior. Larger values of $L$ can improve power when the sample
size permits, but no single partition dominates across alternatives. Accordingly, the partition should reflect the departures that are relevant in a particular application. The power results in Section~\ref{sec:power} provide a formal interpretation of this choice.

\newpage
\section*{Appendix}
\label{sec:appendix}

\noindent\textbf{Proof of Theorem~\ref{thm:fixed-partition}.}

Recall that $\widehat{\boldsymbol{p}}(\boldsymbol{\theta}_{0})=n^{-1}\widehat{\boldsymbol{O}}(\boldsymbol{\theta}_{0})$,
$\widehat{\boldsymbol{\pi}}(\boldsymbol{\theta})=n^{-1}\sum_{i=1}^{n}\boldsymbol{\Phi}_{\boldsymbol{\theta}}(\boldsymbol{X}_{i})\otimes\boldsymbol{1}_{\mathcal{A}}(\boldsymbol{X}_{i})$,
and $\boldsymbol{\pi}(\boldsymbol{\theta})=P_{\boldsymbol{X}}\left[\boldsymbol{\Phi}_{\boldsymbol{\theta}}\otimes\boldsymbol{1}_{\mathcal{A}}\right]$.
Under $H_{0}$, $\boldsymbol{\pi}(\boldsymbol{\theta}_{0})=\boldsymbol{\pi}_{0}=\boldsymbol{v}\otimes\boldsymbol{q}$.
Write $\widehat{\boldsymbol{\xi}}_{0}=\sqrt{n}\left[\widehat{\boldsymbol{p}}(\boldsymbol{\theta}_{0})-\widehat{\boldsymbol{\pi}}(\boldsymbol{\theta}_{0})\right]$
and $\boldsymbol{\Lambda}_{0}=\operatorname{diag}^{1/2}(\boldsymbol{\pi}_{0})$.
We first record some elementary facts used below. For probability vectors
$\boldsymbol{a},\boldsymbol{b}\in\mathbb{R}^{m}$, let $K(\boldsymbol{a},\boldsymbol{b})=\sum_{r=1}^{m}a_{r}\log(a_r/b_r)$, with the conventions $0\log(0/b)=0$ for $b\geq0$ and $a\log(a/0)=+\infty$ for $a>0$. Pinsker's inequality and $\|\cdot\|\leq\|\cdot\|_{1}$ give
\begin{equation}
\frac{1}{2}\|\boldsymbol{a}-\boldsymbol{b}\|^{2}\leq K(\boldsymbol{a},\boldsymbol{b}).\label{eq:t1-kl-bounds-1}
\end{equation}
If, in addition, $\min_{r}b_{r}\geq\varepsilon>0$, then
\begin{equation}
    K(\boldsymbol{a},\boldsymbol{b})\leq\sum_{r=1}^{m}\frac{(a_{r}-b_{r})^{2}}{b_{r}}\label{eq:t1-kl-bounds-2}
\end{equation}
since $\log x\leq x-1$. Moreover, a third-order Taylor
expansion, uniformly over $b_{r}\geq\varepsilon$, gives, whenever
$\|\boldsymbol{a}-\boldsymbol{b}\|$ is sufficiently small, 
\begin{equation}
\left|2K(\boldsymbol{a},\boldsymbol{b})-\sum_{r=1}^{m}\frac{(a_{r}-b_{r})^{2}}{b_{r}}\right|\leq C_{\varepsilon,m}\|\boldsymbol{a}-\boldsymbol{b}\|^{3}.\label{eq:t1-kl-taylor}
\end{equation}

Second, A1(iv)--A1(v) imply that there exist $\delta>0$ and $0<s\leq S<\infty$
such that 
\begin{equation}
s\|\boldsymbol{\theta}-\boldsymbol{\theta}_{0}\|\leq\|\boldsymbol{\pi}(\boldsymbol{\theta})-\boldsymbol{\pi}_{0}\|\leq S\|\boldsymbol{\theta}-\boldsymbol{\theta}_{0}\|,\label{eq:t1-local-identification}
\end{equation}
whenever $\|\boldsymbol{\theta}-\boldsymbol{\theta}_{0}\|<\delta$.
Indeed, by A1(iv), $\boldsymbol{\pi}(\boldsymbol{\theta})-\boldsymbol{\pi}_{0}=\dot{\boldsymbol{\pi}}_{0}(\boldsymbol{\theta}-\boldsymbol{\theta}_{0})+\boldsymbol{r}(\boldsymbol{\theta})$, $\|\boldsymbol{r}(\boldsymbol{\theta})\|=o(\|\boldsymbol{\theta}-\boldsymbol{\theta}_{0}\|)$,
and A1(v) implies $c_{0}:=\inf_{\|\boldsymbol{t}\|=1}\|\dot{\boldsymbol{\pi}}_{0}\boldsymbol{t}\|>0$.
The lower bound follows from the reverse triangle inequality after
choosing $\delta$ so that $\|\boldsymbol{r}(\boldsymbol{\theta})\|\leq(c_{0}/2)\|\boldsymbol{\theta}-\boldsymbol{\theta}_{0}\|$;
the upper bound follows similarly.

\medskip{}

\noindent\textit{Step 1: Consistency and root-$n$ consistency of
$\widehat{\boldsymbol{\theta}}$.}

Since the global class $\mathcal{M}$ is $P_{\boldsymbol{X}}$-Glivenko--Cantelli
by A1(vi), and $\boldsymbol{1}_{\mathcal{A}}$ is a fixed bounded
vector of indicators, the class $\mathcal{M}_{1}=\left\{ \boldsymbol{x}\mapsto\boldsymbol{\Phi}_{\boldsymbol{\theta}}(\boldsymbol{x})\otimes\boldsymbol{1}_{\mathcal{A}}(\boldsymbol{x}):\boldsymbol{\theta}\in\Theta\right\} $
is also $P_{\boldsymbol{X}}$-Glivenko--Cantelli. Hence 
\begin{equation}
\sup_{\boldsymbol{\theta}\in\Theta}\left\Vert \widehat{\boldsymbol{\pi}}(\boldsymbol{\theta})-\boldsymbol{\pi}(\boldsymbol{\theta})\right\Vert =o_{P}(1).\label{eq:t1-global-gc}
\end{equation}
Moreover, under $H_{0}$, $\widehat{\boldsymbol{p}}(\boldsymbol{\theta}_{0})-\boldsymbol{\pi}_{0}=O_{P}(n^{-1/2})$ and $\widehat{\boldsymbol{\xi}}_0=O_P(1)$ by the multivariate central limit theorem. By A1(iii), all components of $\boldsymbol{\pi}_{0}$ are strictly
positive. Since $\widehat{\boldsymbol{\pi}}(\boldsymbol{\theta}_{0})\to_{P}\boldsymbol{\pi}_{0}$, \eqref{eq:t1-kl-bounds-2} gives 
\[
\widehat{G}_{\boldsymbol{\theta}_{0}}^{2}(\boldsymbol{\theta}_{0})\leq2\widehat{X}_{\boldsymbol{\theta}_{0}}^{2}(\boldsymbol{\theta}_{0})=O_{P}(1).
\]
Because $\widehat{\boldsymbol{\theta}}$ minimizes $\widehat{G}_{\boldsymbol{\theta}_{0}}^{2}(\boldsymbol{\theta})$,
$\widehat{G}_{\boldsymbol{\theta}_{0}}^{2}(\widehat{\boldsymbol{\theta}})\leq\widehat{G}_{\boldsymbol{\theta}_{0}}^{2}(\boldsymbol{\theta}_{0})$. Therefore, by \eqref{eq:t1-kl-bounds-1},
\begin{equation}
n\left\Vert
\widehat{\boldsymbol{p}}(\boldsymbol{\theta}_{0})
-\widehat{\boldsymbol{\pi}}(\widehat{\boldsymbol{\theta}})
\right\Vert^{2}
\leq
\widehat{G}_{\boldsymbol{\theta}_{0}}^{2}(\widehat{\boldsymbol{\theta}})
\leq
\widehat{G}_{\boldsymbol{\theta}_{0}}^{2}(\boldsymbol{\theta}_{0})
=O_{P}(1). \label{eq:t1-criterion-bound}
\end{equation}
Consequently, by the triangle inequality, 
\begin{equation*}
\left\Vert \boldsymbol{\pi}(\widehat{\boldsymbol{\theta}})-\boldsymbol{\pi}_{0}\right\Vert \leq{} \left\Vert \widehat{\boldsymbol{p}}(\boldsymbol{\theta}_{0})-\boldsymbol{\pi}_{0}\right\Vert +\left\Vert \boldsymbol{\pi}(\widehat{\boldsymbol{\theta}})-\widehat{\boldsymbol{\pi}}(\widehat{\boldsymbol{\theta}})\right\Vert  +\left\Vert \widehat{\boldsymbol{\pi}}(\widehat{\boldsymbol{\theta}})-\widehat{\boldsymbol{p}}(\boldsymbol{\theta}_{0})\right\Vert =o_{P}(1),
\end{equation*}
where \eqref{eq:t1-global-gc} controls the second term. By the global
identification condition A1(ii), 
\begin{equation}
\widehat{\boldsymbol{\theta}}\to_{P}\boldsymbol{\theta}_{0}.\label{eq:t1-consistency}
\end{equation}

We now strengthen consistency to the root-$n$ rate. Let $\epsilon>0$
be such that the local class $\mathcal{M}_{\epsilon}$ is $P_{\boldsymbol{X}}$-Donsker,
as in A1(vi). By \eqref{eq:t1-consistency}, $P\{\widehat{\boldsymbol{\theta}}\in\mathcal{N}_{\epsilon}\}\to1$.
The corresponding class $\left\{ \boldsymbol{\Phi}_{\boldsymbol{\theta}}\otimes\boldsymbol{1}_{\mathcal{A}}:\boldsymbol{\theta}\in\mathcal{N}_{\epsilon}\right\} $
is $P_{\boldsymbol{X}}$-Donsker because $\boldsymbol{1}_{\mathcal{A}}$
is fixed and bounded. Since A1(iv) implies $\|\boldsymbol{\Phi}_{\widehat{\boldsymbol{\theta}}}-\boldsymbol{\Phi}_{\boldsymbol{\theta}_{0}}\|_{P_{\boldsymbol{X}},2}=o_{P}(1)$,
the random-index result for Donsker classes \citep[Lemma 19.24]{van2000asymptotic}
yields 
\[
\sqrt{n}\left\{ \left[\widehat{\boldsymbol{\pi}}(\widehat{\boldsymbol{\theta}})-\boldsymbol{\pi}(\widehat{\boldsymbol{\theta}})\right]-\left[\widehat{\boldsymbol{\pi}}(\boldsymbol{\theta}_{0})-\boldsymbol{\pi}_{0}\right]\right\} =o_{P}(1).
\]
Since $\widehat{\boldsymbol{\pi}}(\boldsymbol{\theta}_{0})-\boldsymbol{\pi}_{0}=O_{P}(n^{-1/2})$,
it follows that $\widehat{\boldsymbol{\pi}}(\widehat{\boldsymbol{\theta}})-\boldsymbol{\pi}(\widehat{\boldsymbol{\theta}})=O_{P}(n^{-1/2})$.
Together with \eqref{eq:t1-criterion-bound} and $\widehat{\boldsymbol{p}}(\boldsymbol{\theta}_{0})-\boldsymbol{\pi}_{0}=O_{P}(n^{-1/2})$,
the preceding triangle inequality now gives 
\begin{equation*}
\left\Vert \boldsymbol{\pi}(\widehat{\boldsymbol{\theta}})-\boldsymbol{\pi}_{0}\right\Vert =O_{P}(n^{-1/2}).
\end{equation*}
Finally, \eqref{eq:t1-local-identification} and consistency imply
$\widehat{\boldsymbol{\theta}}-\boldsymbol{\theta}_{0}=O_{P}(n^{-1/2})$.

\medskip{}

\noindent\textit{Step 2: Uniform equivalence of the likelihood-ratio
and Pearson criteria.}

Let $M_{n}\to\infty$ increase sufficiently slowly and define 
\[
B_{n}=\left\{ \boldsymbol{\theta}\in\Theta:\|\boldsymbol{\theta}-\boldsymbol{\theta}_{0}\|\leq\frac{M_{n}}{\sqrt{n}}\right\} .
\]
By Step 1, $M_{n}$ may be chosen so that $P\{\widehat{\boldsymbol{\theta}}\in B_{n}\}\to1$.
By \eqref{eq:t1-local-identification}, 
\[
\begin{aligned}\sup_{\boldsymbol{\theta}\in B_{n}}\left\Vert \boldsymbol{\pi}(\boldsymbol{\theta})-\boldsymbol{\pi}_{0}\right\Vert  & =O(M_{n}/\sqrt{n})=o\left(1\right).\end{aligned}
\]
Furthermore, A1(vi) gives the local Donsker property. Together with
the continuity in $L^{2}(P_{\boldsymbol{X}})$ imposed in A1(iv),
this implies asymptotic equicontinuity, and hence 
\begin{equation}
\begin{aligned}\sup_{\boldsymbol{\theta}\in B_{n}}\sqrt{n}\Big\Vert & \left[\widehat{\boldsymbol{\pi}}(\boldsymbol{\theta})-\boldsymbol{\pi}(\boldsymbol{\theta})\right]-\left[\widehat{\boldsymbol{\pi}}(\boldsymbol{\theta}_{0})-\boldsymbol{\pi}_{0}\right]\Big\Vert=o_{P}\left(1\right)\\[-2pt]\end{aligned}
\label{eq:t1-local-equicontinuity}
\end{equation}
Thus, uniformly over $B_{n}$, $\left\Vert \widehat{\boldsymbol{p}}(\boldsymbol{\theta}_{0})-\widehat{\boldsymbol{\pi}}(\boldsymbol{\theta})\right\Vert =O_{P}(M_{n}/\sqrt{n})$.
Also, by A1(iii), with probability tending to one, $\inf_{\boldsymbol{\theta}\in B_{n}}\min_{\ell,j}\widehat{\pi}_{\ell j}(\boldsymbol{\theta})\geq c$
for some nonrandom $c>0$. Applying \eqref{eq:t1-kl-taylor}, 
\[
\sup_{\boldsymbol{\theta}\in B_{n}}\left|\widehat{G}_{\boldsymbol{\theta}_{0}}^{2}(\boldsymbol{\theta})-\widehat{X}_{\boldsymbol{\theta}_{0}}^{2}(\boldsymbol{\theta})\right|\leq Cn\sup_{\boldsymbol{\theta}\in B_{n}}\left\Vert \widehat{\boldsymbol{p}}(\boldsymbol{\theta}_{0})-\widehat{\boldsymbol{\pi}}(\boldsymbol{\theta})\right\Vert ^{3}.
\]
Choosing $M_{n}$ sufficiently slowly that $M_{n}^{3}/\sqrt{n}\to0$
gives 
\begin{equation}
\sup_{\boldsymbol{\theta}\in B_{n}}\left|\widehat{G}_{\boldsymbol{\theta}_{0}}^{2}(\boldsymbol{\theta})-\widehat{X}_{\boldsymbol{\theta}_{0}}^{2}(\boldsymbol{\theta})\right|=o_{P}(1).\label{eq:t1-lr-pearson}
\end{equation}

\medskip{}

\noindent\textit{Step 3: Uniform quadratic approximation.}

Define 
\[
\widehat{Q}_{0}(\boldsymbol{\theta})=\left\Vert \boldsymbol{\Lambda}_{0}^{-1}\left[\widehat{\boldsymbol{\xi}}_{0}-\dot{\boldsymbol{\pi}}_{0}\sqrt{n}(\boldsymbol{\theta}-\boldsymbol{\theta}_{0})\right]\right\Vert ^{2}.
\]
We show that 
\begin{equation}
\sup_{\boldsymbol{\theta}\in B_{n}}\left|\widehat{X}_{\boldsymbol{\theta}_{0}}^{2}(\boldsymbol{\theta})-\widehat{Q}_{0}(\boldsymbol{\theta})\right|=o_{P}(1).\label{eq:t1-quadratic}
\end{equation}

By differentiability in A1(iv), for $M_{n}\to\infty$ sufficiently slowly, 
\begin{equation}
\sup_{\boldsymbol{\theta}\in B_{n}}\sqrt{n}\left\Vert \boldsymbol{\pi}(\boldsymbol{\theta})-\boldsymbol{\pi}_{0}-\dot{\boldsymbol{\pi}}_{0}(\boldsymbol{\theta}-\boldsymbol{\theta}_{0})\right\Vert =o(1).\label{eq:t1-pop-linearization}
\end{equation}
Combining \eqref{eq:t1-local-equicontinuity} and \eqref{eq:t1-pop-linearization}
gives 
\begin{equation}
\sup_{\boldsymbol{\theta}\in B_{n}}\left\Vert \sqrt{n}\left[\widehat{\boldsymbol{p}}(\boldsymbol{\theta}_{0})-\widehat{\boldsymbol{\pi}}(\boldsymbol{\theta})\right]-\left[\widehat{\boldsymbol{\xi}}_{0}-\dot{\boldsymbol{\pi}}_{0}\sqrt{n}(\boldsymbol{\theta}-\boldsymbol{\theta}_{0})\right]\right\Vert =o_{P}(1).\label{eq:t1-residual-linearization}
\end{equation}
Moreover, $\sup_{\boldsymbol{\theta}\in B_{n}}\|\widehat{\boldsymbol{\pi}}(\boldsymbol{\theta})-\boldsymbol{\pi}_{0}\|=o_{P}(1)$,
and A1(iii) therefore implies
\begin{equation*}
    \sup_{\boldsymbol{\theta}\in B_{n}}\allowbreak\lVert\allowbreak\operatorname{diag}^{-1}\left[\widehat{\boldsymbol{\pi}}(\boldsymbol{\theta})\right]\allowbreak-\boldsymbol{\Lambda}_{0}^{-2}\rVert\allowbreak=o_{P}(1).
\end{equation*}
Since the vectors in \eqref{eq:t1-residual-linearization} are uniformly
$O_{P}(M_{n})$, reducing the growth rate of $M_{n}$ if necessary
yields \eqref{eq:t1-quadratic}. Together with \eqref{eq:t1-lr-pearson},
\begin{equation}
\sup_{\boldsymbol{\theta}\in B_{n}}\left|\widehat{G}_{\boldsymbol{\theta}_{0}}^{2}(\boldsymbol{\theta})-\widehat{Q}_{0}(\boldsymbol{\theta})\right|=o_{P}(1).\label{eq:t1-lr-quadratic}
\end{equation}

\medskip{}

\noindent\textit{Step 4: Approximation of the minimum.}

Use $\boldsymbol{A}_{0}$, $\boldsymbol{R}_{0}$, and $\boldsymbol{H}_{0}$
as defined in \eqref{eq:fixed-projection-matrices}, and let
$\boldsymbol{J}_{0}=\dot{\boldsymbol{\pi}}_{0}^{\text{\textsc{t}}}\boldsymbol{\Lambda}_{0}^{-2}\dot{\boldsymbol{\pi}}_{0}$.
By A1(v), $\boldsymbol{J}_{0}$ is nonsingular. The unrestricted minimizer
of $\widehat{Q}_{0}$ over $\mathbb{R}^{p}$ is 
\[
\boldsymbol{\theta}^{*}=\boldsymbol{\theta}_{0}+\frac{1}{\sqrt{n}}\boldsymbol{J}_{0}^{-1}\dot{\boldsymbol{\pi}}_{0}^{\text{\textsc{t}}}\boldsymbol{\Lambda}_{0}^{-2}\widehat{\boldsymbol{\xi}}_{0},
\]
so that $\sqrt{n}(\boldsymbol{\theta}^{*}-\boldsymbol{\theta}_{0})=O_{P}(1)$
and 
\begin{equation*}
\inf_{\boldsymbol{\theta}\in\mathbb{R}^{p}}\widehat{Q}_{0}(\boldsymbol{\theta})=\left\Vert \boldsymbol{H}_{0}\boldsymbol{\Lambda}_{0}^{-1}\widehat{\boldsymbol{\xi}}_{0}\right\Vert ^{2}.
\end{equation*}
By A1(i), $P\{\boldsymbol{\theta}^{*}\in\Theta\}\to1$. The sequence
$M_{n}$ may be chosen so that $P\{\widehat{\boldsymbol{\theta}}\in B_{n},\,\boldsymbol{\theta}^{*}\in B_{n}\}\to1$.
Since $\widehat{\boldsymbol{\theta}}$ minimizes the likelihood-ratio
criterion, \eqref{eq:t1-lr-pearson}, \eqref{eq:t1-quadratic}, and
\eqref{eq:t1-lr-quadratic} imply 
\[
\widehat{X}_{\boldsymbol{\theta}_{0}}^{2}(\widehat{\boldsymbol{\theta}})=\inf_{\boldsymbol{\theta}\in\mathbb{R}^{p}}\widehat{Q}_{0}(\boldsymbol{\theta})+o_{P}(1)=\left\Vert \boldsymbol{H}_{0}\boldsymbol{\Lambda}_{0}^{-1}\widehat{\boldsymbol{\xi}}_{0}\right\Vert ^{2}+o_{P}(1).
\]
The same representation holds for $\widehat{G}_{\boldsymbol{\theta}_{0}}^{2}(\widehat{\boldsymbol{\theta}})$.

\medskip{}

\noindent\textit{Step 5: Limiting distribution.}

Under $H_{0}$, 
\begin{equation}
\boldsymbol{\Lambda}_{0}^{-1}\widehat{\boldsymbol{\xi}}_{0}\ \rightarrow_{d}\ N_{LJ}(\boldsymbol{0},\boldsymbol{\Omega})\;\text{and }\;\boldsymbol{\Omega}=\left(\boldsymbol{I}_{L}-\sqrt{\boldsymbol{v}}\sqrt{\boldsymbol{v}}^{\text{\textsc{t}}}\right)\otimes\boldsymbol{I}_{J}.\label{eq:t1-clt}
\end{equation}
For each $j$, $\sum_{\ell=1}^{L}\pi_{\ell j}(\boldsymbol{\theta})=q_{j}$,
which does not depend on $\boldsymbol{\theta}$. Differentiating at
$\boldsymbol{\theta}_{0}$ gives $\sum_{\ell=1}^{L}\dot{\pi}_{0,\ell j}=\boldsymbol{0}^{\text{\textsc{t}}}$,
$j=1,\ldots,J$. Let $\boldsymbol{\Upsilon}=\sqrt{\boldsymbol{v}}\sqrt{\boldsymbol{v}}^{\text{\textsc{t}}}\otimes\boldsymbol{I}_{J}$, so that
$\boldsymbol{\Omega}=\boldsymbol{I}_{LJ}-\boldsymbol{\Upsilon}$.
The preceding adding-up restrictions imply $\boldsymbol{\Upsilon}\boldsymbol{A}_{0}=\boldsymbol{0}$,
so the column space of $\boldsymbol{A}_{0}$ is contained in $\operatorname{ran}(\boldsymbol{\Omega})$.
Consequently, $\boldsymbol{\Omega}\boldsymbol{R}_{0}=\boldsymbol{R}_{0}\boldsymbol{\Omega}=\boldsymbol{R}_{0}$,
and hence $\boldsymbol{H}_{0}\boldsymbol{\Omega}\boldsymbol{H}_{0}=\boldsymbol{\Omega}-\boldsymbol{R}_{0}$.
The matrix $\boldsymbol{\Omega}-\boldsymbol{R}_{0}$ is symmetric
and idempotent, with $\operatorname{rank}(\boldsymbol{\Omega}-\boldsymbol{R}_{0})=J(L-1)-p$,
because $\operatorname{rank}(\boldsymbol{\Omega})=J(L-1)$ and, by
A1(v), $\operatorname{rank}(\boldsymbol{R}_{0})=p$. It follows from \eqref{eq:t1-clt} and the continuous mapping theorem
that 
\[
\left\Vert \boldsymbol{H}_{0}\boldsymbol{\Lambda}_{0}^{-1}\widehat{\boldsymbol{\xi}}_{0}\right\Vert ^{2}\ \rightarrow_{d}\ \chi_{J(L-1)-p}^{2}.
\]
Combining this with Step 4 gives
\[
\widehat{X}_{\boldsymbol{\theta}_{0}}^{2}(\widehat{\boldsymbol{\theta}})\ \rightarrow_{d}\ \chi_{J(L-1)-p}^{2},\qquad\widehat{G}_{\boldsymbol{\theta}_{0}}^{2}(\widehat{\boldsymbol{\theta}})\ \rightarrow_{d}\ \chi_{J(L-1)-p}^{2}.
\]

\hfill{}$\square$

\noindent\textbf{Proof of Theorem~\ref{thm:random-partition}.}

Under $H_{0}$, the residual process defined in
\eqref{eq:frequency-residual-definition} has the marked empirical-process
representation in \eqref{eq:marked-process}. Throughout the proof,
we use the population probability and weighting matrices in
\eqref{eq:population-probability-weighting}, the derivative
$\dot{\boldsymbol{\pi}}_{0}(\boldsymbol{\gamma})$ defined after
\eqref{eq:23}, and the projection matrices in
\eqref{eq:projection-matrices}.

\medskip{}

\noindent\textit{Step 1: Root-$n$ consistency of $\widehat{\boldsymbol{\theta}}$.}

By A2(viii), $\widehat{\boldsymbol{\theta}}\to_{P}\boldsymbol{\theta}_{0}$
and $\widehat{G}_{\boldsymbol{\theta}_{0}}^{\,2}(\widehat{\boldsymbol{\Gamma}},\widehat{\boldsymbol{\theta}})\leq\inf_{\boldsymbol{\theta}\in\Theta}\widehat{G}_{\boldsymbol{\theta}_{0}}^{\,2}(\widehat{\boldsymbol{\Gamma}},\boldsymbol{\theta})+o_{P}(1)$.
In particular, $\widehat{G}_{\boldsymbol{\theta}_{0}}^{\,2}(\widehat{\boldsymbol{\Gamma}},\widehat{\boldsymbol{\theta}})\leq\widehat{G}_{\boldsymbol{\theta}_{0}}^{\,2}(\widehat{\boldsymbol{\Gamma}},\boldsymbol{\theta}_{0})+o_{P}(1)$.
By A2(ii)--A2(iii), $\widehat{\boldsymbol{\Gamma}}\to_{P}\boldsymbol{\Gamma}$
and every component of $\boldsymbol{q}(\boldsymbol{\Gamma})$ is strictly
positive. Hence, with probability approaching one, all components
of $\widehat{\boldsymbol{\pi}}_{\boldsymbol{\theta}_{0}}(\widehat{\boldsymbol{\Gamma}},\boldsymbol{\theta}_{0})$
are bounded away from zero. The elementary likelihood-ratio/Pearson
bounds used in the proof of Theorem~\ref{thm:fixed-partition} therefore give $\widehat{G}_{\boldsymbol{\theta}_{0}}^{\,2}(\widehat{\boldsymbol{\Gamma}},\boldsymbol{\theta}_{0})=O_{P}(1)$,
because $\widehat{\boldsymbol{\xi}}_{0}(\widehat{\boldsymbol{\Gamma}})=O_{P}(1)$.
The latter follows from the Donsker property of the marked class 
\[
\mathcal{Z}=\left\{ (y,\boldsymbol{x})\mapsto\boldsymbol{w}(y,\boldsymbol{x})\otimes\boldsymbol{1}_{\boldsymbol{\gamma}}(\boldsymbol{x}):\boldsymbol{\gamma}\in\mathcal{G}\right\} ,
\]
which follows from A2(vii), since the marks are bounded and $J$ and
$L$ are fixed. Consequently, 
\begin{equation}
\left\Vert \widehat{\boldsymbol{p}}(\widehat{\boldsymbol{\Gamma}},\boldsymbol{\theta}_{0})-\widehat{\boldsymbol{\pi}}_{\boldsymbol{\theta}_{0}}(\widehat{\boldsymbol{\Gamma}},\widehat{\boldsymbol{\theta}})\right\Vert =O_{P}(n^{-1/2}).\label{eq:t2-1}
\end{equation}

We next control the empirical model probabilities. By A2(vi) and A2(vii),
both $\mathcal{M}_{\epsilon}=\{\boldsymbol{\Phi}_{\boldsymbol{\theta}}:\boldsymbol{\theta}\in\mathcal{N}_{\epsilon}\}$ and $\mathcal{H}=\{\boldsymbol{1}_{\boldsymbol{\gamma}}:\boldsymbol{\gamma}\in\mathcal{G}\}$
are Donsker and uniformly bounded. Hence the product class 
\[
\mathcal{K_{\epsilon}}=\left\{ \boldsymbol{x}\mapsto\boldsymbol{\Phi}_{\boldsymbol{\theta}}(\boldsymbol{x})\otimes\boldsymbol{1}_{\boldsymbol{\gamma}}(\boldsymbol{x}):\boldsymbol{\theta}\in\mathcal{N}_{\epsilon},\ \boldsymbol{\gamma}\in\mathcal{G}\right\} 
\]
is $P_{\boldsymbol{X}}$-Donsker by the product-preservation property;
see \citet[Section~2.10]{van1996weak}. Therefore, 
\begin{equation}
\sup_{\boldsymbol{\theta}\in\mathcal{N}_{\epsilon},\,\boldsymbol{\gamma}\in\mathcal{G}}\left\Vert \widehat{\boldsymbol{\pi}}_{\boldsymbol{\theta}_{0}}(\boldsymbol{\gamma},\boldsymbol{\theta})-\boldsymbol{\pi}_{\boldsymbol{\theta}_{0}}(\boldsymbol{\gamma},\boldsymbol{\theta})\right\Vert =O_{P}(n^{-1/2}).\label{eq:t2-2}
\end{equation}

By definition, $\boldsymbol{\pi}_{\boldsymbol{\theta}_{0}}(\boldsymbol{\gamma},\boldsymbol{\theta})-\boldsymbol{\pi}_{\boldsymbol{\theta}_{0}}(\boldsymbol{\gamma},\boldsymbol{\theta}_{0})=P_{\boldsymbol{X}}\left[\left(\boldsymbol{\Phi}_{\boldsymbol{\theta}_{0},\boldsymbol{\theta}}-\boldsymbol{\Phi}_{\boldsymbol{\theta}_{0},\boldsymbol{\theta}_{0}}\right)\otimes\boldsymbol{1}_{\boldsymbol{\gamma}}\right]$. By A2(iv), uniformly in $\boldsymbol{\gamma}\in\mathcal{G}$, 
\begin{equation*}    
\boldsymbol{\pi}_{\boldsymbol{\theta}_{0}}(\boldsymbol{\gamma},\boldsymbol{\theta})-\boldsymbol{\pi}_{\boldsymbol{\theta}_{0}}(\boldsymbol{\gamma},\boldsymbol{\theta}_{0})=\dot{\boldsymbol{\pi}}_{0}(\boldsymbol{\gamma})(\boldsymbol{\theta}-\boldsymbol{\theta}_{0})+o(\|\boldsymbol{\theta}-\boldsymbol{\theta}_{0}\|),
\end{equation*}
because $\left\Vert P_{\boldsymbol{X}}\left[\boldsymbol{r}_{\boldsymbol{\theta}}\otimes\boldsymbol{1}_{\boldsymbol{\gamma}}\right]\right\Vert \leq\|\boldsymbol{r}_{\boldsymbol{\theta}}\|_{P_{\boldsymbol{X}},2}$.
Moreover, A2(ii), A2(iv), and the absolute continuity of the integral
give 
\begin{equation*}  
\dot{\boldsymbol{\pi}}_{0}(\widehat{\boldsymbol{\Gamma}})=\dot{\boldsymbol{\pi}}_{0}(\boldsymbol{\Gamma})+o_{P}(1).
\end{equation*}
By A2(v), $\dot{\boldsymbol{\pi}}_{0}(\boldsymbol{\Gamma})$ has column
rank $p$. Hence there exist $s>0$ and a neighborhood of $\boldsymbol{\theta}_{0}$
such that, with probability approaching one, 
\begin{equation}  
s\|\boldsymbol{\theta}-\boldsymbol{\theta}_{0}\|\leq\left\Vert \boldsymbol{\pi}_{\boldsymbol{\theta}_{0}}(\widehat{\boldsymbol{\Gamma}},\boldsymbol{\theta})-\boldsymbol{\pi}_{\boldsymbol{\theta}_{0}}(\widehat{\boldsymbol{\Gamma}},\boldsymbol{\theta}_{0})\right\Vert \label{eq:t2-5}
\end{equation}
throughout that neighborhood. Since $\widehat{\boldsymbol{\theta}}\to_{P}\boldsymbol{\theta}_{0}$, \eqref{eq:t2-5} applies at $\widehat{\boldsymbol{\theta}}$. Using \eqref{eq:t2-1}, \eqref{eq:t2-2}, and 
\[
\widehat{\boldsymbol{p}}(\widehat{\boldsymbol{\Gamma}},\boldsymbol{\theta}_{0})-\widehat{\boldsymbol{\pi}}_{\boldsymbol{\theta}_{0}}(\widehat{\boldsymbol{\Gamma}},\boldsymbol{\theta}_{0})=n^{-1/2}\widehat{\boldsymbol{\xi}}_{0}(\widehat{\boldsymbol{\Gamma}})=O_{P}(n^{-1/2}),
\]
we obtain
\begin{equation}    
\widehat{\boldsymbol{\theta}}-\boldsymbol{\theta}_{0}=O_{P}(n^{-1/2}).\label{eq:t2-6}
\end{equation}

\medskip{}

\noindent\textit{Step 2: Reduction to the limiting partition.}

The class $\mathcal{Z}$ is $P$-Donsker. Its canonical semimetric
is dominated by the product symmetric-difference semimetric. Indeed,
since $\boldsymbol{w}$ is bounded, for $\boldsymbol{\gamma}_{1},\boldsymbol{\gamma}_{2}\in\mathcal{G}$,
\[
\begin{aligned} & P\left\Vert \boldsymbol{w}(Y,\boldsymbol{X})\otimes\left[\boldsymbol{1}_{\boldsymbol{\gamma}_{1}}(\boldsymbol{X})-\boldsymbol{1}_{\boldsymbol{\gamma}_{2}}(\boldsymbol{X})\right]\right\Vert ^{2}\leq K\sum_{j=1}^{J}P_{\boldsymbol{X}}(\gamma_{1j}\triangle\gamma_{2j})\end{aligned}
\]
for some finite $K$. By A2(ii), $d_{P_{\boldsymbol{X}}}(\widehat{\boldsymbol{\Gamma}},\boldsymbol{\Gamma})\to_{P}0$.
Therefore, $P\left\Vert \boldsymbol{z}_{\widehat{\boldsymbol{\Gamma}}}-\boldsymbol{z}_{\boldsymbol{\Gamma}}\right\Vert ^{2}=o_{P}(1)$,
where $\boldsymbol{z}_{\boldsymbol{\gamma}}=\boldsymbol{w}\otimes\boldsymbol{1}_{\boldsymbol{\gamma}}$. Thus,
\citet[Lemma~19.24]{van2000asymptotic} yields 
\begin{equation}
\widehat{\boldsymbol{\xi}}_{0}(\widehat{\boldsymbol{\Gamma}})=\widehat{\boldsymbol{\xi}}_{0}(\boldsymbol{\Gamma})+o_{P}(1).\label{eq:t2-7}    
\end{equation}
The same argument, using A2(ii), A2(iv) and the absolute continuity of the
integral, gives 
\begin{equation}
\boldsymbol{\Lambda}_{0}(\widehat{\boldsymbol{\Gamma}})=\boldsymbol{\Lambda}_{0}(\boldsymbol{\Gamma})+o_{P}(1)\quad\text{ and }\quad\dot{\boldsymbol{\pi}}_{0}(\widehat{\boldsymbol{\Gamma}})=\dot{\boldsymbol{\pi}}_{0}(\boldsymbol{\Gamma})+o_{P}(1).\label{eq:t2-8}    
\end{equation}
By A2(iii), all components of $\boldsymbol{q}(\boldsymbol{\Gamma})$
are strictly positive, so the inverse weighting matrices are continuous
at $\boldsymbol{\Gamma}$.

\medskip{}

\noindent\textit{Step 3: Uniform local quadratic approximation.}

Let $M_{n}\to\infty$ increase sufficiently slowly and set $B_{n}=\left\{ \boldsymbol{\theta}:\|\boldsymbol{\theta}-\boldsymbol{\theta}_{0}\|\leq M_{n}/\sqrt{n}\right\}$.
By \eqref{eq:t2-6}, $M_{n}$ may be chosen so that $P\{\widehat{\boldsymbol{\theta}}\in B_{n}\}\to1$.

The product class $\mathcal{K_{\epsilon}}$ is Donsker, as shown in
Step~1. A2(iv) implies 
\[
\|\boldsymbol{\Phi}_{\boldsymbol{\theta}}-\boldsymbol{\Phi}_{\boldsymbol{\theta}_{0}}\|_{P_{\boldsymbol{X}},2}\to 0\qquad\text{as }\boldsymbol{\theta}\to\boldsymbol{\theta}_{0}.
\]
Consequently, asymptotic equicontinuity of the empirical process indexed
by functions in the class $\mathcal{K}_{\epsilon}$ gives 
\begin{equation}
\sup_{\boldsymbol{\gamma}\in\mathcal{G},\,\boldsymbol{\theta}\in B_{n}}\sqrt{n}\left\Vert \left[\widehat{\boldsymbol{\pi}}_{\boldsymbol{\theta}_{0}}(\boldsymbol{\gamma},\boldsymbol{\theta})-\boldsymbol{\pi}_{\boldsymbol{\theta}_{0}}(\boldsymbol{\gamma},\boldsymbol{\theta})\right]-\left[\widehat{\boldsymbol{\pi}}_{\boldsymbol{\theta}_{0}}(\boldsymbol{\gamma},\boldsymbol{\theta}_{0})-\boldsymbol{\pi}_{\boldsymbol{\theta}_{0}}(\boldsymbol{\gamma},\boldsymbol{\theta}_{0})\right]\right\Vert =o_{P}(1).\label{eq:t2-9}
\end{equation}
The differentiability remainder in A2(iv) satisfies, uniformly in
$\boldsymbol{\gamma}$, 
\[
\begin{aligned}\left\Vert P_{\boldsymbol{X}}\!\left[\left\{\boldsymbol{\Phi}_{\boldsymbol{\theta}_{0},\boldsymbol{\theta}}\!-\!\boldsymbol{\Phi}_{\boldsymbol{\theta}_{0},\boldsymbol{\theta}_{0}}\!-\!\dot{\boldsymbol{\Phi}}_{0}(\boldsymbol{\theta}\!-\!\boldsymbol{\theta}_{0})\right\}\!\otimes\!\boldsymbol{1}_{\boldsymbol{\gamma}}\right]\right\Vert  & \!\leq\!\left\Vert \boldsymbol{\Phi}_{\boldsymbol{\theta}_{0},\boldsymbol{\theta}}\!-\!\boldsymbol{\Phi}_{\boldsymbol{\theta}_{0},\boldsymbol{\theta}_{0}}\!-\!\dot{\boldsymbol{\Phi}}_{0}(\boldsymbol{\theta}\!-\!\boldsymbol{\theta}_{0})\right\Vert _{P_{\boldsymbol{X}},2}.\end{aligned}
\]
Choosing $M_{n}$ sufficiently slowly therefore gives 
\begin{equation}    
\sup_{\boldsymbol{\gamma}\in\mathcal{G},\,\boldsymbol{\theta}\in B_{n}}\left\Vert \sqrt{n}\left[\boldsymbol{\pi}_{\boldsymbol{\theta}_{0}}(\boldsymbol{\gamma},\boldsymbol{\theta})-\boldsymbol{\pi}_{\boldsymbol{\theta}_{0}}(\boldsymbol{\gamma},\boldsymbol{\theta}_{0})\right]-\dot{\boldsymbol{\pi}}_{0}(\boldsymbol{\gamma})\sqrt{n}(\boldsymbol{\theta}-\boldsymbol{\theta}_{0})\right\Vert =o(1).\label{eq:t2-10}
\end{equation}
Combining \eqref{eq:t2-9}--\eqref{eq:t2-10}, 
\begin{equation}    
\sup_{\boldsymbol{\gamma}\in\mathcal{G},\,\boldsymbol{\theta}\in B_{n}}\Big\| \sqrt{n}\left[\widehat{\boldsymbol{p}}(\boldsymbol{\gamma},\boldsymbol{\theta}_{0})-\widehat{\boldsymbol{\pi}}_{\boldsymbol{\theta}_{0}}(\boldsymbol{\gamma},\boldsymbol{\theta})\right]-\left[\widehat{\boldsymbol{\xi}}_{0}(\boldsymbol{\gamma})-\dot{\boldsymbol{\pi}}_{0}(\boldsymbol{\gamma})\sqrt{n}(\boldsymbol{\theta}-\boldsymbol{\theta}_{0})\right]\Big\|=o_{P}(1).
\label{eq:t2-11}
\end{equation}

A2(ii)--A2(iv), together with the Glivenko--Cantelli property implied
by A2(vi)--A2(vii), ensure that, on a shrinking neighborhood of $\boldsymbol{\Gamma}$
and uniformly for $\boldsymbol{\theta}\in B_{n}$, the estimated cell
probabilities remain bounded away from zero with probability approaching
one. Hence the Taylor expansion of the likelihood-ratio criterion
used in Theorem~\ref{thm:fixed-partition} applies uniformly. Reducing the growth rate of
$M_{n}$ if necessary, \eqref{eq:t2-11} gives 
\begin{equation}    
\sup_{\substack{d_{P_{\boldsymbol{X}}}(\boldsymbol{\gamma},\boldsymbol{\Gamma})\leq\delta_{n}\\
\boldsymbol{\theta}\in B_{n}
}
}\left|\widehat{G}_{\boldsymbol{\theta}_{0}}^{\,2}(\boldsymbol{\gamma},\boldsymbol{\theta})-\widehat{Q}_{0}(\boldsymbol{\gamma},\boldsymbol{\theta})\right|=o_{P}(1),\label{eq:t2-12}
\end{equation}
and 
\begin{equation}    
\sup_{\substack{d_{P_{\boldsymbol{X}}}(\boldsymbol{\gamma},\boldsymbol{\Gamma})\leq\delta_{n}\\
\boldsymbol{\theta}\in B_{n}
}
}\left|\widehat{X}_{\boldsymbol{\theta}_{0}}^{\,2}(\boldsymbol{\gamma},\boldsymbol{\theta})-\widehat{Q}_{0}(\boldsymbol{\gamma},\boldsymbol{\theta})\right|=o_{P}(1),\label{eq:t2-13}
\end{equation}
for some deterministic $\delta_{n}\downarrow0$ such that $P\left\{ d_{P_{\boldsymbol{X}}}(\widehat{\boldsymbol{\Gamma}},\boldsymbol{\Gamma})\leq\delta_{n}\right\} \to1$,
where 
\[
\widehat{Q}_{0}(\boldsymbol{\gamma},\boldsymbol{\theta})=\left\Vert \boldsymbol{\Lambda}_{0}^{-1}(\boldsymbol{\gamma})\left[\widehat{\boldsymbol{\xi}}_{0}(\boldsymbol{\gamma})-\dot{\boldsymbol{\pi}}_{0}(\boldsymbol{\gamma})\sqrt{n}(\boldsymbol{\theta}-\boldsymbol{\theta}_{0})\right]\right\Vert ^{2}.
\]

\medskip{}

\noindent\textit{Step 4: Projection representation at the random
partition.}

For each $\boldsymbol{\gamma}$ in a neighborhood of $\boldsymbol{\Gamma}$,
the unrestricted minimizer of $\widehat{Q}_{0}(\boldsymbol{\gamma},\cdot)$
is 
\[
\boldsymbol{\theta}^{*}(\boldsymbol{\gamma})=\boldsymbol{\theta}_{0}+\frac{1}{\sqrt{n}}\boldsymbol{J}_{0}^{-1}(\boldsymbol{\gamma})\dot{\boldsymbol{\pi}}_{0}^{\text{\textsc{t}}}(\boldsymbol{\gamma})\boldsymbol{\Lambda}_{0}^{-2}(\boldsymbol{\gamma})\widehat{\boldsymbol{\xi}}_{0}(\boldsymbol{\gamma}).
\]
By A2(iii)--A2(v) and continuity in $\boldsymbol{\gamma}$, $\boldsymbol{J}_{0}(\boldsymbol{\gamma})$
is nonsingular throughout a sufficiently small neighborhood of $\boldsymbol{\Gamma}$.
Moreover, 
\begin{equation*}    
\inf_{\boldsymbol{\theta}\in\mathbb{R}^{p}}\widehat{Q}_{0}(\boldsymbol{\gamma},\boldsymbol{\theta})=\left\Vert \boldsymbol{H}_{0}(\boldsymbol{\gamma})\boldsymbol{\Lambda}_{0}^{-1}(\boldsymbol{\gamma})\widehat{\boldsymbol{\xi}}_{0}(\boldsymbol{\gamma})\right\Vert ^{2}.
\end{equation*}
The approximate-minimization condition in A2(viii), together with
\eqref{eq:t2-12}--\eqref{eq:t2-13}, therefore yields 
\begin{equation*}  
\widehat{X}_{\boldsymbol{\theta}_{0}}^{\,2}(\widehat{\boldsymbol{\Gamma}},\widehat{\boldsymbol{\theta}})=\left\Vert \boldsymbol{H}_{0}(\widehat{\boldsymbol{\Gamma}})\boldsymbol{\Lambda}_{0}^{-1}(\widehat{\boldsymbol{\Gamma}})\widehat{\boldsymbol{\xi}}_{0}(\widehat{\boldsymbol{\Gamma}})\right\Vert ^{2}+o_{P}(1),
\end{equation*}
and the same representation holds for the likelihood-ratio statistic. By \eqref{eq:t2-7}--\eqref{eq:t2-8} and continuity of matrix inversion, 
\begin{equation*}  
\boldsymbol{H}_{0}(\widehat{\boldsymbol{\Gamma}})\boldsymbol{\Lambda}_{0}^{-1}(\widehat{\boldsymbol{\Gamma}})\widehat{\boldsymbol{\xi}}_{0}(\widehat{\boldsymbol{\Gamma}})=\boldsymbol{H}_{0}(\boldsymbol{\Gamma})\boldsymbol{\Lambda}_{0}^{-1}(\boldsymbol{\Gamma})\widehat{\boldsymbol{\xi}}_{0}(\boldsymbol{\Gamma})+o_{P}(1).
\end{equation*}
Hence 
\begin{equation}  
\widehat{X}_{\boldsymbol{\theta}_{0}}^{\,2}(\widehat{\boldsymbol{\Gamma}},\widehat{\boldsymbol{\theta}})=\left\Vert \boldsymbol{H}_{0}(\boldsymbol{\Gamma})\boldsymbol{\Lambda}_{0}^{-1}(\boldsymbol{\Gamma})\widehat{\boldsymbol{\xi}}_{0}(\boldsymbol{\Gamma})\right\Vert ^{2}+o_{P}(1).\label{eq:t2-17}
\end{equation}

\medskip{}

\noindent\textit{Step 5: Limiting distribution.}

For the fixed limiting partition $\boldsymbol{\Gamma}$, the multivariate
central limit theorem gives 
\begin{equation}  
\boldsymbol{\Lambda}_{0}^{-1}(\boldsymbol{\Gamma})\widehat{\boldsymbol{\xi}}_{0}(\boldsymbol{\Gamma})\rightarrow_{d} N_{LJ}(\boldsymbol{0},\boldsymbol{\Omega}),\qquad\boldsymbol{\Omega}=\left(\boldsymbol{I}_{L}-\sqrt{\boldsymbol{v}}\sqrt{\boldsymbol{v}}^{\text{\textsc{t}}}\right)\otimes\boldsymbol{I}_{J}.\label{eq:t2-18}
\end{equation}
For every $\boldsymbol{\gamma}$, $\sum_{\ell=1}^{L}\pi_{\boldsymbol{\theta}_{0},\ell j}(\boldsymbol{\gamma},\boldsymbol{\theta})=q_{j}(\boldsymbol{\gamma})$,
which does not depend on $\boldsymbol{\theta}$. Differentiating at
$\boldsymbol{\theta}_{0}$ shows that the columns of $\boldsymbol{A}_{0}(\boldsymbol{\Gamma})$
belong to the range of $\boldsymbol{\Omega}$. Therefore, $\boldsymbol{H}_{0}(\boldsymbol{\Gamma})\boldsymbol{\Omega}\boldsymbol{H}_{0}(\boldsymbol{\Gamma})=\boldsymbol{\Omega}-\boldsymbol{R}_{0}(\boldsymbol{\Gamma})$.
This matrix is symmetric and idempotent. By A2(v), $\operatorname{rank}\left[\boldsymbol{\Omega}-\boldsymbol{R}_{0}(\boldsymbol{\Gamma})\right]=J(L-1)-p$.
It follows from \eqref{eq:t2-17}--\eqref{eq:t2-18} and the continuous mapping theorem
that 
\[
\widehat{X}_{\boldsymbol{\theta}_{0}}^{\,2}(\widehat{\boldsymbol{\Gamma}},\widehat{\boldsymbol{\theta}})\rightarrow_{d}\chi_{J(L-1)-p}^{2}.
\]
The same conclusion follows for the corresponding likelihood-ratio
statistic from \eqref{eq:t2-12}. \hfill{}$\square$

\noindent\textbf{Proof of Theorem~\ref{thm:feasible}.}

We need the following lemmas.

\begin{lemma}\label{lem:transformation-equivalence} Let $\boldsymbol{\theta}_{n}$ be any possibly random
sequence satisfying $\sqrt{n}(\boldsymbol{\theta}_{n}-\boldsymbol{\theta}_{0})=O_{P}(1)$.
Then, under $H_{0}$, 
\begin{equation*}
\widehat{\boldsymbol{\xi}}_{\boldsymbol{\theta}_{n}}(\widehat{\boldsymbol{\Gamma}},\boldsymbol{\theta}_{n})-\widehat{\boldsymbol{\xi}}_{\boldsymbol{\theta}_{0}}(\widehat{\boldsymbol{\Gamma}},\boldsymbol{\theta}_{n})=o_{P}(1).
\end{equation*}
\end{lemma} 
\begin{proof}
For $\boldsymbol{\theta}\in\mathcal{N}_{\epsilon}$ and $\boldsymbol{\gamma}\in\mathcal{G}$,
define 
\begin{equation*}
\boldsymbol{h}_{\boldsymbol{\theta},\boldsymbol{\gamma}}(y,\boldsymbol{x})={}
\Big\{\boldsymbol{1}_{\mathcal{U}}[F_{\boldsymbol{\theta}}(y\mid\boldsymbol{x})]-\boldsymbol{1}_{\mathcal{U}}[F_{\boldsymbol{\theta}_{0}}(y\mid\boldsymbol{x})]-\left[\boldsymbol{\Phi}_{\boldsymbol{\theta},\boldsymbol{\theta}}(\boldsymbol{x})-\boldsymbol{\Phi}_{\boldsymbol{\theta}_{0},\boldsymbol{\theta}}(\boldsymbol{x})\right]\Big\}\otimes\boldsymbol{1}_{\boldsymbol{\gamma}}(\boldsymbol{x}).
\end{equation*}
By definition, 
\[
\begin{aligned} & \widehat{\boldsymbol{\xi}}_{\boldsymbol{\theta}}(\boldsymbol{\gamma},\boldsymbol{\theta})-\widehat{\boldsymbol{\xi}}_{\boldsymbol{\theta}_{0}}(\boldsymbol{\gamma},\boldsymbol{\theta})=\sqrt{n}\,P_{n}\boldsymbol{h}_{\boldsymbol{\theta},\boldsymbol{\gamma}}=\mathbb{G}_{n}\boldsymbol{h}_{\boldsymbol{\theta},\boldsymbol{\gamma}}+\sqrt{n}\,P\boldsymbol{h}_{\boldsymbol{\theta},\boldsymbol{\gamma}}\end{aligned}
\]
It therefore suffices to show that $\mathbb{G}_{n}\boldsymbol{h}_{\boldsymbol{\theta}_{n},\widehat{\boldsymbol{\Gamma}}}=o_{P}(1)$
and $\sqrt{n}\left\Vert P\boldsymbol{h}_{\boldsymbol{\theta}_{n},\widehat{\boldsymbol{\Gamma}}}\right\Vert =o_{P}(1)$.

\textit{First, consider the empirical-process term.} Let $\mathcal{H}_{\epsilon}=\left\{ \boldsymbol{h}_{\boldsymbol{\theta},\boldsymbol{\gamma}}:\boldsymbol{\theta}\in\mathcal{N}_{\epsilon},\ \boldsymbol{\gamma}\in\mathcal{G}\right\}$.
We first show that $\mathcal{H}_{\epsilon}$ is $P$-Donsker.

For each $\ell=1,\ldots,L$, 
\[
1_{I_{\ell}}\!\left(F_{\boldsymbol{\theta}}(y\mid\boldsymbol{x})\right)=1\!\left\{ F_{\boldsymbol{\theta}}(y\mid\boldsymbol{x})\leq\vartheta_{\ell}\right\} -1\!\left\{ F_{\boldsymbol{\theta}}(y\mid\boldsymbol{x})\leq\vartheta_{\ell-1}\right\} ,
\]
up to the immaterial convention at the cell boundaries. By A3(vi),
the classes on the right-hand side are $P$-Donsker. Since $L$ is
fixed and Donskerity is preserved under finite sums and differences,
it follows that 
\[
\mathcal{W}_{\epsilon}=\left\{ (y,\boldsymbol{x})\mapsto\boldsymbol{1}_{\mathcal{U}}[F_{\boldsymbol{\theta}}(y\mid\boldsymbol{x})]-\boldsymbol{1}_{\mathcal{U}}[F_{\boldsymbol{\theta}_{0}}(y\mid\boldsymbol{x})]:\boldsymbol{\theta}\in\mathcal{N}_{\epsilon}\right\} 
\]
is $P$-Donsker; see, e.g., \citet[Section 2.10]{van1996weak}. Moreover, by A2(vii), the indicator class generated
by $\mathcal{C}$ is $P_{\boldsymbol{X}}$-Donsker. Since both classes
are uniformly bounded, the permanence properties of Donsker classes
under products imply that 
\[
\left\{ (y,\boldsymbol{x})\mapsto\left[\boldsymbol{1}_{\mathcal{U}}(F_{\boldsymbol{\theta}}(y\mid\boldsymbol{x}))-\boldsymbol{1}_{\mathcal{U}}(F_{\boldsymbol{\theta}_{0}}(y\mid\boldsymbol{x}))\right]\otimes\boldsymbol{1}_{\boldsymbol{\gamma}}(\boldsymbol{x}):\boldsymbol{\theta}\in\mathcal{N}_{\epsilon},\ \boldsymbol{\gamma}\in\mathcal{G}\right\} 
\]
is $P$-Donsker.

Since $\boldsymbol{\Phi}_{\boldsymbol{\theta},\boldsymbol{\theta}}(\boldsymbol{X})=\boldsymbol{v}$ $P_{\boldsymbol{X}}$-a.s.,
the remaining term in $\boldsymbol{h}_{\boldsymbol{\theta},\boldsymbol{\gamma}}$
is $\left[\boldsymbol{\Phi}_{\boldsymbol{\theta}_{0},\boldsymbol{\theta}}-\boldsymbol{v}\right]\otimes\boldsymbol{1}_{\boldsymbol{\gamma}}$.
The local class $\mathcal{M}_{\epsilon}$ is $P_{\boldsymbol{X}}$-Donsker
by A2(vi). Because its elements and the partition indicators are bounded,
the same product-preservation result implies that 
\[
\left\{ (y,\boldsymbol{x})\mapsto\left[\boldsymbol{\Phi}_{\boldsymbol{\theta}_{0},\boldsymbol{\theta}}(\boldsymbol{x})-\boldsymbol{v}\right]\otimes\boldsymbol{1}_{\boldsymbol{\gamma}}(\boldsymbol{x}):\boldsymbol{\theta}\in\mathcal{N}_{\epsilon},\ \boldsymbol{\gamma}\in\mathcal{G}\right\} 
\]
is $P$-Donsker. Hence $\mathcal{H}_{\epsilon}$ is $P$-Donsker.

We next show that $\sup_{\boldsymbol{\gamma}\in\mathcal{G}}\|\boldsymbol{h}_{\boldsymbol{\theta},\boldsymbol{\gamma}}\|_{P,2}\to 0$ as $\boldsymbol{\theta}\to\boldsymbol{\theta}_{0}$.
Write $U_{\boldsymbol{\theta}}=F_{\boldsymbol{\theta}}(Y\mid\boldsymbol{X})$, $U_{0}=F_{\boldsymbol{\theta}_{0}}(Y\mid\boldsymbol{X})$.
By A3(ii), $\|U_{\boldsymbol{\theta}}-U_{0}\|_{P,2}\to0$, and hence
$U_{\boldsymbol{\theta}}\to_{P}U_{0}$ as $\boldsymbol{\theta}\to\boldsymbol{\theta}_{0}$. For every $\eta>0$, 
\[
\begin{aligned}\left\{ \boldsymbol{1}_{\mathcal{U}}(U_{\boldsymbol{\theta}})\neq\boldsymbol{1}_{\mathcal{U}}(U_{0})\right\} \subseteq{} & \left\{ |U_{\boldsymbol{\theta}}-U_{0}|>\eta\right\} \cup\bigcup_{\ell=1}^{L-1}\left\{ |U_{0}-\vartheta_{\ell}|\leq\eta\right\} .\end{aligned}
\]
Consequently, 
\[
\begin{aligned} & P\left\{ \boldsymbol{1}_{\mathcal{U}}(U_{\boldsymbol{\theta}})\neq\boldsymbol{1}_{\mathcal{U}}(U_{0})\right\} \leq P\left\{ |U_{\boldsymbol{\theta}}-U_{0}|>\eta\right\} +\sum_{\ell=1}^{L-1}P\left\{ |U_{0}-\vartheta_{\ell}|\leq\eta\right\} .\end{aligned}
\]
For fixed $\eta>0$, the first term converges to zero by A3(ii). Moreover,
A3(v) implies $P\left\{ |U_{0}-\vartheta_{\ell}|\leq\eta\right\} \to 0$, $\ell=1,\ldots,L-1$,
as $\eta\downarrow0$. Therefore, $P\left\{ \boldsymbol{1}_{\mathcal{U}}(U_{\boldsymbol{\theta}})\neq\boldsymbol{1}_{\mathcal{U}}(U_{0})\right\} \to 0$. Since $\mathcal{U}=\{I_{1},\ldots,I_{L}\}$ is a partition, 
\[
\begin{aligned}\left\Vert \boldsymbol{1}_{\mathcal{U}}(U_{\boldsymbol{\theta}})-\boldsymbol{1}_{\mathcal{U}}(U_{0})\right\Vert _{P,2}^{2} & =2P\left\{ U_{\boldsymbol{\theta}}\text{ and }U_{0}\text{ belong to different cells of }\mathcal{U}\right\} \\
 & =2P\left\{ \boldsymbol{1}_{\mathcal{U}}(U_{\boldsymbol{\theta}})\neq\boldsymbol{1}_{\mathcal{U}}(U_{0})\right\} \to 0.
\end{aligned}
\]
Thus, $\left\Vert \boldsymbol{1}_{\mathcal{U}}[F_{\boldsymbol{\theta}}(Y\mid\boldsymbol{X})]-\boldsymbol{1}_{\mathcal{U}}[F_{\boldsymbol{\theta}_{0}}(Y\mid\boldsymbol{X})]\right\Vert _{P,2}\to 0$. In addition, A3(iii) implies 
\[
\left\Vert \boldsymbol{\Phi}_{\boldsymbol{\theta}_{0},\boldsymbol{\theta}}-\boldsymbol{\Phi}_{\boldsymbol{\theta}_{0},\boldsymbol{\theta}_{0}}\right\Vert _{P_{\boldsymbol{X}},2}\to 0.
\]
Since the cells of $\boldsymbol{\gamma}$ form a partition, $\|\boldsymbol{1}_{\boldsymbol{\gamma}}(\boldsymbol{X})\|=1$,
and hence 
\[
\begin{aligned}\sup_{\boldsymbol{\gamma}\in\mathcal{G}}\|\boldsymbol{h}_{\boldsymbol{\theta},\boldsymbol{\gamma}}\|_{P,2}\leq{} & \left\Vert \boldsymbol{1}_{\mathcal{U}}[F_{\boldsymbol{\theta}}(Y\mid\boldsymbol{X})]-\boldsymbol{1}_{\mathcal{U}}[F_{\boldsymbol{\theta}_{0}}(Y\mid\boldsymbol{X})]\right\Vert _{P,2}\\
 & +\left\Vert \boldsymbol{\Phi}_{\boldsymbol{\theta}_{0},\boldsymbol{\theta}}-\boldsymbol{\Phi}_{\boldsymbol{\theta}_{0},\boldsymbol{\theta}_{0}}\right\Vert _{P_{\boldsymbol{X}},2}=o(1).
\end{aligned}
\]

Since $\sqrt{n}(\boldsymbol{\theta}_{n}-\boldsymbol{\theta}_{0})=O_{P}(1)$,
we have $\boldsymbol{\theta}_{n}\to_{P}\boldsymbol{\theta}_{0}$ and,
for every fixed $\epsilon>0$, $P\{\boldsymbol{\theta}_{n}\in\mathcal{N}_{\epsilon}\}\to1$.
Therefore, $\|\boldsymbol{h}_{\boldsymbol{\theta}_{n},\widehat{\boldsymbol{\Gamma}}}\|_{P,2}\to_{P}0$.
Since $\mathcal{H}_{\epsilon}$ is $P$-Donsker and $\|\boldsymbol{h}_{\boldsymbol{\theta}_{n},\widehat{\boldsymbol{\Gamma}}}\|_{P,2}\to_{P}0$,
\citet[Lemma 19.24]{van2000asymptotic} yields $\mathbb{G}_{n}\boldsymbol{h}_{\boldsymbol{\theta}_{n},\widehat{\boldsymbol{\Gamma}}}=o_{P}(1)$.

\textit{Second, consider the population term.} Under $H_{0}$, 
\[
P\boldsymbol{h}_{\boldsymbol{\theta},\boldsymbol{\gamma}}=P\left[\left\{ \boldsymbol{\Phi}_{\boldsymbol{\theta},\boldsymbol{\theta}_{0}}+\boldsymbol{\Phi}_{\boldsymbol{\theta}_{0},\boldsymbol{\theta}}-2\boldsymbol{v}\right\} \otimes\boldsymbol{1}_{\boldsymbol{\gamma}}\right].
\]
By the joint $L^{2}(P_{\boldsymbol{X}})$-differentiability in A3(iii),
$\boldsymbol{\Phi}_{\boldsymbol{\theta},\boldsymbol{\theta}_{0}}=\boldsymbol{v}+\dot{\boldsymbol{\Phi}}_{1,0}(\boldsymbol{\theta}-\boldsymbol{\theta}_{0})+\boldsymbol{r}_{1,\boldsymbol{\theta}}$,
and 
\[
\boldsymbol{\Phi}_{\boldsymbol{\theta}_{0},\boldsymbol{\theta}}=\boldsymbol{v}+\dot{\boldsymbol{\Phi}}_{2,0}(\boldsymbol{\theta}-\boldsymbol{\theta}_{0})+\boldsymbol{r}_{2,\boldsymbol{\theta}},
\]
where $\|\boldsymbol{r}_{j,\boldsymbol{\theta}}\|_{P_{\boldsymbol{X}},2}=o(\|\boldsymbol{\theta}-\boldsymbol{\theta}_{0}\|)$, $j=1,2$. Since $\boldsymbol{\Phi}_{\boldsymbol{\theta},\boldsymbol{\theta}}=\boldsymbol{v}$ for every $\boldsymbol{\theta}$,
differentiation along the diagonal gives $\dot{\boldsymbol{\Phi}}_{1,0}+\dot{\boldsymbol{\Phi}}_{2,0}=\boldsymbol{0}$.
It follows that 
\[
\sup_{\boldsymbol{\gamma}\in\mathcal{G}}\left\Vert P\boldsymbol{h}_{\boldsymbol{\theta},\boldsymbol{\gamma}}\right\Vert \leq\|\boldsymbol{r}_{1,\boldsymbol{\theta}}+\boldsymbol{r}_{2,\boldsymbol{\theta}}\|_{P_{\boldsymbol{X}},2}=o(\|\boldsymbol{\theta}-\boldsymbol{\theta}_{0}\|).
\]
Therefore, 
\begin{equation*}
\sqrt{n}\left\Vert P\boldsymbol{h}_{\boldsymbol{\theta}_{n},\widehat{\boldsymbol{\Gamma}}}\right\Vert  \leq\sqrt{n}\sup_{\boldsymbol{\gamma}\in\mathcal{G}}\left\Vert P\boldsymbol{h}_{\boldsymbol{\theta}_{n},\boldsymbol{\gamma}}\right\Vert =\sqrt{n}\,o_{P}(\|\boldsymbol{\theta}_{n}-\boldsymbol{\theta}_{0}\|)=o_{P}(1).
\end{equation*}
Combining the two terms proves the result. 
\end{proof}
\begin{lemma}\label{lem:model-probability-expansion} Let $\boldsymbol{\theta}_{n}$ be any possibly random
sequence satisfying $\sqrt{n}(\boldsymbol{\theta}_{n}-\boldsymbol{\theta}_{0})=O_{P}(1)$.
Then, under $H_{0}$, 
\[
\widehat{\boldsymbol{\xi}}_{\boldsymbol{\theta}_{0}}(\widehat{\boldsymbol{\Gamma}},\boldsymbol{\theta}_{n})=\widehat{\boldsymbol{\xi}}_{0}(\widehat{\boldsymbol{\Gamma}})-\dot{\boldsymbol{\pi}}_{0}(\widehat{\boldsymbol{\Gamma}})\sqrt{n}(\boldsymbol{\theta}_{n}-\boldsymbol{\theta}_{0})+o_{P}(1).
\]
\end{lemma} 
\begin{proof}
By definition, 
\[
\begin{aligned}\widehat{\boldsymbol{\xi}}_{\boldsymbol{\theta}_{0}}(\boldsymbol{\gamma},\boldsymbol{\theta})-\widehat{\boldsymbol{\xi}}_{0}(\boldsymbol{\gamma}) & =-\sqrt{n}\left[\widehat{\boldsymbol{\pi}}_{\boldsymbol{\theta}_{0}}(\boldsymbol{\gamma},\boldsymbol{\theta})-\widehat{\boldsymbol{\pi}}_{\boldsymbol{\theta}_{0}}(\boldsymbol{\gamma},\boldsymbol{\theta}_{0})\right].\end{aligned}
\]
By the $L^{2}(P_{\boldsymbol{X}})$-differentiability in A3(iii),
\[
\boldsymbol{\Phi}_{\boldsymbol{\theta}_{0},\boldsymbol{\theta}}=\boldsymbol{\Phi}_{\boldsymbol{\theta}_{0},\boldsymbol{\theta}_{0}}+\dot{\boldsymbol{\Phi}}_{0}(\boldsymbol{\theta}-\boldsymbol{\theta}_{0})+\boldsymbol{r}_{\boldsymbol{\theta}},\qquad\|\boldsymbol{r}_{\boldsymbol{\theta}}\|_{P_{\boldsymbol{X}},2}=o(\|\boldsymbol{\theta}-\boldsymbol{\theta}_{0}\|).
\]
Hence 
\begin{align*}
\sqrt{n}\left[\widehat{\boldsymbol{\pi}}_{\boldsymbol{\theta}_{0}}(\boldsymbol{\gamma},\boldsymbol{\theta})-\widehat{\boldsymbol{\pi}}_{\boldsymbol{\theta}_{0}}(\boldsymbol{\gamma},\boldsymbol{\theta}_{0})\right] & = \dot{\boldsymbol{\pi}}_{0}(\boldsymbol{\gamma})\sqrt{n}(\boldsymbol{\theta}-\boldsymbol{\theta}_{0})+\sqrt{n}P_{n}\left(\boldsymbol{r}_{\boldsymbol{\theta}}\otimes\boldsymbol{1}_{\boldsymbol{\gamma}}\right)\\
 & \quad + \mathbb{G}_{n}\left(\dot{\boldsymbol{\Phi}}_{0}\otimes\boldsymbol{1}_{\boldsymbol{\gamma}}\right)(\boldsymbol{\theta}-\boldsymbol{\theta}_{0}),
\end{align*}
where $\dot{\boldsymbol{\pi}}_{0}(\boldsymbol{\gamma})=P\left(\dot{\boldsymbol{\Phi}}_{0}\otimes\boldsymbol{1}_{\boldsymbol{\gamma}}\right)$.
By A2(vii), the partition-indicator class is $P_{\boldsymbol{X}}$-Donsker,
while A3(iii) implies $\dot{\boldsymbol{\Phi}}_{0}\in L^{2}(P_{\boldsymbol{X}})$.
Hence the class $\left\{ \dot{\boldsymbol{\Phi}}_{0}\otimes\boldsymbol{1}_{\boldsymbol{\gamma}}:\boldsymbol{\gamma}\in\mathcal{G}\right\} $
is $P$-Donsker by the permanence property for multiplication of a
Donsker class by a fixed square-integrable function; see \citet[Section 2.10]{van1996weak}.
Therefore, $\sup_{\boldsymbol{\gamma}\in\mathcal{G}}\left\Vert \mathbb{G}_{n}\left[\dot{\boldsymbol{\Phi}}_{0}\otimes\boldsymbol{1}_{\boldsymbol{\gamma}}\right]\right\Vert =O_{P}(1)$,
and the empirical-process term involving $\dot{\boldsymbol{\Phi}}_{0}$
is $o_{P}(1)$ uniformly in $\boldsymbol{\gamma}$, since $\|\boldsymbol{\theta}_{n}-\boldsymbol{\theta}_{0}\|=O_{P}(n^{-1/2})$.

For the remainder, first note that the class $\mathcal{R}_{\epsilon}=\left\{ \boldsymbol{x}\mapsto\boldsymbol{r}_{\boldsymbol{\theta}}(\boldsymbol{x}):\boldsymbol{\theta}\in\mathcal{N}_{\epsilon}\right\}$,
where $\boldsymbol{r}_{\boldsymbol{\theta}}=\boldsymbol{\Phi}_{\boldsymbol{\theta}_{0},\boldsymbol{\theta}}-\boldsymbol{\Phi}_{\boldsymbol{\theta}_{0},\boldsymbol{\theta}_{0}}-\dot{\boldsymbol{\Phi}}_{0}(\boldsymbol{\theta}-\boldsymbol{\theta}_{0})$,
is $P_{\boldsymbol{X}}$-Donsker. Indeed, $\{\boldsymbol{\Phi}_{\boldsymbol{\theta}_{0},\boldsymbol{\theta}}:\boldsymbol{\theta}\in\mathcal{N}_{\epsilon}\}$
is $P_{\boldsymbol{X}}$-Donsker by A2(vi), and A3(iii) implies $\dot{\boldsymbol{\Phi}}_{0}\in L^{2}(P_{\boldsymbol{X}})$.
Hence the finite-dimensional linear class 
\[
\left\{ \boldsymbol{x}\mapsto\dot{\boldsymbol{\Phi}}_{0}(\boldsymbol{x})(\boldsymbol{\theta}-\boldsymbol{\theta}_{0}):\boldsymbol{\theta}\in\mathcal{N}_{\epsilon}\right\} 
\]
is $P_{\boldsymbol{X}}$-Donsker. Preservation under finite sums and
differences therefore yields the Donsker property of $\mathcal{R}_{\epsilon}$.
By A2(vii), the partition-indicator class is bounded and $P_{\boldsymbol{X}}$-Donsker;
consequently, the product class $\left\{ \boldsymbol{r}_{\boldsymbol{\theta}}\otimes\boldsymbol{1}_{\boldsymbol{\gamma}}:\boldsymbol{\theta}\in\mathcal{N}_{\epsilon},\ \boldsymbol{\gamma}\in\mathcal{G}\right\} $
is $P$-Donsker by the product-preservation property; see \citet[Section 2.10]{van1996weak}. Now, 
\[
\sqrt{n}P_{n}[\boldsymbol{r}_{\boldsymbol{\theta}_{n}}\otimes\boldsymbol{1}_{\boldsymbol{\gamma}}]=\sqrt{n}P[\boldsymbol{r}_{\boldsymbol{\theta}_{n}}\otimes\boldsymbol{1}_{\boldsymbol{\gamma}}]+\mathbb{G}_{n}[\boldsymbol{r}_{\boldsymbol{\theta}_{n}}\otimes\boldsymbol{1}_{\boldsymbol{\gamma}}].
\]
Uniformly in $\boldsymbol{\gamma}$, 
\[
\sqrt{n}\left\Vert P[\boldsymbol{r}_{\boldsymbol{\theta}_{n}}\otimes\boldsymbol{1}_{\boldsymbol{\gamma}}]\right\Vert \leq\sqrt{n}\|\boldsymbol{r}_{\boldsymbol{\theta}_{n}}\|_{P_{\boldsymbol{X}},2}=o_{P}(1),
\]
where the last equality follows from the remainder condition in A3(iii)
and $\sqrt{n}(\boldsymbol{\theta}_{n}-\boldsymbol{\theta}_{0})=O_{P}(1)$. Moreover, 
\[
\sup_{\boldsymbol{\gamma}\in\mathcal{G}}\|\boldsymbol{r}_{\boldsymbol{\theta}_{n}}\otimes\boldsymbol{1}_{\boldsymbol{\gamma}}\|_{P,2}\leq\|\boldsymbol{r}_{\boldsymbol{\theta}_{n}}\|_{P_{\boldsymbol{X}},2}=o_{P}(1).
\]
Thus, by asymptotic equicontinuity of this Donsker class (equivalently,
by \citet[Lemma 19.24]{van2000asymptotic} applied to the corresponding
random indices), $\sup_{\boldsymbol{\gamma}\in\mathcal{G}}\left\Vert \mathbb{G}_{n}[\boldsymbol{r}_{\boldsymbol{\theta}_{n}}\otimes\boldsymbol{1}_{\boldsymbol{\gamma}}]\right\Vert =o_{P}(1)$. Consequently,
\[
\sup_{\boldsymbol{\gamma}\in\mathcal{G}}\left\Vert \sqrt{n}\left[\widehat{\boldsymbol{\pi}}_{\boldsymbol{\theta}_{0}}(\boldsymbol{\gamma},\boldsymbol{\theta}_{n})-\widehat{\boldsymbol{\pi}}_{\boldsymbol{\theta}_{0}}(\boldsymbol{\gamma},\boldsymbol{\theta}_{0})\right]-\dot{\boldsymbol{\pi}}_{0}(\boldsymbol{\gamma})\sqrt{n}(\boldsymbol{\theta}_{n}-\boldsymbol{\theta}_{0})\right\Vert =o_{P}(1).
\]
Evaluating at $\boldsymbol{\gamma}=\widehat{\boldsymbol{\Gamma}}$
and using the definition of $\widehat{\boldsymbol{\xi}}_{\boldsymbol{\theta}_{0}}$
gives the result.
\end{proof}

\begin{lemma}\label{lem:sample-matrix-consistency} Suppose that $\widetilde{\boldsymbol{\theta}}\to_{P}\boldsymbol{\theta}_{0}$
and $\widehat{\boldsymbol{\Gamma}}\to_{P}\boldsymbol{\Gamma}$. Then
$\widehat{\dot{\boldsymbol{\pi}}}(\widehat{\boldsymbol{\Gamma}},\widetilde{\boldsymbol{\theta}})=\dot{\boldsymbol{\pi}}_{0}(\boldsymbol{\Gamma})+o_{P}(1)$,
and $\widehat{\boldsymbol{\Lambda}}_{0}(\widehat{\boldsymbol{\Gamma}})=\boldsymbol{\Lambda}_{0}(\boldsymbol{\Gamma})+o_{P}(1)$.
Consequently, $\widehat{\boldsymbol{J}}(\widehat{\boldsymbol{\Gamma}},\widetilde{\boldsymbol{\theta}})=\boldsymbol{J}_{0}(\boldsymbol{\Gamma})+o_{P}(1)$,
where $\boldsymbol{J}_{0}(\boldsymbol{\Gamma})=\dot{\boldsymbol{\pi}}_{0}^{\text{\textsc{t}}}(\boldsymbol{\Gamma})\boldsymbol{\Lambda}_{0}^{-2}(\boldsymbol{\Gamma})\dot{\boldsymbol{\pi}}_{0}(\boldsymbol{\Gamma})$.
By A3(vii), $\lambda_{\min}\{\boldsymbol{J}_{0}(\boldsymbol{\Gamma})\}>0$.
Consequently, with probability approaching one, $\widehat{\boldsymbol{J}}(\widehat{\boldsymbol{\Gamma}},\widetilde{\boldsymbol{\theta}})$
is nonsingular and $\widehat{\boldsymbol{J}}^{-1}(\widehat{\boldsymbol{\Gamma}},\widetilde{\boldsymbol{\theta}})=\boldsymbol{J}_{0}^{-1}(\boldsymbol{\Gamma})+o_{P}(1)$.
\end{lemma} 
\begin{proof}
Decompose 
\[
\begin{aligned}
\left\Vert \widehat{\dot{\boldsymbol{\pi}}}(\widehat{\boldsymbol{\Gamma}},\widetilde{\boldsymbol{\theta}})-\dot{\boldsymbol{\pi}}_{0}(\boldsymbol{\Gamma})\right\Vert &\leq\left\Vert \widehat{\dot{\boldsymbol{\pi}}}(\widehat{\boldsymbol{\Gamma}},\widetilde{\boldsymbol{\theta}})-\dot{\boldsymbol{\pi}}(\widehat{\boldsymbol{\Gamma}},\widetilde{\boldsymbol{\theta}})\right\Vert
+\left\Vert \dot{\boldsymbol{\pi}}(\widehat{\boldsymbol{\Gamma}},\widetilde{\boldsymbol{\theta}})-\dot{\boldsymbol{\pi}}(\widehat{\boldsymbol{\Gamma}},\boldsymbol{\theta}_{0})\right\Vert \\
&\quad +\left\Vert \dot{\boldsymbol{\pi}}_{0}(\widehat{\boldsymbol{\Gamma}})-\dot{\boldsymbol{\pi}}_{0}(\boldsymbol{\Gamma})\right\Vert .
\end{aligned}
\]
The first term is $o_{P}(1)$ by the uniform law of large numbers
for the product class $\left\{ \dot{\boldsymbol{\Phi}}_{\boldsymbol{\theta}}\otimes\boldsymbol{1}_{\boldsymbol{\gamma}}:\boldsymbol{\theta}\in\mathcal{N}_{\epsilon},\ \boldsymbol{\gamma}\in\mathcal{G}\right\}$, which
is Glivenko--Cantelli by A3(iv), A2(vii), and the preservation property
for products of a Glivenko--Cantelli class with a bounded Glivenko--Cantelli
class. For the second term, 
\[
\left\Vert \dot{\boldsymbol{\pi}}(\widehat{\boldsymbol{\Gamma}},\widetilde{\boldsymbol{\theta}})-\dot{\boldsymbol{\pi}}(\widehat{\boldsymbol{\Gamma}},\boldsymbol{\theta}_{0})\right\Vert \leq\left\Vert \dot{\boldsymbol{\Phi}}_{\widetilde{\boldsymbol{\theta}}}-\dot{\boldsymbol{\Phi}}_{\boldsymbol{\theta}_{0}}\right\Vert _{P_{\boldsymbol{X}},2}=o_{P}(1),
\]
by the continuity of the derivative in A3(iii). For the third term,
A2(ii), together with the absolute continuity of the integral, gives
$P_{\boldsymbol{X}}(\widehat{\Gamma}_{j}\triangle\Gamma_{j})\to_{P}0$
and hence $\dot{\boldsymbol{\pi}}_{0}(\widehat{\boldsymbol{\Gamma}})-\dot{\boldsymbol{\pi}}_{0}(\boldsymbol{\Gamma})=o_{P}(1)$.
Hence $\widehat{\dot{\boldsymbol{\pi}}}(\widehat{\boldsymbol{\Gamma}},\widetilde{\boldsymbol{\theta}})\to_{P}\dot{\boldsymbol{\pi}}_{0}(\boldsymbol{\Gamma})$.
Next, 
\[
\widehat{\boldsymbol{q}}(\widehat{\boldsymbol{\Gamma}})-\boldsymbol{q}(\boldsymbol{\Gamma})=(P_{n}-P)\boldsymbol{1}_{\widehat{\boldsymbol{\Gamma}}}+P\left[\boldsymbol{1}_{\widehat{\boldsymbol{\Gamma}}}-\boldsymbol{1}_{\boldsymbol{\Gamma}}\right]=o_{P}(1),
\]
by the Glivenko--Cantelli property of the indicator class, which
is VC by A2(vii), and the partition consistency in A2(ii). Therefore,
$\widehat{\boldsymbol{\Lambda}}_{0}(\widehat{\boldsymbol{\Gamma}})\to_{P}\boldsymbol{\Lambda}_{0}(\boldsymbol{\Gamma})$.
Since the components of $\boldsymbol{q}(\boldsymbol{\Gamma})$ are
strictly positive, $\widehat{\boldsymbol{\Lambda}}_{0}^{-2}(\widehat{\boldsymbol{\Gamma}})\to_{P}\boldsymbol{\Lambda}_{0}^{-2}(\boldsymbol{\Gamma})$.
It follows that $\widehat{\boldsymbol{J}}(\widehat{\boldsymbol{\Gamma}},\widetilde{\boldsymbol{\theta}})\to_{P}\boldsymbol{J}_{0}(\boldsymbol{\Gamma})$.

Finally, A3(vii) gives $\lambda_{\min}\{\boldsymbol{J}_{0}(\boldsymbol{\Gamma})\}>0$,
where $\lambda_{\min}\{\boldsymbol{A}\}$ is the minimum eigenvalue
of the matrix $\boldsymbol{A}$. Continuity of the eigenvalues then
gives $\lambda_{\min}\left\{ \widehat{\boldsymbol{J}}(\widehat{\boldsymbol{\Gamma}},\widetilde{\boldsymbol{\theta}})\right\} \to_{P}\lambda_{\min}\{\boldsymbol{J}_{0}(\boldsymbol{\Gamma})\}>0$.
Thus the sample matrix is nonsingular with probability approaching
one, and continuity of matrix inversion yields $\widehat{\boldsymbol{J}}^{-1}(\widehat{\boldsymbol{\Gamma}},\widetilde{\boldsymbol{\theta}})\to_{P}\boldsymbol{J}_{0}^{-1}(\boldsymbol{\Gamma})$. 
\end{proof}
Lemmas~\ref{lem:transformation-equivalence}--\ref{lem:sample-matrix-consistency}
provide the ingredients needed to complete the proof.

\textit{Root-$n$ consistency and first-order representation of the
one-step estimator.} By construction, the one-step estimator satisfies
\[
\widehat{\boldsymbol{\theta}}^{*}=\widetilde{\boldsymbol{\theta}}+\widehat{\boldsymbol{J}}^{-1}(\widehat{\boldsymbol{\Gamma}},\widetilde{\boldsymbol{\theta}})\cdot\widehat{\dot{\boldsymbol{\pi}}}^{\,\text{\textsc{t}}}(\widehat{\boldsymbol{\Gamma}},\widetilde{\boldsymbol{\theta}})\cdot\widehat{\boldsymbol{\Lambda}}_{0}^{-2}(\widehat{\boldsymbol{\Gamma}})\cdot n^{-1/2}\widehat{\boldsymbol{\xi}}_{\widetilde{\boldsymbol{\theta}}}(\widehat{\boldsymbol{\Gamma}},\widetilde{\boldsymbol{\theta}}),
\]
where the matrices on the right-hand side are evaluated at the preliminary
estimator.

Since A3(i) gives $\sqrt{n}(\widetilde{\boldsymbol{\theta}}-\boldsymbol{\theta}_{0})=O_{P}(1)$,
Lemma~\ref{lem:transformation-equivalence} yields
\[
\widehat{\boldsymbol{\xi}}_{\widetilde{\boldsymbol{\theta}}}(\widehat{\boldsymbol{\Gamma}},\widetilde{\boldsymbol{\theta}})=\widehat{\boldsymbol{\xi}}_{\boldsymbol{\theta}_{0}}(\widehat{\boldsymbol{\Gamma}},\widetilde{\boldsymbol{\theta}})+o_{P}(1).
\]
Lemma~\ref{lem:model-probability-expansion} then yields 
\begin{equation*}
\widehat{\boldsymbol{\xi}}_{\widetilde{\boldsymbol{\theta}}}(\widehat{\boldsymbol{\Gamma}},\widetilde{\boldsymbol{\theta}})=\widehat{\boldsymbol{\xi}}_{0}(\widehat{\boldsymbol{\Gamma}})-\dot{\boldsymbol{\pi}}_{0}(\widehat{\boldsymbol{\Gamma}})\cdot\sqrt{n}(\widetilde{\boldsymbol{\theta}}-\boldsymbol{\theta}_{0})+o_{P}(1).
\end{equation*}

Let $\boldsymbol{z}_{\boldsymbol{\gamma}}$ denote the summand in \eqref{eq:marked-process}. The corresponding
class $\mathcal{Z}$ is defined in \eqref{eq:marked-class}. By the partition consistency in A2(ii), for some $K<\infty$, 
\[
P\left\Vert \boldsymbol{z}_{\widehat{\boldsymbol{\Gamma}}}-\boldsymbol{z}_{\boldsymbol{\Gamma}}\right\Vert ^{2}=P\left\Vert \boldsymbol{w}\otimes\left[\boldsymbol{1}_{\widehat{\boldsymbol{\Gamma}}}-\boldsymbol{1}_{\boldsymbol{\Gamma}}\right]\right\Vert ^{2}\leq K\sum_{j=1}^{J}P_{\boldsymbol{X}}(\widehat{\Gamma}_{j}\triangle\Gamma_{j})=o_{P}(1).
\]
Since $\mathcal{Z}$ is $P$-Donsker, \citet[Lemma 19.24]{van2000asymptotic}
implies that $\widehat{\boldsymbol{\xi}}_{0}(\widehat{\boldsymbol{\Gamma}})-\widehat{\boldsymbol{\xi}}_{0}(\boldsymbol{\Gamma})=\mathbb{G}_{n}\left(\boldsymbol{z}_{\widehat{\boldsymbol{\Gamma}}}-\boldsymbol{z}_{\boldsymbol{\Gamma}}\right)=o_{P}(1)$. Continuity in $\mathcal{G}$ of the vector measure defining $\dot{\boldsymbol{\pi}}_{0}(\cdot)$ gives
\[
\dot{\boldsymbol{\pi}}_{0}(\widehat{\boldsymbol{\Gamma}})=\dot{\boldsymbol{\pi}}_{0}(\boldsymbol{\Gamma})+o_{P}(1).
\]
Consequently, 
\[
\widehat{\boldsymbol{\xi}}_{\widetilde{\boldsymbol{\theta}}}(\widehat{\boldsymbol{\Gamma}},\widetilde{\boldsymbol{\theta}})=\widehat{\boldsymbol{\xi}}_{0}(\boldsymbol{\Gamma})-\dot{\boldsymbol{\pi}}_{0}(\boldsymbol{\Gamma})\sqrt{n}(\widetilde{\boldsymbol{\theta}}-\boldsymbol{\theta}_{0})+o_{P}(1).
\]

By Lemma~\ref{lem:sample-matrix-consistency}, $\widehat{\dot{\boldsymbol{\pi}}}(\widehat{\boldsymbol{\Gamma}},\widetilde{\boldsymbol{\theta}})=\dot{\boldsymbol{\pi}}_{0}(\boldsymbol{\Gamma})+o_{P}(1)$,
and the corresponding weighting and information matrices converge
to their population counterparts. In particular, $\widehat{\boldsymbol{J}}^{-1}(\widehat{\boldsymbol{\Gamma}},\widetilde{\boldsymbol{\theta}})=\boldsymbol{J}_{0}^{-1}(\boldsymbol{\Gamma})+o_{P}(1)$
with probability tending to one. It follows immediately that $\sqrt{n}(\widehat{\boldsymbol{\theta}}^{*}-\boldsymbol{\theta}_{0})=O_{P}(1)$. More precisely, 
\begin{align*}
\sqrt{n}(\widehat{\boldsymbol{\theta}}^{*}-\boldsymbol{\theta}_{0})={} & \sqrt{n}(\widetilde{\boldsymbol{\theta}}-\boldsymbol{\theta}_{0})\\
 & +\boldsymbol{J}_{0}^{-1}(\boldsymbol{\Gamma})\dot{\boldsymbol{\pi}}_{0}^{\text{\textsc{t}}}(\boldsymbol{\Gamma})\boldsymbol{\Lambda}_{0}^{-2}(\boldsymbol{\Gamma})\left[\widehat{\boldsymbol{\xi}}_{0}(\boldsymbol{\Gamma})-\dot{\boldsymbol{\pi}}_{0}(\boldsymbol{\Gamma})\sqrt{n}(\widetilde{\boldsymbol{\theta}}-\boldsymbol{\theta}_{0})\right]+o_{P}(1).
\end{align*}
Since $\boldsymbol{J}_{0}(\boldsymbol{\Gamma})=\dot{\boldsymbol{\pi}}_{0}^{\text{\textsc{t}}}(\boldsymbol{\Gamma})\boldsymbol{\Lambda}_{0}^{-2}(\boldsymbol{\Gamma})\dot{\boldsymbol{\pi}}_{0}(\boldsymbol{\Gamma})$,
the term involving the preliminary estimator cancels. Hence 
\begin{equation}
\sqrt{n}(\widehat{\boldsymbol{\theta}}^{*}-\boldsymbol{\theta}_{0})=\boldsymbol{J}_{0}^{-1}(\boldsymbol{\Gamma})\dot{\boldsymbol{\pi}}_{0}^{\text{\textsc{t}}}(\boldsymbol{\Gamma})\boldsymbol{\Lambda}_{0}^{-2}(\boldsymbol{\Gamma})\widehat{\boldsymbol{\xi}}_{0}(\boldsymbol{\Gamma})+o_{P}(1).\label{eq:89}
\end{equation}
Thus, the one-step estimator has the same first-order representation
as the infeasible grouped-data estimator in Theorem~\ref{thm:random-partition}. In particular,
the first-order distribution does not depend on the preliminary estimator.

\textit{Projection representation.} Use the projection matrices defined
in \eqref{eq:projection-matrices}, evaluated at $\boldsymbol{\Gamma}$.
Since $\boldsymbol{A}_{0}^{\text{\textsc{t}}}(\boldsymbol{\Gamma})\boldsymbol{A}_{0}(\boldsymbol{\Gamma})=\boldsymbol{J}_{0}(\boldsymbol{\Gamma})$,
the first-order representation obtained in \eqref{eq:89} implies
\begin{align*}
 & \boldsymbol{\Lambda}_{0}^{-1}(\boldsymbol{\Gamma})\left[\widehat{\boldsymbol{\xi}}_{0}(\boldsymbol{\Gamma})-\dot{\boldsymbol{\pi}}_{0}(\boldsymbol{\Gamma})\sqrt{n}(\widehat{\boldsymbol{\theta}}^{*}-\boldsymbol{\theta}_{0})\right]\\
 & \qquad=\left\{ \boldsymbol{I}_{LJ}-\boldsymbol{A}_{0}(\boldsymbol{\Gamma})\left[\boldsymbol{A}_{0}^{\text{\textsc{t}}}(\boldsymbol{\Gamma})\boldsymbol{A}_{0}(\boldsymbol{\Gamma})\right]^{-1}\boldsymbol{A}_{0}^{\text{\textsc{t}}}(\boldsymbol{\Gamma})\right\} \boldsymbol{\Lambda}_{0}^{-1}(\boldsymbol{\Gamma})\widehat{\boldsymbol{\xi}}_{0}(\boldsymbol{\Gamma})+o_{P}(1)\\
 & \qquad=\boldsymbol{H}_{0}(\boldsymbol{\Gamma})\boldsymbol{\Lambda}_{0}^{-1}(\boldsymbol{\Gamma})\widehat{\boldsymbol{\xi}}_{0}(\boldsymbol{\Gamma})+o_{P}(1).
\end{align*}
Thus, the effect of parameter estimation is the orthogonal projection
onto the complement of the column space of $\boldsymbol{A}_{0}(\boldsymbol{\Gamma})$,
exactly as in Theorem~\ref{thm:random-partition}.

\textit{Asymptotic equivalence and limiting distribution.} Applying
Lemmas~\ref{lem:transformation-equivalence}--\ref{lem:sample-matrix-consistency} once more, now with $\boldsymbol{\theta}_{n}=\widehat{\boldsymbol{\theta}}^{*}$,
which is permitted by the root-$n$ consistency established above,
and using $\widehat{\boldsymbol{\Gamma}}\to_{P}\boldsymbol{\Gamma}$,
gives 
\[
\widehat{X}_{\widehat{\boldsymbol{\theta}}^{*}}^{2}(\widehat{\boldsymbol{\Gamma}},\widehat{\boldsymbol{\theta}}^{*})=\left\Vert \boldsymbol{H}_{0}(\boldsymbol{\Gamma})\boldsymbol{\Lambda}_{0}^{-1}(\boldsymbol{\Gamma})\widehat{\boldsymbol{\xi}}_{0}(\boldsymbol{\Gamma})\right\Vert ^{2}+o_{P}(1).
\]
This is the same first-order representation as that of the infeasible
statistic in Theorem~\ref{thm:random-partition}. Hence the feasible and infeasible statistics
are asymptotically equivalent.

Under $H_{0}$, $\boldsymbol{\Lambda}_{0}^{-1}(\boldsymbol{\Gamma})\widehat{\boldsymbol{\xi}}_{0}(\boldsymbol{\Gamma})\rightarrow_{d}\boldsymbol{Z}_{0}$,
with the covariance structure established in Theorem~\ref{thm:random-partition}. Moreover,
$\boldsymbol{H}_{0}(\boldsymbol{\Gamma})$ is the same projection
matrix as in that theorem. Since $\dot{\boldsymbol{\pi}}_{0}(\boldsymbol{\Gamma})$
has column rank $p$, the projection removes $p$ directions from
the $J(L-1)$-dimensional multinomial tangent space. Therefore, by
Theorem~\ref{thm:random-partition} and the continuous mapping theorem, 
\[
\widehat{X}_{\widehat{\boldsymbol{\theta}}^{*}}^{2}(\widehat{\boldsymbol{\Gamma}},\widehat{\boldsymbol{\theta}}^{*})\rightarrow_{d}\chi_{J(L-1)-p}^{2}.
\]
This completes the proof. \hfill{}$\square$
\medskip{}

\noindent\textbf{Proof of Theorem~\ref{thm:wald}.}

Write $\widehat{\boldsymbol{\theta}}^{\dagger}=\boldsymbol{\theta}_{0}+\boldsymbol{\Delta}_{n}/\sqrt{n}$,
and $\boldsymbol{\Delta}_n=\sqrt{n}(\widehat{\boldsymbol{\theta}}^{\dagger}-\boldsymbol{\theta}_{0})$.
By A4(i), 
\begin{equation}
\boldsymbol{\Delta}_{n}=\frac{1}{\sqrt{n}}\sum_{i=1}^{n}\mathbb{I}^{-1}(\boldsymbol{\theta}_{0})\boldsymbol{s}_{\boldsymbol{\theta}_{0}}(Y_{i},\boldsymbol{X}_{i})+o_{P}(1),\label{eq:t4-mle-linear}
\end{equation}
and therefore $\boldsymbol{\Delta}_{n}=O_{P}(1)$.

\medskip{}

\noindent\textit{Step 1: First-order representation of the frequency
residuals.}

Since $\sqrt{n}(\widehat{\boldsymbol{\theta}}^{\dagger}-\boldsymbol{\theta}_{0})=O_{P}(1)$,
the parameter-replacement and local-expansion arguments used in the
proof of Theorem~\ref{thm:feasible} apply with $\boldsymbol{\theta}_{n}=\widehat{\boldsymbol{\theta}}^{\dagger}$.
Under A3(ii), A3(iii), A3(v), and A3(vi), 
\begin{equation}
\widehat{\boldsymbol{\xi}}_{\widehat{\boldsymbol{\theta}}^{\dagger}}(\widehat{\boldsymbol{\Gamma}},\widehat{\boldsymbol{\theta}}^{\dagger})=\widehat{\boldsymbol{\xi}}_{\boldsymbol{\theta}_{0}}(\widehat{\boldsymbol{\Gamma}},\widehat{\boldsymbol{\theta}}^{\dagger})+o_{P}(1),\label{eq:t4-replacement}
\end{equation}
while A3(iii), A2(vi), and A2(vii) give
\begin{equation}
\widehat{\boldsymbol{\xi}}_{\boldsymbol{\theta}_{0}}(\widehat{\boldsymbol{\Gamma}},\widehat{\boldsymbol{\theta}}^{\dagger})=\widehat{\boldsymbol{\xi}}_{0}(\widehat{\boldsymbol{\Gamma}})-\dot{\boldsymbol{\pi}}_{0}(\widehat{\boldsymbol{\Gamma}})\boldsymbol{\Delta}_{n}+o_{P}(1).\label{eq:t4-model-expansion}
\end{equation}
Combining \eqref{eq:t4-mle-linear}--\eqref{eq:t4-model-expansion},
\begin{equation}
\widehat{\boldsymbol{\xi}}_{\widehat{\boldsymbol{\theta}}^{\dagger}}(\widehat{\boldsymbol{\Gamma}},\widehat{\boldsymbol{\theta}}^{\dagger})=\frac{1}{\sqrt{n}}\sum_{i=1}^{n}\boldsymbol{\zeta}_{\widehat{\boldsymbol{\Gamma}}}(Y_{i},\boldsymbol{X}_{i})+o_{P}(1),\label{eq:t4-if-random}
\end{equation}
where, for $\boldsymbol{\gamma}\in\mathcal{G}$, 
\[
\boldsymbol{\zeta}_{\boldsymbol{\gamma}}(y,\boldsymbol{x})=\boldsymbol{w}(y,\boldsymbol{x})\otimes\boldsymbol{1}_{\boldsymbol{\gamma}}(\boldsymbol{x})-\dot{\boldsymbol{\pi}}_{0}(\boldsymbol{\gamma})\cdot\mathbb{I}^{-1}(\boldsymbol{\theta}_{0})\cdot\boldsymbol{s}_{\boldsymbol{\theta}_{0}}(y,\boldsymbol{x}).
\]

\medskip{}

\noindent\textit{Step 2: Reduction to the limiting partition.}

We show that $\mathcal{Z}_{2}=\left\{ \boldsymbol{\zeta}_{\boldsymbol{\gamma}}:\boldsymbol{\gamma}\in\mathcal{G}\right\} $
is $P$-Donsker. The first component, $\left\{ \boldsymbol{w}\otimes\boldsymbol{1}_{\boldsymbol{\gamma}}:\boldsymbol{\gamma}\in\mathcal{G}\right\}$, is
the marked Donsker class used in Theorem~\ref{thm:random-partition}, by A2(vii), because the
marks are bounded and $J$ and $L$ are fixed.

For the second component, note first that $\dot{\boldsymbol{\Phi}}_{0}\in L^{2}(P_{\boldsymbol{X}})$
by A3(iii), and $\dot{\boldsymbol{\pi}}_{0}(\boldsymbol{\gamma})=P_{\boldsymbol{X}}\left[\dot{\boldsymbol{\Phi}}_{0}\otimes\boldsymbol{1}_{\boldsymbol{\gamma}}\right]$.
Hence, for $\boldsymbol{\gamma}_{1},\boldsymbol{\gamma}_{2}\in\mathcal{G}$,
the Cauchy--Schwarz inequality gives 
\begin{equation}
\left\Vert \dot{\boldsymbol{\pi}}_{0}(\boldsymbol{\gamma}_{1})-\dot{\boldsymbol{\pi}}_{0}(\boldsymbol{\gamma}_{2})\right\Vert \leq C\left\{ \sum_{j=1}^{J}P_{\boldsymbol{X}}(\gamma_{1j}\triangle\gamma_{2j})\right\} ^{1/2}\label{eq:t4-dotpi-continuity}
\end{equation}
for some finite $C$. Thus $\boldsymbol{\gamma}\mapsto\dot{\boldsymbol{\pi}}_{0}(\boldsymbol{\gamma})$
is continuous with respect to the product symmetric-difference semimetric.
Since the indicator class in A2(vii) is $P_{\boldsymbol{X}}$-Donsker,
it is totally bounded in its $L^{2}(P_{\boldsymbol{X}})$ semimetric;
because $J$ is fixed, $\mathcal{G}$ is totally bounded in the corresponding
product semimetric. It follows from \eqref{eq:t4-dotpi-continuity}
that $\left\{ \dot{\boldsymbol{\pi}}_{0}(\boldsymbol{\gamma}):\boldsymbol{\gamma}\in\mathcal{G}\right\} $
is a totally bounded subset of a finite-dimensional Euclidean space.

By A4(i), $\mathbb{I}^{-1}(\boldsymbol{\theta}_{0})\boldsymbol{s}_{\boldsymbol{\theta}_{0}}\in L^{2}(P)$.
Consequently, the class 
\[
\left\{ (y,\boldsymbol{x})\mapsto\dot{\boldsymbol{\pi}}_{0}(\boldsymbol{\gamma})\mathbb{I}^{-1}(\boldsymbol{\theta}_{0})\boldsymbol{s}_{\boldsymbol{\theta}_{0}}(y,\boldsymbol{x}):\boldsymbol{\gamma}\in\mathcal{G}\right\} 
\]
is a finite-dimensional $P$-Donsker class. Preservation under finite
sums and differences therefore implies that $\mathcal{Z}_{2}$ is
$P$-Donsker; see \citet[Section 2.10]{van1996weak}.

Next, A2(ii) and \eqref{eq:t4-dotpi-continuity} imply $\left\Vert \dot{\boldsymbol{\pi}}_{0}(\widehat{\boldsymbol{\Gamma}})-\dot{\boldsymbol{\pi}}_{0}(\boldsymbol{\Gamma})\right\Vert =o_{P}(1)$.
Moreover, since the marks are bounded, $P\left\Vert \boldsymbol{w}\otimes\left[\boldsymbol{1}_{\widehat{\boldsymbol{\Gamma}}}-\boldsymbol{1}_{\boldsymbol{\Gamma}}\right]\right\Vert ^{2}\leq K\sum_{j=1}^{J}P_{\boldsymbol{X}}(\widehat{\Gamma}_{j}\triangle\Gamma_{j})=o_{P}(1)$
by A2(ii). Therefore, $P\left\Vert \boldsymbol{\zeta}_{\widehat{\boldsymbol{\Gamma}}}-\boldsymbol{\zeta}_{\boldsymbol{\Gamma}}\right\Vert ^{2}=o_{P}(1)$.
Since $\mathcal{Z}_{2}$ is $P$-Donsker, \citet[Lemma~19.24]{van2000asymptotic}
gives 
\begin{equation}
\frac{1}{\sqrt{n}}\sum_{i=1}^{n}\left[\boldsymbol{\zeta}_{\widehat{\boldsymbol{\Gamma}}}(Y_{i},\boldsymbol{X}_{i})-\boldsymbol{\zeta}_{\boldsymbol{\Gamma}}(Y_{i},\boldsymbol{X}_{i})\right]=o_{P}(1).\label{eq:t4-random-index}
\end{equation}
Combining \eqref{eq:t4-if-random} and \eqref{eq:t4-random-index},
\begin{equation}
\widehat{\boldsymbol{\xi}}_{\widehat{\boldsymbol{\theta}}^{\dagger}}(\widehat{\boldsymbol{\Gamma}},\widehat{\boldsymbol{\theta}}^{\dagger})=\frac{1}{\sqrt{n}}\sum_{i=1}^{n}\boldsymbol{\zeta}_{\boldsymbol{\Gamma}}(Y_{i},\boldsymbol{X}_{i})+o_{P}(1).\label{eq:t4-if-fixed}
\end{equation}

\medskip{}

\noindent\textit{Step 3: Limiting covariance matrix.}

For fixed $\boldsymbol{\gamma}$, under $H_{0}$, $P\left[\boldsymbol{w}(Y,\boldsymbol{X})\otimes\boldsymbol{1}_{\boldsymbol{\gamma}}(\boldsymbol{X})\right]=\boldsymbol{0}$.
Moreover, 
\[
\operatorname{Var}\left[\boldsymbol{w}(Y,\boldsymbol{X})\otimes\boldsymbol{1}_{\boldsymbol{\gamma}}(\boldsymbol{X})\right]=\boldsymbol{\Lambda}_{0}(\boldsymbol{\gamma})\boldsymbol{\Omega}\boldsymbol{\Lambda}_{0}(\boldsymbol{\gamma}).
\]
The score identity gives 
\begin{equation}
P\left[\left\{ \boldsymbol{w}(Y,\boldsymbol{X})\otimes\boldsymbol{1}_{\boldsymbol{\gamma}}(\boldsymbol{X})\right\} \boldsymbol{s}_{\boldsymbol{\theta}_{0}}^{\text{\textsc{t}}}(Y,\boldsymbol{X})\right]=\dot{\boldsymbol{\pi}}_{0}(\boldsymbol{\gamma}).\label{eq:t4-score-identity}
\end{equation}
Indeed, the left-hand side is the derivative at $\boldsymbol{\theta}_{0}$
of the vector of model-implied cell probabilities with the transformation
parameter fixed at $\boldsymbol{\theta}_{0}$. Using $P\left[\boldsymbol{s}_{\boldsymbol{\theta}_{0}}\boldsymbol{s}_{\boldsymbol{\theta}_{0}}^{\text{\textsc{t}}}\right]=\mathbb{I}(\boldsymbol{\theta}_{0})$,
\eqref{eq:t4-score-identity} yields 
\begin{equation*}
    \operatorname{Var}\left[\boldsymbol{\zeta}_{\boldsymbol{\gamma}}(Y,\boldsymbol{X})\right]=\boldsymbol{\Lambda}_{0}(\boldsymbol{\gamma})\boldsymbol{\Omega}\boldsymbol{\Lambda}_{0}(\boldsymbol{\gamma})-\dot{\boldsymbol{\pi}}_{0}(\boldsymbol{\gamma})\mathbb{I}^{-1}(\boldsymbol{\theta}_{0})\dot{\boldsymbol{\pi}}_{0}^{\text{\textsc{t}}}(\boldsymbol{\gamma}).
\end{equation*}
Therefore, $\operatorname{Var}\left[\boldsymbol{\Lambda}_{0}^{-1}(\boldsymbol{\gamma})\boldsymbol{\zeta}_{\boldsymbol{\gamma}}\right]=\boldsymbol{\Sigma}_{0}(\boldsymbol{\gamma})=\boldsymbol{\Omega}-\boldsymbol{S}_{0}(\boldsymbol{\gamma})$, where 
\[
\boldsymbol{S}_{0}(\boldsymbol{\gamma})=\boldsymbol{B}_{0}(\boldsymbol{\gamma})\mathbb{I}^{-1}(\boldsymbol{\theta}_{0})\boldsymbol{B}_{0}^{\text{\textsc{t}}}(\boldsymbol{\gamma}),\qquad\boldsymbol{B}_{0}(\boldsymbol{\gamma})=\boldsymbol{\Lambda}_{0}^{-1}(\boldsymbol{\gamma})\dot{\boldsymbol{\pi}}_{0}(\boldsymbol{\gamma}).
\]
By the multivariate central limit theorem, \eqref{eq:t4-if-fixed},
A2(ii)--A2(iii), and continuity of $\boldsymbol{\Lambda}_{0}^{-1}(\cdot)$
at $\boldsymbol{\Gamma}$, 
\begin{equation}
\widehat{\boldsymbol{\Lambda}}_{0}^{-1}(\widehat{\boldsymbol{\Gamma}})\widehat{\boldsymbol{\xi}}_{\widehat{\boldsymbol{\theta}}^{\dagger}}(\widehat{\boldsymbol{\Gamma}},\widehat{\boldsymbol{\theta}}^{\dagger})\rightarrow_{d}N_{LJ}\left(\boldsymbol{0},\boldsymbol{\Sigma}_{0}(\boldsymbol{\Gamma})\right).\label{eq:t4-normal}
\end{equation}

\medskip{}

\noindent\textit{Step 4: Consistency of the estimated covariance
matrix and the Wald limit.}

By A3(iv), A2(vii), and the product Glivenko--Cantelli preservation
property, $\widehat{\dot{\boldsymbol{\pi}}}(\widehat{\boldsymbol{\Gamma}},\widehat{\boldsymbol{\theta}}^{\dagger})\to_{P}\dot{\boldsymbol{\pi}}_{0}(\boldsymbol{\Gamma})$.
Also, $\widehat{\boldsymbol{\Lambda}}_{0}(\widehat{\boldsymbol{\Gamma}})\to_{P}\boldsymbol{\Lambda}_{0}(\boldsymbol{\Gamma})$
by A2(ii)--A2(iii). Since $\widehat{\mathbb{I}}(\widehat{\boldsymbol{\theta}}^{\dagger})\to_{P}\mathbb{I}(\boldsymbol{\theta}_{0})$,
we have $\widehat{\boldsymbol{S}}(\widehat{\boldsymbol{\Gamma}})\to_{P}\boldsymbol{S}_{0}(\boldsymbol{\Gamma})$,
$\widehat{\boldsymbol{\Sigma}}(\widehat{\boldsymbol{\Gamma}})\to_{P}\boldsymbol{\Sigma}_{0}(\boldsymbol{\Gamma})$.

It remains to justify the generalized inverse used in the main text.
For brevity, write $\boldsymbol{B}=\boldsymbol{B}_{0}(\boldsymbol{\Gamma})$,
$\boldsymbol{S}=\boldsymbol{S}_{0}(\boldsymbol{\Gamma})$, and $\boldsymbol{\Sigma}=\boldsymbol{\Sigma}_{0}(\boldsymbol{\Gamma})$.
By A4(ii), $\mathbb{I}(\boldsymbol{\theta}_{0})-\boldsymbol{B}^{\text{\textsc{t}}}\boldsymbol{B}\succ0$.
Hence every nonzero eigenvalue of $\boldsymbol{S}=\boldsymbol{B}\mathbb{I}^{-1}(\boldsymbol{\theta}_{0})\boldsymbol{B}^{\text{\textsc{t}}}$
is strictly smaller than one, and $\boldsymbol{I}_{LJ}-\boldsymbol{S}$
is positive definite. Moreover, \eqref{eq:wald-orthogonality} implies
$\boldsymbol{S}\boldsymbol{\Omega}=\boldsymbol{\Omega}\boldsymbol{S}=\boldsymbol{S}$,
so 
\[
\boldsymbol{\Sigma}=\boldsymbol{\Omega}-\boldsymbol{S}=(\boldsymbol{I}_{LJ}-\boldsymbol{S})\boldsymbol{\Omega}=\boldsymbol{\Omega}(\boldsymbol{I}_{LJ}-\boldsymbol{S}).
\]
Since $\boldsymbol{I}_{LJ}-\boldsymbol{S}$ is nonsingular, $\operatorname{null}(\boldsymbol{\Sigma})=\operatorname{null}(\boldsymbol{\Omega})$
and therefore $\operatorname{rank}(\boldsymbol{\Sigma})=J(L-1)$.
Finally, using $\boldsymbol{\Omega}^{2}=\boldsymbol{\Omega}$, 
\[
\begin{aligned}\boldsymbol{\Sigma}(\boldsymbol{I}_{LJ}-\boldsymbol{S})^{-1}\boldsymbol{\Sigma} & =(\boldsymbol{I}_{LJ}-\boldsymbol{S})\boldsymbol{\Omega}(\boldsymbol{I}_{LJ}-\boldsymbol{S})^{-1}(\boldsymbol{I}_{LJ}-\boldsymbol{S})\boldsymbol{\Omega}\\
 & =(\boldsymbol{I}_{LJ}-\boldsymbol{S})\boldsymbol{\Omega}^{2}=\boldsymbol{\Sigma}.
\end{aligned}
\]
Thus $\boldsymbol{\Sigma}^{-}=(\boldsymbol{I}_{LJ}-\boldsymbol{S})^{-1}$
is a generalized inverse, as asserted in \eqref{eq:wald-generalized-inverse}.

Hence, by \eqref{eq:t4-normal}, the preceding consistency results,
Slutsky's theorem, and the continuous mapping theorem, 
\[
\widehat{W}(\widehat{\boldsymbol{\Gamma}})\rightarrow_{d}\chi_{J(L-1)}^{2}.
\]
\hfill{}$\square$

\noindent\textbf{Proof of Theorem~\ref{thm:local-power}.}
Under A5 and (\ref{eq:pi-local-expansion}), a triangular-array central limit theorem gives
\begin{equation}
\boldsymbol\Lambda_0^{-1}(\boldsymbol\Gamma)
\widehat{\boldsymbol\xi}_0(\boldsymbol\Gamma)
\to_d
N_{LJ}\left(
\boldsymbol\Lambda_0^{-1}(\boldsymbol\Gamma)
\boldsymbol\delta^1(\boldsymbol\Gamma),
\boldsymbol\Omega
\right)
\qquad\text{under }P_n.
\label{eq:local-normal}
\end{equation}
The adding-up restriction (\ref{eq:local-adding-up}) implies that the standardized mean
shift belongs to the range of $\boldsymbol\Omega$. The quadratic approximation used in
Theorem~\ref{thm:random-partition}, together with the projection matrices in
(\ref{eq:projection-matrices}), therefore yields
\[
\widehat X^2_{\boldsymbol\theta_0}
\left(\boldsymbol\Gamma,\widehat{\boldsymbol\theta}(\boldsymbol\Gamma)\right)
=
\left\|
\boldsymbol H_0(\boldsymbol\Gamma)
\boldsymbol\Lambda_0^{-1}(\boldsymbol\Gamma)
\widehat{\boldsymbol\xi}_0(\boldsymbol\Gamma)
\right\|^2
+o_{P_n}(1).
\]
As in the proof of Theorem~\ref{thm:random-partition},
$\boldsymbol H_0(\boldsymbol\Gamma)\boldsymbol\Omega
\boldsymbol H_0(\boldsymbol\Gamma)=
\boldsymbol\Omega-\boldsymbol R_0(\boldsymbol\Gamma)$, which is symmetric and idempotent
with rank $J(L-1)-p$. Hence (\ref{eq:local-normal}) gives the noncentral chi-squared limit
with noncentrality parameter (\ref{eq:lambda-projected}).

Finally, contiguity transfers the $o_P(1)$ asymptotic equivalences established under the
null in Theorem~\ref{thm:feasible} to $o_{P_n}(1)$. In particular, replacing the
limiting partition by $\widehat{\boldsymbol\Gamma}$ and the infeasible grouped estimator by
the feasible one-step estimator does not alter the first-order limit. This proves the result.
\hfill{}$\square$

 \bibliographystyle{plainnat}
\bibliography{References}

\end{document}